\documentclass[11pt]{article}

\usepackage{times}
\usepackage[margin=3.5cm]{geometry}

\usepackage{amsmath,amssymb}
\usepackage{graphicx}
\usepackage[colorlinks=True, citecolor=blue]{hyperref}
\usepackage{subfigure}
\makeatletter
\@addtoreset{subfigure}{figure}
\makeatother

\usepackage{notations}

\newcommand{\unn}[1]{\textcolor{magenta}{#1}}

\usepackage{myAlgorithm}
\usepackage[table]{xcolor}
\usepackage{natbib}

\def\figdir{.}

\def\twofig{0.48\textwidth}

\def\p{\wp}
\def\bx{{\bf x}}
\def\tP{\tilde{P}}

\newtheorem{theo}{Theorem}[section]
\newtheorem{prop}[theo]{Proposition}

\begin{document}

\title{On Markov chain Monte Carlo methods for tall data}

\author{
R\'emi Bardenet$^{1,}$\footnote{Corresponding author: \href{mailto:remi.bardenet@gmail.com}{remi.bardenet@gmail.com}}, Arnaud Doucet$^2$, Chris
  Holmes$^2$\\
$^1$ CNRS \& CRIStAL, Universit\'e de Lille, 59651 Villeneuve d'Ascq,
France \\
$^2$ Department of Statistics, University of Oxford, Oxford OX1 3TG,
UK
}
\maketitle

\begin{abstract}
Markov chain Monte Carlo methods are often deemed  too computationally
intensive to be of any practical use for big data applications, and
in particular for inference on datasets containing a large number
$n$ of individual data points, also known as tall datasets. In scenarios
where data are assumed independent, various approaches to scale up
the Metropolis-Hastings algorithm in a Bayesian inference context
have been recently proposed in machine learning and computational
statistics. These approaches can be grouped into two categories: divide-and-conquer
approaches and, subsampling-based algorithms. The aims of this article
are as follows. First, we present a comprehensive review of the existing
literature, commenting on the underlying assumptions and theoretical
guarantees of each method. Second, by leveraging our understanding
of these limitations, we propose an original subsampling-based approach
which samples from a distribution provably close to the posterior
distribution of interest, yet can require less than $\cO(n)$ data
point likelihood evaluations at each iteration for certain statistical
models in favourable scenarios. Finally, we have only been able so far
to propose subsampling-based methods
which display good performance in scenarios where the Bernstein-von
Mises approximation of the target posterior distribution is
excellent. It remains an open challenge to develop such methods in
scenarios where the Bernstein-von Mises approximation is poor.
\end{abstract}

\tableofcontents{}

\section{Introduction}

\label{s:intro} Performing inference on tall datasets, that is datasets
containing a large number $n$ of individual data points, is a major
aspect of the big data challenge. Statistical models, and Bayesian methods in particular, commonly 
demand Markov chain Monte Carlo
(MCMC) algorithms to make inference, yet running MCMC  on such tall datasets is often far too computationally
intensive to be of any practical use. Indeed, MCMC algorithms such
as the Metropolis-Hastings (MH)~algorithm require at each iteration
to sweep over the whole dataset. Frequentist or variational Bayes
approaches are thus usually preferred to a fully Bayesian analysis
in the tall data context on computational grounds. However, they might be difficult to put
in practice or justify in scenarios where the likelihood function is complex;
e.g. non-differentiable \citep{ChHo03}. Moreover, some applications
require precise quantification of uncertainties and a full Bayesian
approach might be preferable in those instances. This is the case
for example for applications from experimental sciences, such as
cosmology \citep{Tro06} or genomics \citep{Wri14}, were such big data problems abound.
 Consequently, much efforts have been devoted
over recent years to develop scalable MCMC algorithms. These approaches
can be broadly classified into two groups: divide-and-conquer approaches
and subsampling-based algorithms. Divide-and-conquer approaches divide
the initial dataset into batches, run MCMC on each batch separately, and then
combine these results to obtain an approximation of the posterior:
Subsampling approaches aim at reducing the number of individual data
point likelihood evaluations necessary at each iteration of the MH
algorithm.

After briefly reviewing the limitations of MCMC for tall data, introducing our notation and two running examples
in Section~\ref{s:MH}, we first review the divide-and-conquer literature
in Section~\ref{s:reviewDivide}. The rest of the paper is devoted
to subsampling approaches. In Section~\ref{s:reviewPseudomarginal},
we discuss pseudo-marginal MH algorithms. These approaches are exact
in the sense that they target the right posterior distribution. In
Section~\ref{s:reviewExact}, we review other exact approaches, before
relaxing exactness in Section~\ref{s:reviewSubsampling}. Throughout,
we focus on the assumptions and guarantees of each method. We also
illustrate key methods on two running examples. Finally, in Section~\ref{s:concentration},
we improve over our so-called confidence sampler in \citep{BaDoHo14},
which samples from a controlled approximation of the target. We demonstrate
these improvements yield significant reductions in computational complexity
at each iteration in Section~\ref{s:experiments}. In particular,
our improved confidence sampler can break the $\cO(n)$ barrier of
number of individual data point likelihood evaluations per iteration
in favourable cases. Its main limitation is the requirement for
cheap-to-evaluate proxies for the log-likelihood, with a known error.
We provide examples of such proxies relying on Taylor expansions. 

All examples can be rerun or modified using the companion IPython
notebook\footnote{The IPython notebook and a static html render of it
  can both be
  found at
  \href{http://www.2020science.net/research/scaling-mcmc-methods}{http://www.2020science.net/research/scaling-mcmc-methods}.}
to the paper, available as supplementary material. 

\section{Bayesian inference, MCMC, and tall data}

\label{s:MH}

In this section, we describe the inference problem of interest and
the associated MH algorithm. We also detail the two running examples
on which we benchmark key methods in Section~\ref{s:reviewPseudomarginal},
\ref{s:reviewExact} and \ref{s:reviewSubsampling}.

\subsection{Bayesian inference}

Consider a dataset 
\begin{equation}
\cX=\{x_{1},...,x_{n}\}\subset\Xset\subset\mathbb{R}^{d},\label{e:defDataset}
\end{equation}
and a parameter space $\Theta$. We assume the data are conditionally
independent with associated likelihood $\prod_{i=1}^{n}p(x_{i}\vert\theta)$
given a parameter value $\theta$ and we denote $\ell(\theta)$ the
associated average log-likelihood 
\begin{equation}
\ell(\theta)=\frac{1}{n}\sum_{i=1}^{n}\log p(x_{i}\vert\theta)=\frac{1}{n}\sum_{i=1}^{n}\ell_{i}(\theta).\label{eq:loglikelihood}
\end{equation}
We follow a Bayesian approach where one assigns a prior $p(\theta)$
to the unknown parameter, so that inference relies on the posterior
distribution 
\begin{equation}
\pi(\theta)=p(\theta\vert x)\propto\gamma(\theta)\triangleq p(\theta)e^{n\ell(\theta)},\label{e:posterior}
\end{equation}
where $\gamma$ denotes an unnormalized version of $\pi$. In most
applications, $\pi$ is intractable and we will focus here on Markov
chain Monte Carlo methods (MCMC; \citealp{RoCa04}) and, in particular,
on the Metropolis-Hastings (MH) algorithm to approximate it.

\subsection{The Metropolis-Hastings algorithm}
\label{ss:MH} 
A standard approach to sample approximately from $\pi(\theta)$
is to use MCMC algorithms. To illustrate the limitation of MCMC in the
tall data context, we focus here on the MH algorithm (\citealp[Chapter 7.3]{RoCa04}). The MH algorithm
simulates a Markov chain $(\theta_{k})_{k\geq0}$ of invariant distribution
$\pi$. Then, under weak assumptions, see e.g. \citep[Theorem 7.32]{DoMoSt14},
the following central limit theorem holds for suitable test functions $h$
\begin{equation}
\sqrt{N_{\text{iter}}}\left[\frac{1}{N_{\text{iter}}}\sum_{k=0}^{N_{\text{iter}}}h(\theta_{k})-\int h\left(\theta\right)\pi(\theta)d\theta\right]\rightarrow\cN(0,\sigma_{\text{lim}}^{2}(h)),\label{e:CLT}
\end{equation}
where convergence is in distribution.

\setlength{\algowidth}{\columnwidth} \addtolength{\algowidth}{-0in}
\setlength{\algoitemsep}{3pt} \global\long\def\algotab{\hspace{0.2\labelsep}}

\begin{figure}[!ht]
\centerline{ 
\scalebox{0.9}{
\begin{algorithm}{$\Algo{MH}\big(\gamma(\cdot),\, q(\cdot\vert\cdot),\,\theta_{0},\, N_{\text{iter}}\big)$}
\vspace{-.1cm}
\Aitem \For $k\setto1$ \To $N_{\text{iter}}$ 
\Aitem \mtt $\theta\setto\theta_{k-1}$, 
\label{ai:theta}
\Aitem \mtt $\theta'\sim q(.\vert\theta)$,
\label{ai:thetaPrime}
\Aitem \mtt $u\sim\cU_{(0,1)}$
\Aitem \mtt $\alpha(\theta,\theta')\setto \frac{\gamma(\theta')}{\gamma(\theta)} \times
\frac{q(\theta\vert\theta')}{q(\theta'\vert\theta)}$
\label{ai:alpha}
\Aitem \mtt \If $u < \alpha(\theta,\theta')$
\label{ai:acceptanceBeginning} 
\Aitem \mtttt $\theta_{k}\setto\theta'$
\mtt \algoremark{\unn{Accept}} 
\Aitem \mtt \Else $\theta_{k}\setto\theta$
\mtt \algoremark{\unn{Reject}} \label{ai:acceptanceEnd} 
\Aitem
\Return $(\theta_{k})_{k=1,\dots,N_{\text{iter}}}$
\end{algorithm}
}
}
\caption{The pseudocode of the MH algorithm targeting the distribution $\pi$.
Note that $\pi$ is only involved in ratios, so that one only needs
to know an unnormalized version $\gamma$ of $\pi$.}
\label{f:MH} 
\end{figure}

%

The pseudocode of MH targeting a generic distribution $\pi$ is given
in Figure~\ref{f:MH}. In the case of Bayesian inference with independent
data \eqref{e:posterior}, Step~\ref{ai:alpha} is equivalent to
setting 
\begin{eqnarray}
\log \alpha(\theta,\theta') & = & \log\left[\frac{p\left(\theta'\right)}{p\left(\theta\right)}\frac{q(\theta\vert\theta')}{q(\theta'\vert\theta)}\right]+n\left[\ell(\theta')-\ell(\theta)\right].\label{e:MHAcceptanceRatio}
\end{eqnarray}
When the dataset is tall ($n\gg1$), evaluating the log likelihood
ratio in \eqref{e:MHAcceptanceRatio} is too costly an operation and
rules out the applicability of such a method. As we shall see, two
possible options are to either divide the dataset into tractable batches, or
approximate the acceptance ratio in \eqref{e:MHAcceptanceRatio} using
only part of the dataset.

\subsection{Running examples}

We will evaluate some of the described approaches on two illustrative simple running
examples. We fit a one-dimensional normal distribution $p(\cdot\vert\mu,\sigma)=\cN(\cdot\vert\mu,\sigma^{2})$
to $10^{5}$ i.i.d. points drawn according to $X_{i}\sim\cN(0,1)$
and lognormal observations $X_{i}\sim\log\cN(0,1)$, respectively. The latter example illustrates
a misspecification of the model. We assign a flat prior $p(\mu,\log\sigma) \propto 1$.
For all algorithms, we start the chain at the maximum a posteriori (MAP) estimate. The MH proposal is an isotropic Gaussian random walk, whose
stepsize is first set proportional to $1/\sqrt{n}$ and then adapted
during the first $1\,000$ iterations so as to reach $50\%$ acceptance.
When applicable, we also display the number of likelihood evaluations
per iteration, and compare it to the $n$ evaluations required at
each iteration by the MH algorithm.

\begin{figure}
\subfigure[Chain histograms, $X_{i}\sim\cN(0,1)$]{
\includegraphics[width=\twofig]{\figdir/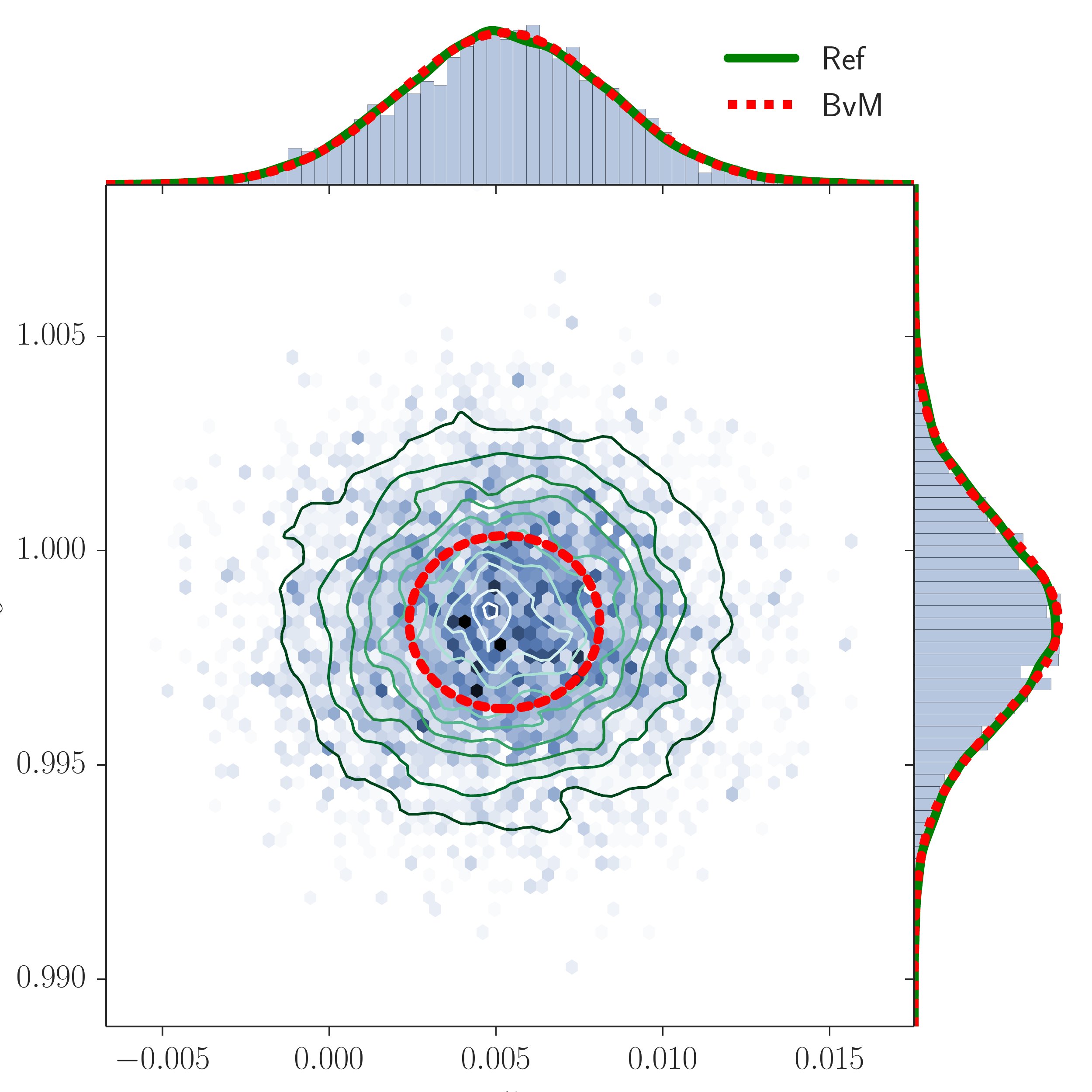}
\label{f:resultsMH:Gaussian:histos}
}
\subfigure[Chain histograms, $X_{i}\sim\log\cN(0,1)$]{
\includegraphics[width=\twofig]{\figdir/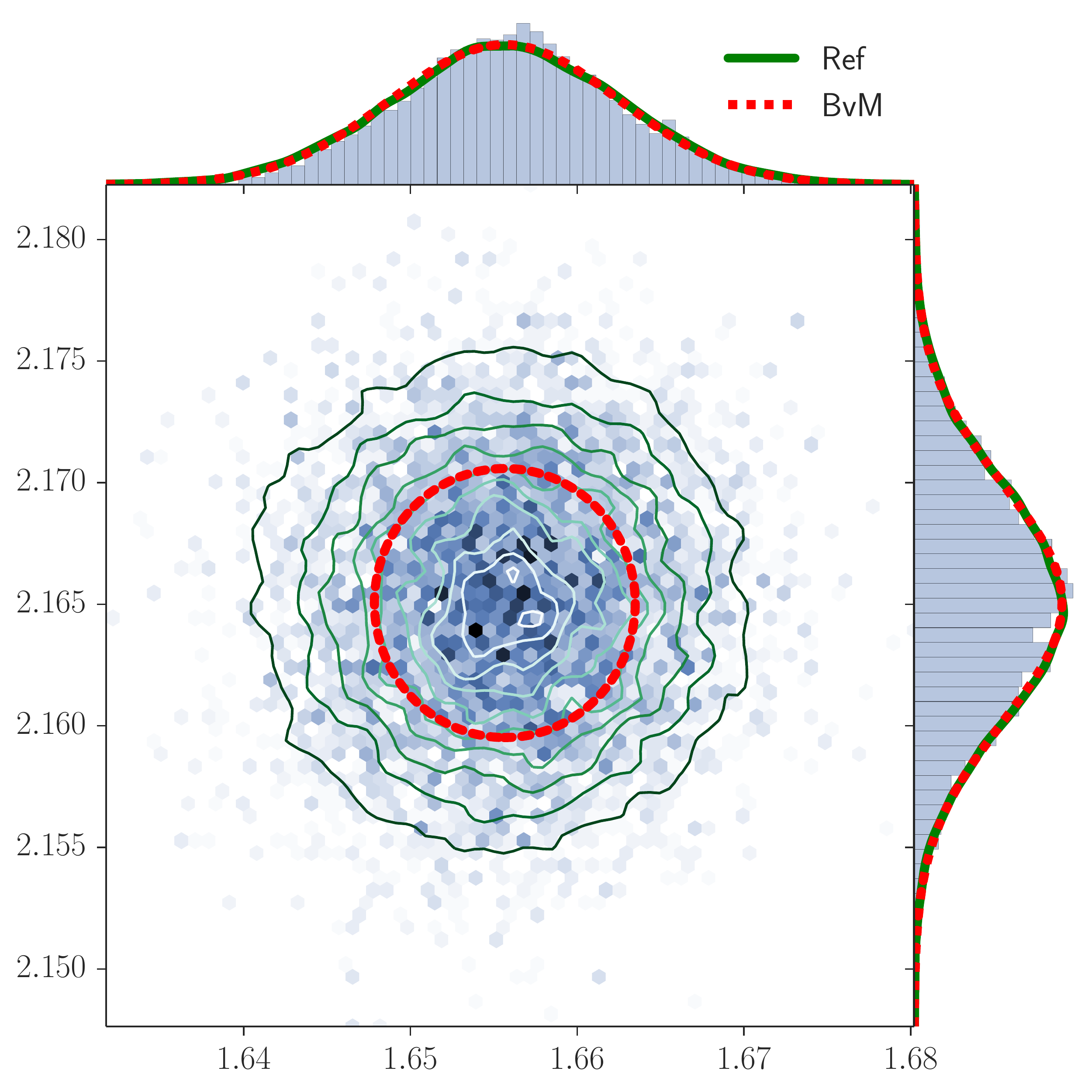}
\label{f:resultsMH:logNormal:histos}
}\\
\subfigure[Autocorr. of $\log\sigma$, $X_{i}\sim\cN(0,1)$]{
\includegraphics[width=\twofig]{\figdir/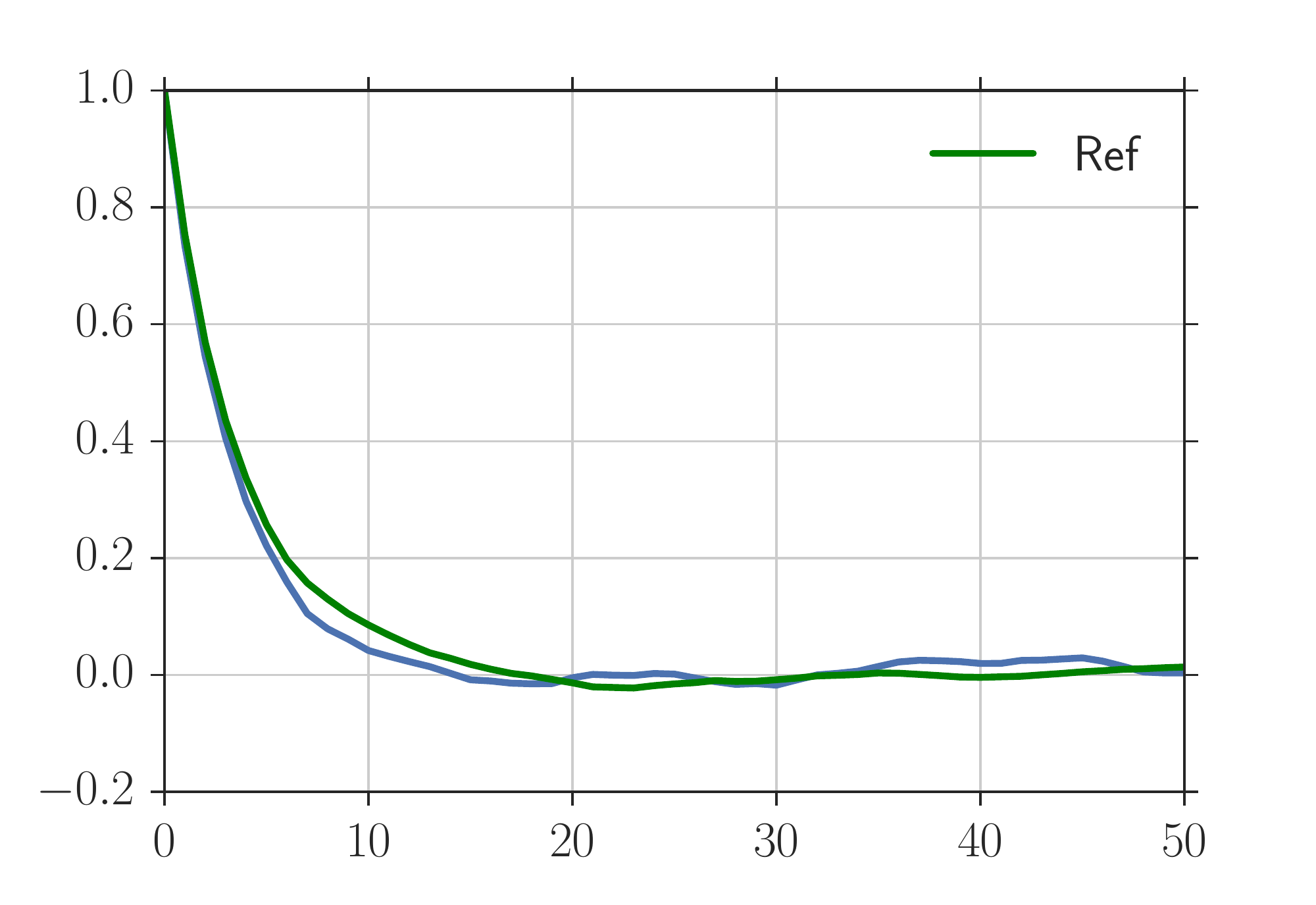}
\label{f:resultsMH:Gaussian:autocorr}
}
\subfigure[Autocorr. of $\log\sigma$, $X_{i}\sim\log\cN(0,1)$]{
\includegraphics[width=\twofig]{\figdir/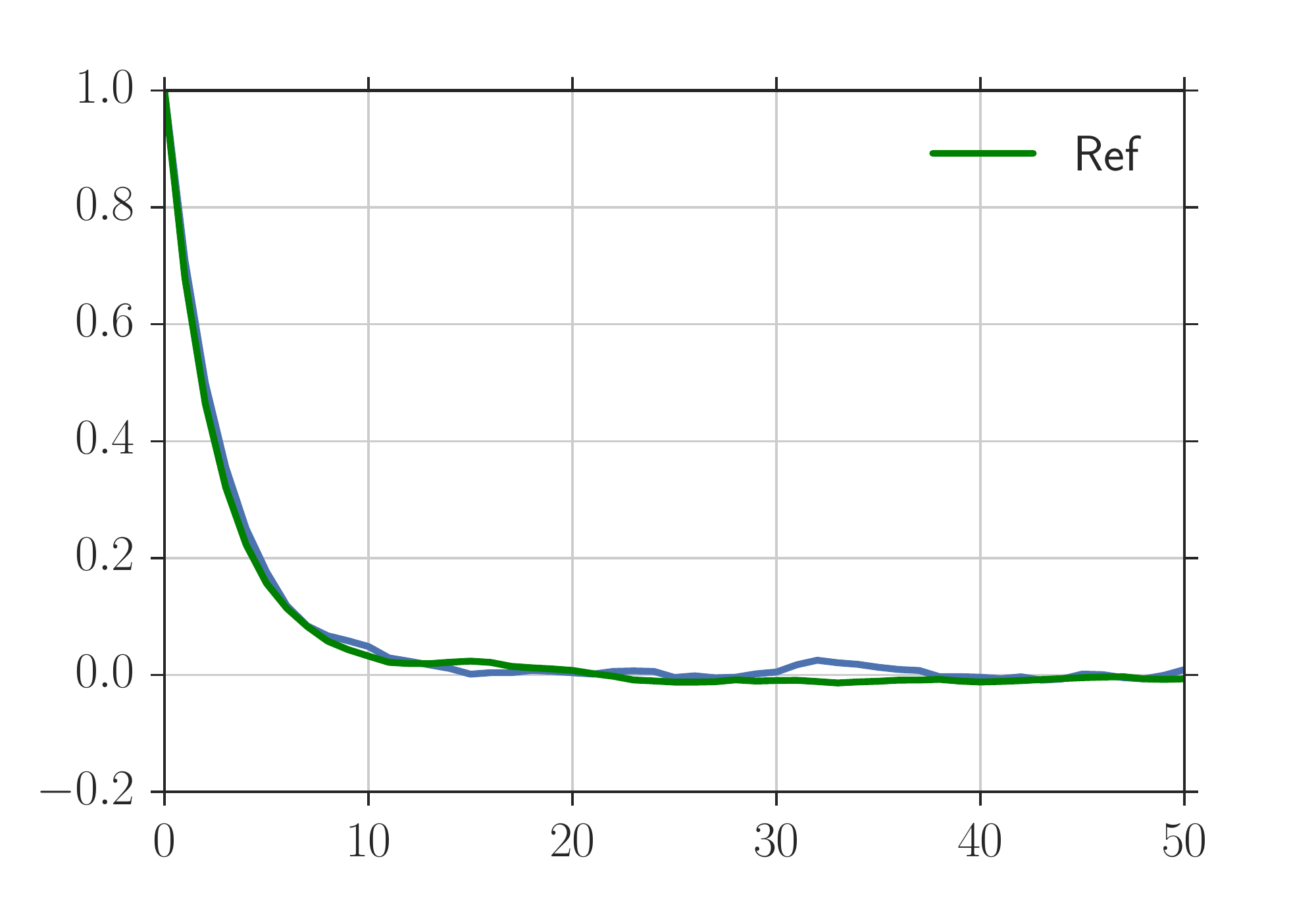}
\label{f:resultsMH:logNormal:autocorr}
}
\caption{Results of 10\,000 iterations of vanilla MH fitting a Gaussian
  model to one-dimensional Gaussian and lognormal synthetic data, on the left and right panel,
  respectively. Figures~\ref{f:resultsMH:Gaussian:histos} and
  \ref{f:resultsMH:logNormal:histos} show the chain histograms, joint
  and marginals; the x-axis corresponds to the mean of the fitted
  Gaussian, the y-axis to the standard deviation. We have superimposed a kernel density estimator of
  a long MH chain for reference in green and the Bernstein-von Mises
  Gaussian approximation in
  red. Figures~\ref{f:resultsMH:Gaussian:autocorr} and
  \ref{f:resultsMH:logNormal:autocorr} show the marginal autocorrelation of
  $\log\sigma$ in blue. The green curves are baselines that correspond to the long MH reference
  run depicted in green in the top panel; although the green
  autocorrelation functions are of limited interest
  when comparing them to vanilla MH, we use them as reference in all similar later figures.}
\label{f:resultsMH}
\end{figure}

In Figure~\ref{f:resultsMH}, we illustrate the results of $10\,000$
iterations of vanilla MH on each of the two datasets. MH does well,
as the posterior coincides with that of a longer reference run  of
$50\,000$ iterations in each case, and the autocorrelations show a fast exponential decrease. The
Bernstein-von Mises approximation \cite[Chapter 10.2]{Vaa00}, a Gaussian
centered at the true value, with covariance minus the scaled inverse
Fisher information, is a very good approximation to the posterior in both
cases. We are thus in simple cases of heavy concentration of the posterior,
where subsampling should help a lot if it is to be of any help in
tackling tall data problems.

%
%
%


\section{Divide-and-conquer approaches}

\label{s:reviewDivide} A natural way to tackle tall data problems
is to divide the data into batches, run MH on each batch separately,
and then combine the results.

\subsection{Representing the posterior as a combination of batch posteriors}

Assume data $\cX$ are divided in $B$ batches $\bx_{1},\dots,\bx_{B}$.
Relying on the equality 
\begin{eqnarray}
p(\theta\vert\cX)\propto\prod_{i=1}^{B}p(\theta)^{1/B}p(\bx_{i}\vert\theta),\label{e:subsetPosteriors}
\end{eqnarray}
\cite{HuGe05} propose to combine the batch posterior approximations
using Gaussian approximations or importance sampling. \cite{ScBlBo13}
propose to average samples across batches, noting this is exact under
Gaussian assumptions. \cite{NeWaXi14} propose to
run an MCMC chain on each batch $\bx_{i}$ targeting an artificial
batch posterior 
\[
\pi_{i}(\theta)\propto p(\theta)^{1/B}p(\bx_{i}\vert\theta),
\]
fit a smooth approximation to each batch posterior, and multiply them.
These methods are however theoretically justified only when batch posteriors are
Gaussian, or when the size \textit{of each batch} goes to infinity, to
guarantee that the used smooth approximation of each batch posterior is accurate.

There are few results available on how the properties of combined estimators
scale with the number of batches $B$. \cite{NeWaXi14} fit a kernel
density estimator to the samples of each batchwise chain, and multiply
the resulting kernel density estimators. A sample from this mixture
approximation to $\pi$ is then obtained through an additional MCMC
step. Under simplifying assumptions (all MCMC chains are assumed being
independent draws from their targets, for example), a bound on the
MSE of the final estimator is obtained. However, this bound explodes
as the kernel bandwidth goes to zero, and more importantly, it is
exponential in the number of batches $B$. In a tall data context,
the number of batches is expected to grow with $n$ to ensure that
the size of each batch is less than $\cO(n)$. Thus, the proposed
bound is currently not informative for tall data.

As pointed out by \cite{WaDuSub}, if the supports of the $\pi_i$ are
almost disjoint, then the product of their approximations will be a poor approximation to
$\pi$. To improve the overlap between the approximations of the $\pi_i$'s, 
\cite{WaDuSub} propose to replace the posterior in \eqref{e:subsetPosteriors}
by the product of the Weierstrass transforms of each batch
posterior. When the approximation of $\pi_i$ is an empirical measure,
its Weierstrass transform corresponds to a kernel density
estimator. The product of the
Weierstrass transforms can be interpreted as the marginal distribution
of an extended distribution on $\Theta^{B+1}$, where the first $B$
copies of $\theta$ are associated to the $B$ batches, and the
remaining copy is conditionally Gaussian around a weighted mean of
the first $B$ copies. Unfortunately, sampling from the posterior of this artificial model is
difficult when one only has access to approximate samples of each
$\pi_i$. 

Although it is not strictly speaking a Monte Carlo method, we note
that  \cite{XLTZZ14} and \cite{GVJRCCSub} propose an expectation-propagation-like algorithm that similarly
tackles the issue of disjoint approximate batch posterior
supports. Each batch of data points is represented by its individual
likelihood times a {\it cavity} distribution. The cavity distribution is itself the
product of the prior and a number of terms that represent the
contributions of other batches to the likelihood. The algorithm
iterates between 1) simulating from each batchwise
likelihood times a batch-specific cavity distribution, and 2) fitting each
batch-specific cavity component. Again, while these
approaches are computationally feasible and appear to perform well
experimentally, it is difficult to assert the characteristics of the
proposed approximation of the posterior and there is no convergence
guarantee available for this iterative algorithm. 

\subsection{Replacing the posterior by a geometric combination of
  batch posteriors}

Another avenue of research consists in avoiding multiplying the batch
posteriors by replacing the target by a different combination of the latter. 

By introducing a suitable metric on the space of probability measures such as the Wasserstein metric, it is
for example possible to define the barycenter or the median of a set of
probability measures. \cite{MSLD14} propose to output the median of the
batch posteriors, while \cite{SCTDSub} use the Wasserstein barycenter, which can be computed efficiently in practice 
using the techniques developed by \cite{CuDo14}. While this idea has
some appeal, the statistical meaning of these median or mean measures is unclear, and the robustness of the median
estimate advocated in \citep{MSLD14} may also be a drawback, as in some circumstances valuable
information contained in some batches may be lost.

To conclude, divide-and-conquer techniques appear as a natural approach to handle tall data. 
However, the crux is how to efficiently combine the batch posterior
approximations. The main issues are that the batch posterior
approximations potentially have disjoint supports, that the
multiplicative structure of the posterior \eqref{e:posterior} leads to
poor scaling with the number of batches, that theoretical guarantees
are often asymptotic in the batch size, and that cheap-to-sample
combinations of batch posteriors are difficult to interpret.


\section{Exact subsampling approaches: Pseudo-marginal MH}

\label{s:reviewPseudomarginal} 
\textit{Pseudo-marginal} MH \citep{LiLiSl00, Bea03,AnRo09}
is a variant of MH, which relies on unbiased estimators of an unnormalized version of the target.
Pseudo-marginal MH is useful to help understand several potential
approaches to scale up MCMC.  We start by describing pseudo-marginal MH in Section~\ref{ss:pseudomarginal}.
Then, we present two pseudo-marginal approaches to tall data in Section~\ref{ss:pseudomarginalSubsampling}
and Section \ref{ss:firefly}.

\subsection{Pseudo-marginal Metropolis-Hastings}
\label{ss:pseudomarginal}  
Assume that instead of being able to
evaluate $\gamma(\theta)$, we have access to an unbiased, almost-surely
\textit{non-negative} estimator $\hat{\gamma}(\theta)$ of the unnormalized
target $\gamma(\theta)$. Pseudo-marginal MH substitutes a realization of 
$\hat{\gamma}(\theta')$ to $\gamma(\theta')$ in
Step~\ref{ai:alpha}. Similarly, it replaces $\gamma(\theta)$ in Step~\ref{ai:alpha} by the realization
of $\hat{\gamma}(\theta)$ that was computed when the parameter value $\theta$ was last proposed.
Pseudo-marginal MH is of considerable practical importance,
with applications such as particle marginal MH \citep{AnDoHo10} and
MCMC versions of the approximate Bayesian computation paradigm
\citep{MPRR12}. It is thus worth investigating its use in the context of tall data problems.

The possibility to use an unbiased estimator of $\gamma$ comes
at a price: first, the asymptotic variance $\sigma_{\text{lim}}^{2}$
in \eqref{e:CLT} of an MCMC estimator based on a pseudo-marginal
chain will always be larger than that of an estimator based on the
underlying ``marginal'' MH \citep{AnViToApp}. Second, the qualitative
properties of the underlying MH may not be preserved, meaning that
the rate of convergence to the invariant distribution may go from
geometric to subgeometric, for instance; see \cite{AnRo09} and \cite{AnViToApp}
for a detailed discussion. In practice, if the variance of $\hat{\gamma}(\vartheta)$
is large for some value $\vartheta\in\Theta$, then an MH move to
$\vartheta$ might be accepted while $\hat{\gamma}(\vartheta)$ largely
overestimates $\gamma(\vartheta)$. In that case, it is difficult
for the chain to leave $\vartheta$, and pseudo-marginal MH chains
thus tend to get stuck if the variance of the involved estimators
is not controlled. When some tunable parameter allows to control this
variance, \cite{DPDK15} show that, in order to minimize the variance
of MCMC estimates for a fixed computational complexity, the variance
of the log-likelihood estimator should be kept around $1.0$ when
the ideal MH having access to the exact likelihood generates quasi-i.i.d samples from $\pi$;
or set to around $3.0$ when it exhibits very large integrated autocorrelation times. 
In practice, the integrated autocorrelation times of averages under the ideal MH are unknown as this algorithm cannot be implemented.
In this common scenario, \cite{DPDK15} recommend keeping the variance
around 1.5 as this is a value which ensures a small penalty in performance
even in scenarios where 1.0 or 3.0 are actually optimal. They also
show that the penalty incurred for having a variance too small (i.e.
inferior to 0.2) or too large (i.e. superior to 10) is very large. When
mentioning pseudo-marginal MH algorithms, we will thus comment on
the variance of the logarithm of the involved estimators
$\hat{\gamma}(\theta)$, or, if not available, of their relative variance.

\subsection{Unbiased estimation of the likelihood using unbiased estimates of
the log-likelihood}
\label{ss:pseudomarginalSubsampling}

As described in Section~\ref{ss:pseudomarginal},
pseudo-marginal MH requires an almost-surely nonnegative unbiased
estimator $\hat{\gamma}(\theta)$ of the unnormalized posterior at
$\theta$, for any $\theta$ in $\Tset$. It is easy to check that,
by sampling $x_{1}^{*},\dots,x_{t}^{*}$ from the dataset $\cX$ with
or without replacement, we obtain the following unbiased estimator
of the log-likelihood $n\ell(\theta)$ 
\begin{equation}
n\hat{\ell}(\theta)=\frac{n}{t}\sum_{i=1}^{t}\log p(x_{i}^{*}\vert\theta).\label{eq:unbiasedestimateloglikelihood}
\end{equation}
We denote by $\hat{\ell}(\theta)$ the subsampling estimate of the
average log-likelihood and denote by $\sigma_{t}(\theta)^{2}$ its
variance. Obviously, exponentiating
\eqref{eq:unbiasedestimateloglikelihood} does not provide an unbiased estimate of the likelihood $e^{n\ell(\theta)}.$
However, an interesting question is whether one can design a procedure
which outputs an unbiased, almost-surely nonnegative estimate of $e^{n\ell(\theta)}$ using
unbiased estimates of $n\ell(\theta)$ such as $n\hat{\ell}(\theta)$.
Without making any further assumption about $n\hat{\ell}(\theta)$,
it was recently shown by \cite{JaThSub} that it is not possible.
However, this can be done if one further assumes, for instance, that
there exists $a(\theta)$ such that $\ell_{i}(\theta)>a(\theta)$ for all $i$, see
\citep[Section
3.1]{JaThSub} who rely on a technique by \cite{RhGl13} generalizing \citep{BhKe85}.
 Unfortunately, as we shall see, the resulting estimator $\hat{\gamma}(\theta)$
typically has a very large relative variance, resulting in very poor performance
of the associated pseudo-marginal chain. 

We apply \cite[Theorem 1]{RhGl13}
to build an unbiased non-negative estimator of
$\gamma(\theta)/p(\theta)$, which is equivalent to defining $\hat{\gamma}(\theta)$.
For $j\geq1$, let 
\begin{equation}
D_{j}^{*}=\frac{n}{t}\sum_{i=1}^{t}\log p(x_{i,j}^{*}\vert\theta)-n a(\theta),\label{e:unbiasedBuildingBlock}
\end{equation}
be an unbiased estimator of $n\left(\ell(\theta)-a(\theta)\right)$,
where the $x_{i,j}^{*}$'s are drawn with replacement from $\cX$
for each $i$, and are further independent across $j$. In \cite[Section
3.1]{JaThSub}, $N$ is an integer-valued random variable whose tails do
not decrease too fast, in the sense that $\mathbb{P}(N\geq k)\geq C(1+\eps)^{-k}$. To ease computations, we take
$N$ to be geometric with parameter $\eps/(1+\eps)$. This corresponds to the
lightest tails allowed by \cite[Section 3.1]{JaThSub}, since
$\mathbb{P}(N\geq k)=(1+\eps)^{-k}$. Finally, let
\begin{equation}
Y\defeq e^{na(\theta)}\left[1+\sum_{k=1}^{N}\frac{1}{\mathbb{P}(N\geq k)}\frac{1}{k!}\prod_{j=1}^{k}D_{j}^{*}\right].
\label{e:defRheeAndGlynn}
\end{equation}

By \cite[Theorem 1]{RhGl13}, $Y$ is a non-negative unbiased estimator of the
likelihood $e^{n\ell(\theta)}$. As mentioned in
Section~\ref{ss:pseudomarginal}, it is crucial, if we want to plug
$\hat{\gamma}(\theta) = Y\times p(\theta)$ in a pseudo-marginal
algorithm, to control the variance of its logarithm. The variance of $\log Y$
is difficult to compute, so we use here the relative variance of $Y$
as a proxy.

\begin{prop}
\label{p:relativeVarianceRheeAndGlynn}
Let $\theta\in\Theta$ and $Y$ be the almost surely non-negative estimator of
$e^{n\ell(\theta)}$ defined in \eqref{e:defRheeAndGlynn}. Then its relative variance
satisfies
\begin{equation}
\frac{\Var Y}{e^{2n\ell(\theta)}} \geq \frac{e^{-2n(\ell(\theta)-a(\theta)) + 2n\sqrt{(1+\eps)[\sigma_{t}(\theta)^{2}+(\ell(\theta)-a(\theta))^{2}]}}}{n\sqrt{(1+\eps)[\sigma_{t}(\theta)^{2}+(\ell(\theta)-a(\theta))^{2}]}} + \cO(1).
\label{e:relativeVarianceRheeAndGlynn}
\end{equation}
\end{prop}

The proof of Proposition~\ref{p:relativeVarianceRheeAndGlynn} can be
found in Appendix~A. We can interpret
\eqref{e:relativeVarianceRheeAndGlynn} as follows: in order for the relative
variance of $Y$ not to increase exponentially with $n$, it is
necessary that $n\sigma_t(\theta)$
is of order $1$. But $\sigma_t(\theta)$ if of order $t^{-1/2}$, so that the batchsize $t$ would have to
be of order $n^2$, which is impractical. It is also necessary that
$\sqrt{1+\eps}$ is of order $1+n^{-1}$ to control
the term in $(\ell(\theta)-a(\theta))$. This means that $\eps$ should
be taken of order $n^{-1}$, but then the mean $(1+\eps)\eps^{-1}$ of the geometric
variable $N$ will be of order $n$. This entails that the number of
terms in the randomly truncated series \eqref{e:defRheeAndGlynn}
should be of order $n$, which defeats the purpose of using this estimator.

Hence for the reasons outlined in Section~\ref{ss:pseudomarginal},
we expect the pseudo-marginal MH relying on $Y$ to be highly inefficient.
Indeed, we have not been able to obtain reasonably mixing chains even
on our Gaussian running example. We have experimented with various
choices of $\eps$, and with various values of $t$, but none yielded
satisfactory results. We conclude that this approach is not a viable
solution to MH for tall data.

We note that \cite{StSeGiSub} have recently proposed a different way to exploit the methodology of \cite{RhGl13} in the context of tall
data. However, their methodology does not provide unbiased estimates of the posterior expectations of interest.
It only provides unbiased estimates of some biased MCMC estimates of
these expectations, these MCMC estimates corresponding indeed to
running an MCMC kernel on the whole dataset for a finite number of
iterations. \cite{StSeGiSub} suggest that it might be possible to combine
their algorithm with the recent scheme of \cite{GlRh14} to obtain
unbiased estimates of the posterior expectations. It is yet unclear
whether this could be achieved under realistic assumptions on the MCMC
kernel.

\subsection{Building $\hat{\gamma}(\theta)$ with auxiliary variables}
\label{ss:firefly} 

In \citep{MaAd14}, the authors propose an alternative
MCMC to sample from $\pi$ which, similarly to the method described previously,
only requires evaluating the likelihood of a subset of the data at
each iteration. Assume a bound $\ell_{i}(\theta)\geq b_{i}(\theta)$
is available for each $i$ and $\theta$. For simplicity, we further assume that $b_i(\theta) =
b(\theta,x_i)$ only depends on $i$ through $x_i$. This is the case in
the experiments of \citep{MaAd14}, as well as ours. Note also that in Section~\ref{ss:pseudomarginalSubsampling},
we used a bound that was uniform in the data index $i$; we could have
used similarly a non-uniform bound, but this would have made the
derivation of Proposition~\ref{p:relativeVarianceRheeAndGlynn}
unnecessarily heavy. 

As noted in \citep{MaAd14},
we can then define the following extended target $\tpi$ distribution
on $\Theta\times\{0,1\}^{n}$ 
\begin{eqnarray}
\tpi(\theta,z) & \propto & p(\theta)\prod_{i=1}^{n}\left[\exp(\ell_{i}(\theta)-\exp(b_{i}(\theta))\right]^{z_{i}}\exp(b_{i}(\theta))^{1-z_{i}}\nonumber \\
 & = & p(\theta)\prod_{i=1}^{n}\exp(b_{i}(\theta))\prod_{i=1}^{n}\left[\exp(\ell_{i}(\theta)-b_{i}(\theta))-1\right]^{z_{i}}.\label{e:fireflyTarget}
\end{eqnarray}
This distribution satisfies two important features: it admits $\pi(\theta)$
as a marginal distribution, and its pointwise evaluation only requires
to evaluate $\ell_{i}(\theta)$ for those $i$'s for which $z_{i}=1$.
Note that evaluating $\tpi(\theta,z)$ however requires to evaluate $\prod_{i=1}^{n}\exp(b_{i}(\theta))$,
and the bounds $b_{i}(\theta)$ thus must be chosen so that this computation
is cheap. This is the case for the lower bound of the logistic regression
log-likelihood model discussed in \citep{MaAd14}, which is a quadratic form
in $t_{i}\theta^{T}x_{i}$, where $t_{i}$ is the $\pm1$ label of
datum $x_{i}$. The idea of replacing the evaluation of the target
by a Bernoulli draw and the evaluation of a lower bound has been exploited previously; see e.g. \citep{Mak05}.

Any MCMC sampler could be used to sample from $\tpi$. \cite{MaAd14}
propose an MH-within-Gibbs sampler that leverages the known conditional
$\tpi(z\vert\theta)$. The expected cost of one conditional MH iteration
on $\theta$ at equilibrium, that is the average number of indices
$i$ such that $z_{i}=1$, is $\cO(n)$, and the constant is related
to the expected relative tightness of the bound, see \cite[Section 3.1]{MaAd14}.
The number of likelihood evaluations for an update of $z$ conditional
on $\theta$ is explicitly controlled in \citep{MaAd14} by either
specifying a maximum number of attempted flips, or implicitly specifying
the fraction of flips to $1$. 

The authors of \cite{MaAd14} remarked that their methodology is related
to pseudo-marginal techniques but did not elaborate. We show here
how it is indeed possible to exploit the extended target distribution
$\tpi$ in (\ref{e:fireflyTarget}) to obtain an unbiased estimate
of an unnormalized version of $\pi$. More precisely, we have 
\[
p(x_{i}\vert\theta)=\sum_{z_{i}\in\left\{ 0,1\right\} }p\left(x_{i},z_{i}\vert\theta\right)
\]
where $p\left(z_{i}\vert\theta,x_{i}\right)=\left\{ 1-\exp(b_{i}(\theta)-\ell_{i}(\theta))\right\} ^{z_{i}}\exp(b_{i}(\theta)-\ell_{i}(\theta))^{1-z_{i}}$.
Hence, the marginal distribution of $z_{i}$ under this extended model
is given by 
\begin{eqnarray}
p(z_{i}=1\vert\theta) & = & \int p(z_{i}=1,x_{i}\vert\theta)dx_{i}\nonumber\\
 & = & \int\left[\exp(\ell_{i}(\theta))-\exp(b_{i}(\theta))\right]dx_{i}\nonumber \\
 & = & 1-I_\theta,\label{e:fireflyDefConditionalZ}
\end{eqnarray}
where $I_\theta\defeq \int \exp(b(\theta,x))dx$. Using Bayes' theorem, we obtain accordingly 
\[
p(x_{i}\vert\theta,z_{i}=1)=\frac{\exp(\ell_{i}(\theta))-\exp(b_{i}(\theta))}{1-I_\theta},~p(x_{i}\vert\theta,z_{i}=0)=\frac{\exp(b_{i}(\theta))}{I_\theta}.
\]

An obvious unbiased estimator of the unnormalized posterior is thus given by 
\begin{equation}
\hat{\gamma}(\theta)=p(\theta) \prod_{i=1}^{n}p(x_{i}\vert\theta,z_{i})\label{e:fireflyUnbiasedEstimate}
\end{equation}
where each $z_{i}$ is drawn independently given $\theta$ from (\ref{e:fireflyDefConditionalZ}).
Note that in the case of logistic regression, if $b_{i}(\theta)$
is chosen to be the quadratic lower bound given in \citep{MaAd14},
its integral $I_\theta$ is a Gaussian integral and can thus be
computed. Finally, similarly to the Firefly algorithm of \cite{MaAd14}, the number of
evaluations of the likelihood per iteration is $nI_\theta $, loosely
speaking.

Although the pseudo-marginal variant of Firefly we propose has the
disadvantage of requiring the integrals $I_\theta$ to be
tractable, it comes with two advantages. First, the sampling of $z$
does not require to evaluate the likelihood at all. If computing all
bounds does not become a bottleneck, this avoids the need to explicitly
state a resampling fraction at the risk of augmenting the variance
of the likelihood estimator. Second, the properties of this variant
are easier to understand, as it is a `standard' pseudo-marginal MH
and hence the results from Section~\ref{ss:pseudomarginal} apply.
In particular, although it has the correct target distribution, the asymptotic variance 
of ergodic averages is inflated compared to the ideal algorithm. 

As explained in Section~\ref{ss:pseudomarginal}, we consider the
variance of the log likelihood estimator.

\begin{prop}
\label{p:varianceFirefly}
Let $\theta\in\Theta$. With the notations introduced in
Section~\ref{ss:firefly}, 
\begin{equation}
\Var_z \left[\sum_{i=1}^n \log p(x_i\vert\theta,z_i)\right] =
I_\theta(1-I_\theta) \sum_{i=1}^n \log^2\left[\frac{I_\theta}{1-I_\theta}\left(e^{\ell_i(\theta)-b_i(\theta)}-1\right)\right]
\label{e:varianceFirefly}
\end{equation}
\end{prop}
The proof of Proposition~\ref{p:varianceFirefly} can be
found in Appendix~B. Proposition~\ref{p:varianceFirefly} can be
interpreted as follows: the variance is related to how tight the bound
is. In general, obtaining a variance of order $1$ will only be
possible if {\it most} bounds $b_i(\theta)$ are very tight, and the
bigger $n$, the tighter the bounds have to be. These conditions will
typically not be met when a fixed fraction of ``outlier'' $x_i$'s correspond
to untight bounds.

\begin{figure}
\subfigure[Chain histograms, $X_{i}\sim\cN(0,1)$]{
\includegraphics[width=\twofig]{\figdir/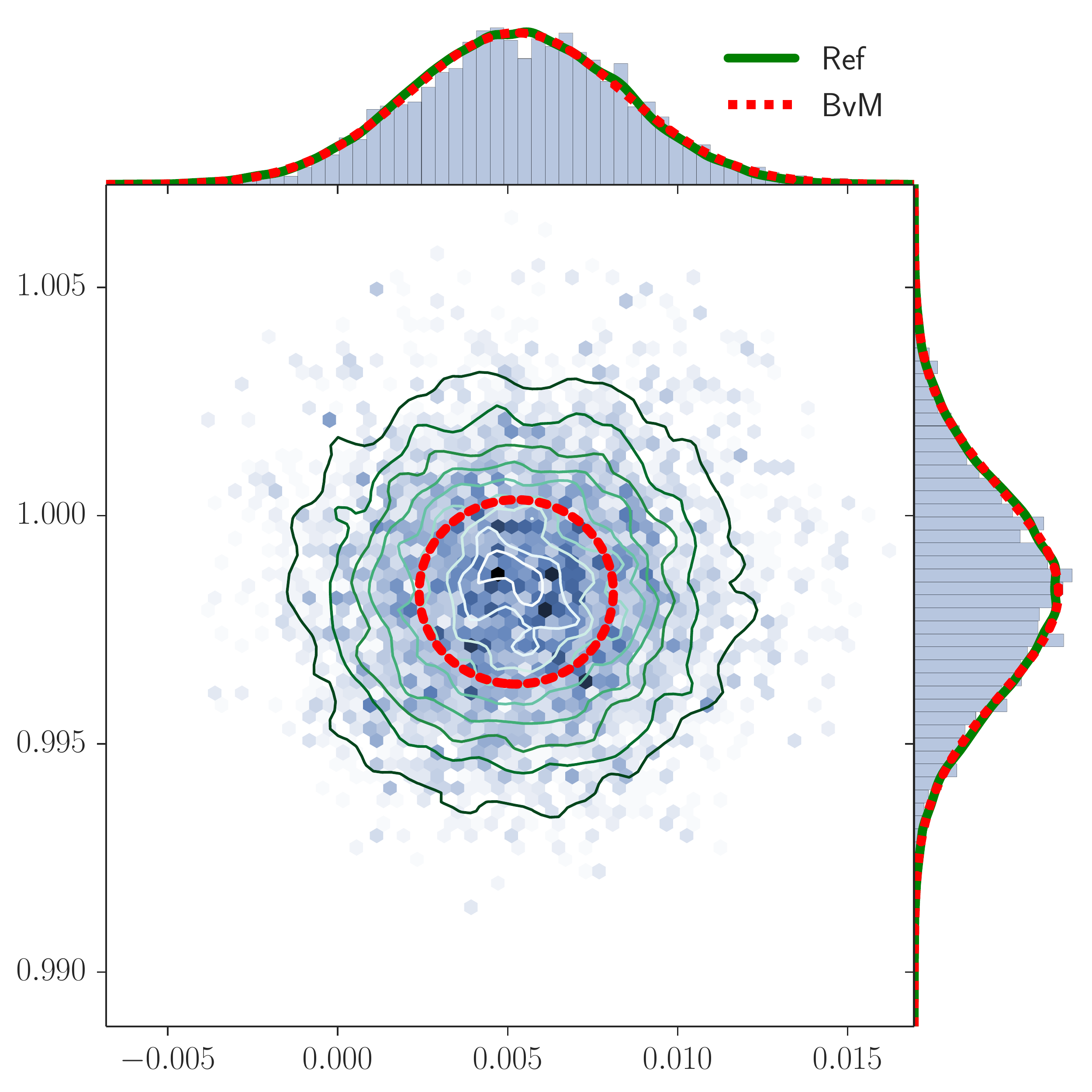}
}
\subfigure[Chain histograms, $X_{i}\sim\log\cN(0,1)$]{
\includegraphics[width=\twofig]{\figdir/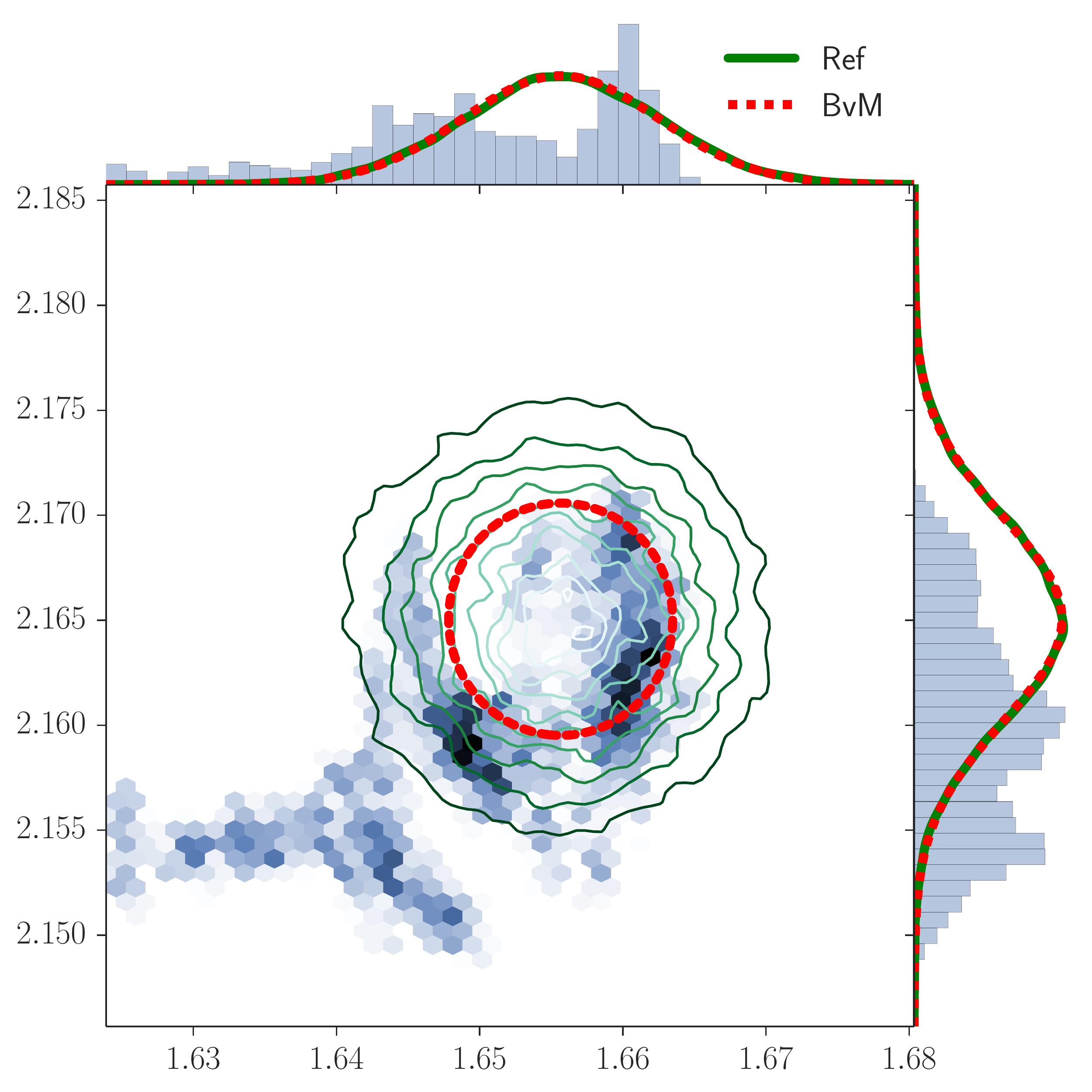}
}\\
\subfigure[Autocorr. of $\log\sigma$, $X_{i}\sim\cN(0,1)$]{
\includegraphics[width=\twofig]{\figdir/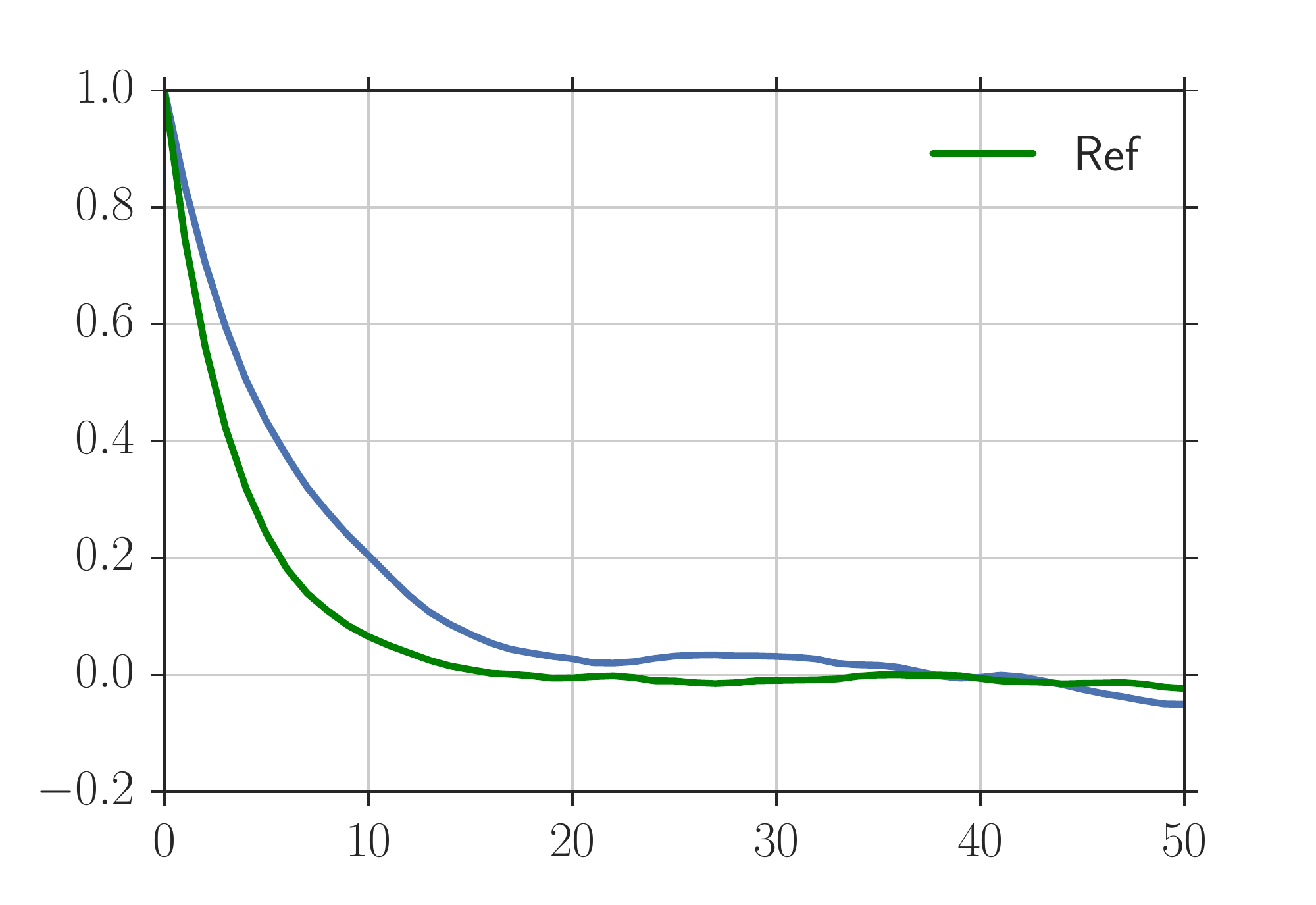}
}
\subfigure[Autocorr. of $\log\sigma$, $X_{i}\sim\log\cN(0,1)$]{
\includegraphics[width=\twofig]{\figdir/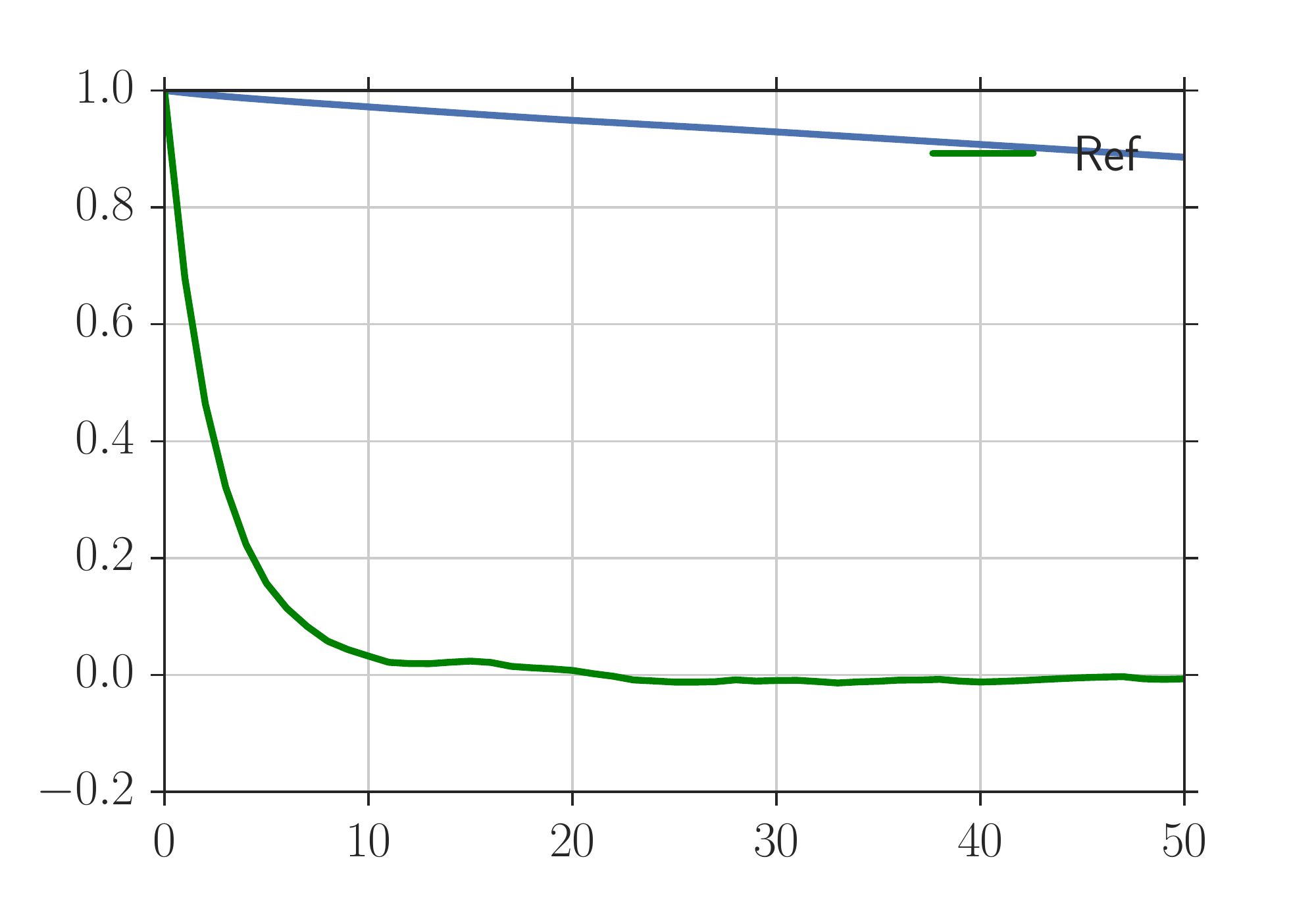}
}\\
\caption{
Results of 10\,000 iterations of Firefly MH \citep{MaAd14} on our
Gaussian and lognormal running examples. See
Section~\ref{ss:firefly} and the caption of Figure~\ref{f:resultsMH} for details.
}
\label{f:resultsFireflyMH}
\end{figure}

We give the results of the original Firefly MH on our running Gaussian
and log normal examples in Figure~\ref{f:resultsFireflyMH}. We bound
each $\ell_{i}(\theta)$ using a 2nd order Taylor expansion at the MLE and the
Taylor-Lagrange inequality, see Section~\ref{ss:Taylor} for further
details. This bound is very tight in both cases, so that we are in
the favourable case where only a few components of $z$ are $1$ at
each iteration, and the number of likelihood evaluations per full
joint iteration is thus roughly the fraction of points for which $z_{i}$
has been resampled. We chose the fraction of resampled points to be $10\%$ here,
and initialized $z$ to have $10\%$ of ones. Trying smaller fractions
led to very slowly mixing chains even for the Gaussian
example. Estimating the number of likelihood evaluations per full
joint iteration as the sum of the number of resampled $z_{i}$'s and the number of ``bright''
points, we obtained in both the Gaussian and lognormal case an almost
constant number of likelihood evaluations close to $10\%$, so that
only a few points are bright. This can be explained by the tightness of the Taylor bound, which leads Firefly MH
to almost exclusively replace the evaluation of the likelihood by that
of the Taylor bound. Finally, unlike the other algorithms we applied, we observed that a bad choice
of the initial value of $z$ can easily take $\theta$ out of the
posterior mode. To be fair, we thus discarded the first $1\,000$
iterations as a burn-in before plotting. 

As expected, the algorithm behaves erratically in the lognormal case, as failure to attempt a
flip of each $z_{i}$ draws the $\mu$-component of the chain towards
the few large values of $(x_{i}-\mu)^{2}$ which are bright. Since
the bright points are rarely updated, the chain mixes very slowly.
%
%
%

\setcounter{subfigure}{0}


\section{Other exact approaches}

\label{s:reviewExact} 
Other exact approaches have been proposed, which do not rely on pseudo-marginal
MH.

\subsection{Forgetting about acceptance: stochastic approximation approaches}

\label{ss:sgld} \cite{WeTe11} proposed an algorithm based on stochastic
gradient Langevin dynamics (SGLD). This is an iterative algorithm
which at iteration $k+1$ uses the following update rule 
\begin{equation}
\theta_{k+1}=\theta_{k}+\frac{\eps_{k+1}}{2}\left[\nabla \log
  p(\theta) + \frac{n}{t}\sum_{i=1}^{t}\nabla\log p(x_{i,k}^{*}\vert\theta)\right]+\sqrt{\eps_{k+1}}\eta_{k+1},\label{e:sgldRecursion}
\end{equation}
$(\eps_{k})$ is a sequence of time steps, $(\eta_{k})$ are
independent $\cN(0,I_{d})$ vectors and
$$\frac{n}{t}\sum_{i=1}^{t}\nabla\log p(x_{i,k}^{*}\vert\theta)$$
is an unbiased estimate of the score computed at each iteration using
a random subsample $\left\{ x_{i,k}^{*}\right\} $ of the observations.
This approach is reminiscent of the Metropolis-adjusted Langevin algorithm
(MALA; \citealt[Section 7.8.5]{RoCa04}), where the proposal given
by
\[
\theta'=\theta+\frac{\eps}{2}\left[\nabla\log p(\theta) + \sum_{i=1}^{n}\nabla\log p(x_{i}\vert\theta)\right]+\sqrt{\eps}\eta,
\]
is used in an MH acceptance step, where $\eps\sim\cN(0,I_d)$. The point of \cite{WeTe11} is that if one suppresses the
MH acceptance step, computes an unbiased estimate of the score but
introduces a sequence of stepsizes $(\eps_{k})$ that decreases to
zero at the right rate, then 
\[
\left(\sum_{k=0}^{N_{\text{iter}}}\eps_{k}\right)^{-1}\sum_{k=0}^{N_{\text{iter}}}\eps_{k}\delta_{\theta_{k}}
\]
is an approximation to $\pi$. The algorithm has been analyzed
recently in \citep{TeThVoSub}, where it has been established that
it provides indeed a consistent estimate of the target. Additionally,
a central limit theorem holds with convergence rate
$N_{\text{iter}}^{-1/3}$, which is slower than the traditional Monte
Carlo rate $N_{\text{iter}}^{-1/2}$.
It is yet unclear how SGLD compares to other subsampling
schemes in theory: it may require a smaller fraction of the dataset
per iteration, but more iterations are needed to reach the same accuracy.

\begin{figure}
\subfigure[Chain hist., $10^4$ iter. at $10\%$, $X_{i}\sim\cN(0,1)$]{
\includegraphics[width=\twofig]{\figdir/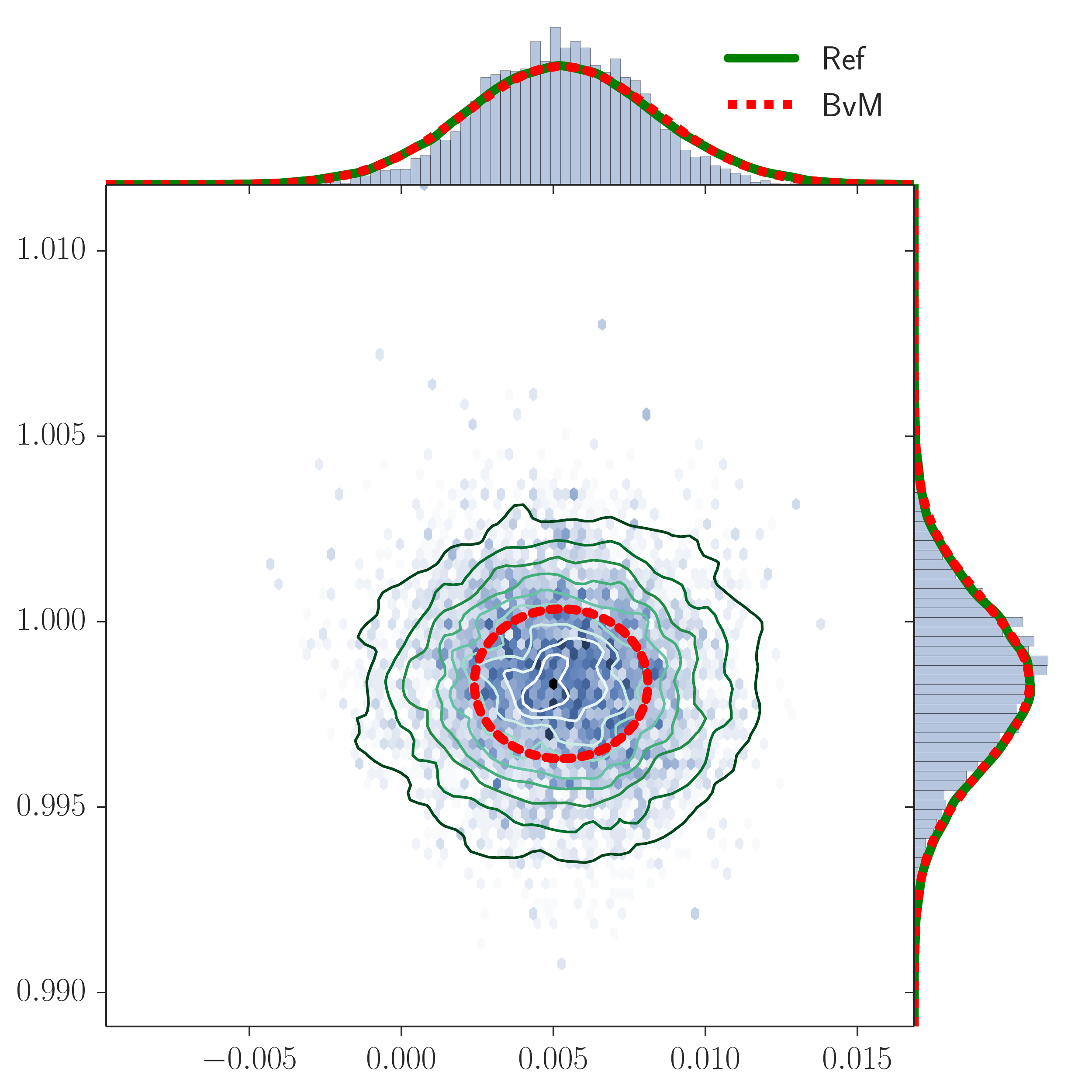}
}
\subfigure[Chain hist., $10^4$ iter. at $10\%$, $X_{i}\sim\log\cN(0,1)$]{
\includegraphics[width=\twofig]{\figdir/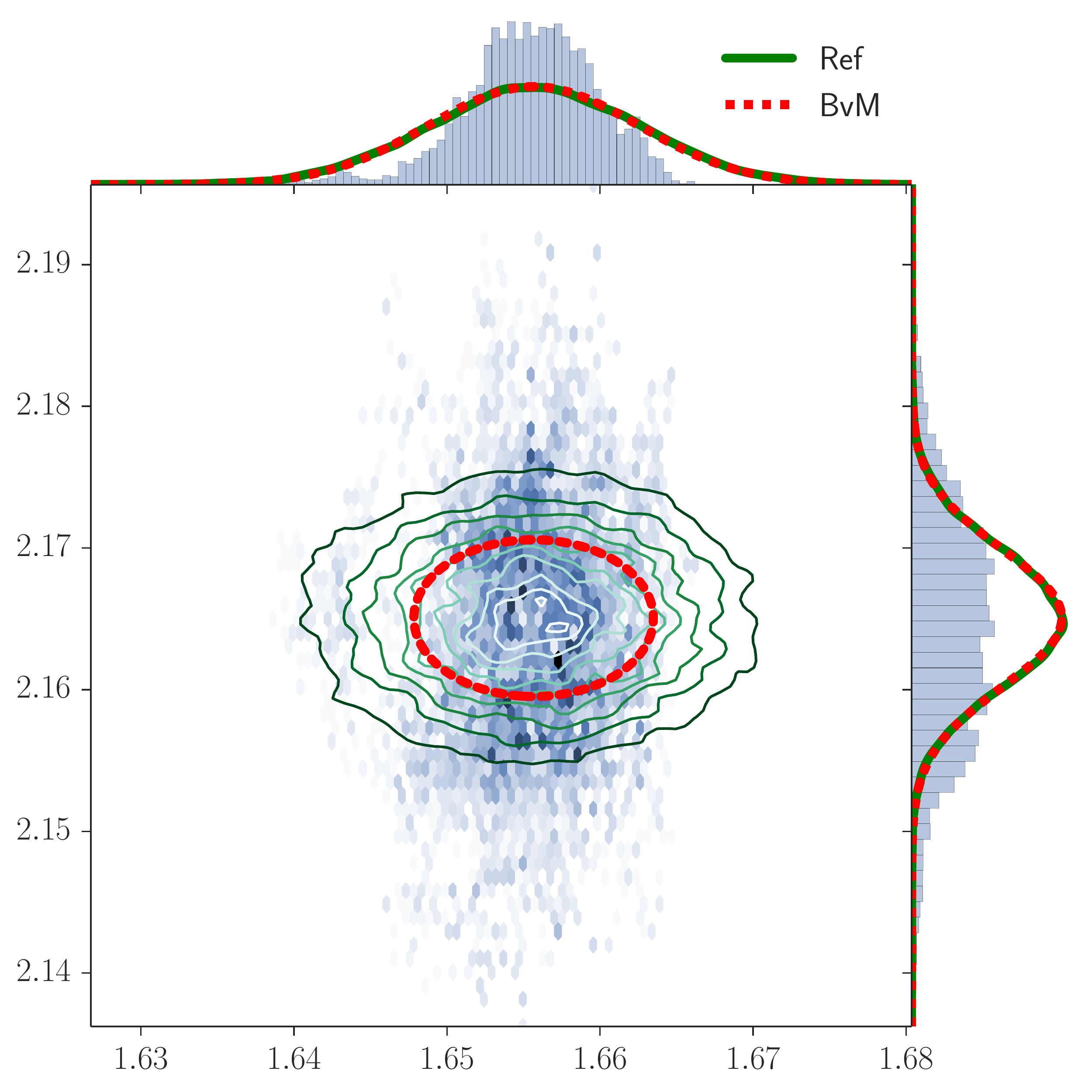}
}\\
\subfigure[Chain hist., $10^5$k iter. at $1\%$, $X_{i}\sim\cN(0,1)$]{
\includegraphics[width=\twofig]{\figdir/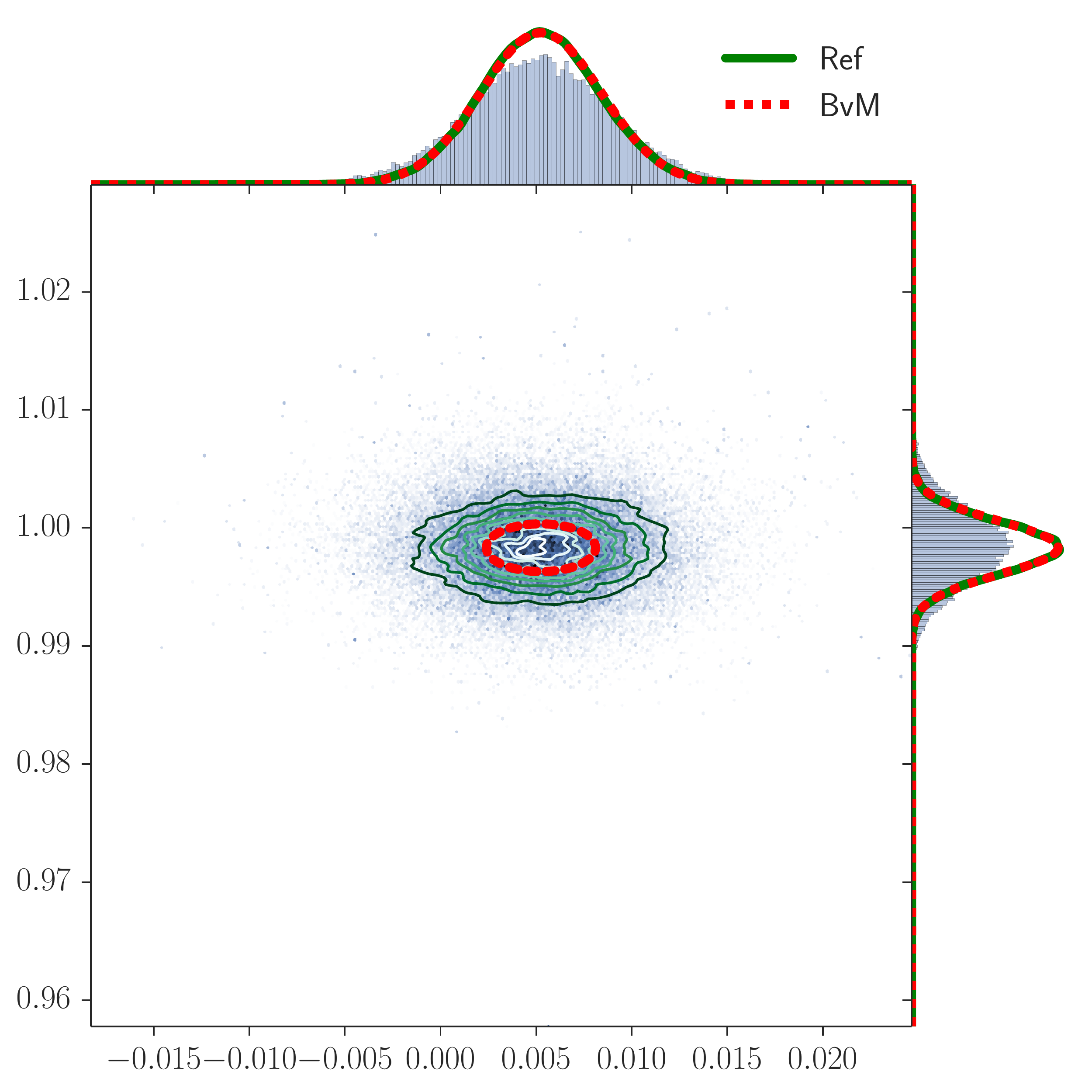}
}
\subfigure[Chain hist., $10^5$k iter. at $1\%$, $X_{i}\sim\log\cN(0,1)$]{
\includegraphics[width=\twofig]{\figdir/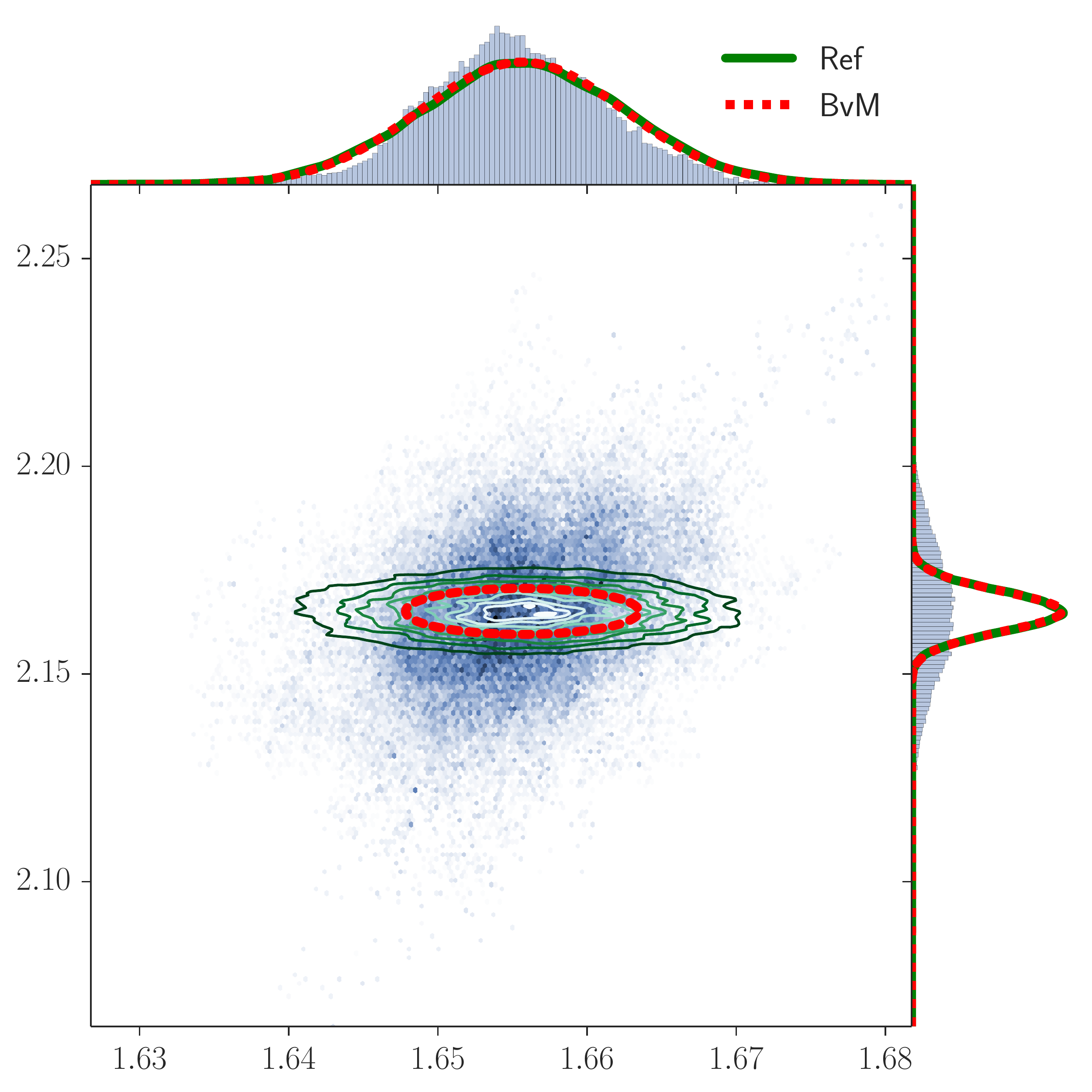}
}\\
\label{f:resultsSGLD}
\caption{
Results of SGLD \citep{WeTe11} on our Gaussian and lognormal running
examples. See Section~\ref{ss:sgld} and the caption of
Figure~\ref{f:resultsMH} for details.
}
\end{figure}

In practice, we show the results of SGLD on our two running examples
in Figure~\ref{f:resultsSGLD}. The stepsize $\eps_{k}$ is chosen
proportional to $k^{-1/3}$, following the recommendations of \cite{TeThVoSub}.
We show the results of two choices for the subsample size $t$: $10\%$
and $1\%$ of the data, with respectively $10\,000$ and $100\,000$
iterations, so that both runs amount to the same $10\%$ fraction
of the budget of the vanilla MH in Figure~\ref{f:resultsMH}. Both
runs are still far from convergence on the lognormal example: subsampling
draws the chain away from the support of the posterior, and one has
to wait for smaller stepsizes to avoid overconfident moves. But then,
the variance of the final estimate gets bigger. Constant stepsizes
lead to comparable results (not shown).

\setcounter{subfigure}{0}
%

Finally, we note that subsampling for Hamiltonian Monte Carlo (HMC;
\citealp{DKPR87}) has also been recently considered. \cite{ChFoGu14} propose a
modification of HMC that is inspired by the SGLD with decreasing
stepsize of \cite{WeTe11}, while \cite{BetSub} explores why naive approaches
suffer from unacceptable biases. The algorithm of \cite{ChFoGu14} is however
a heuristic, which further relies on the subsampling noise being
Gaussian. As demonstrated in \citep{BaDoHo14} and in
Section~\ref{sss:CLT}, relying on a Gaussian noise assumption can
yield arbitrarily poor performance when this assumption is violated.

\subsection{Delayed acceptance}

\cite{BGLRSub} remarked that if we decompose the acceptance ratio
in a product of positive functions 
\[
\alpha(\theta,\theta')=\prod_{i=1}^{B}\rho_{i}(\theta,\theta')
\]
then the MH-like algorithm that accepts the move from $\theta$ to
$\theta'$ with probability 
\[
\prod_{i=1}^{B}\min\left[\rho_{i}(\theta,\theta'),1\right]
\]
still admits $\pi$ as invariant distribution. In practice, in the
case of tall data, we can divide the dataset into $B$ batches and
use for example 
\[
\rho_{i}\left(\theta,\theta'\right)=\frac{p(\theta')^{1/B}p(\bx_{i}\vert\theta')q(\theta\vert\theta')}{p(\theta)^{1/B}p(\bx_{i}\vert\theta)q(\theta'\vert\theta')}.
\]
This allows us to reject candidate $\theta'$ without having to compute
the full likelihoods and the calculations of $\rho_{i}(\theta,\theta')$
can be done in parallel. However, as remarked by \cite{BGLRSub},
the resulting Markov chain has a larger asymptotic variance $\sigma_{\text{lim}}^{2}$
in \eqref{e:CLT} than the original MH, and it does not necessarily
inherit the ergodicity of the original MH. Furthermore, by construction,
every accepted point has to be evaluated on the whole dataset, and
the average proportion of data points used is thus lower bounded by
the acceptance rate of the algorithm, which in practice is often around
$25\%$. Overall, it is an easy-to-implement feature that does not add
any bias, but its benefits are inherently limited, and speed of
convergence might be affected.

\section{Approximate subsampling approaches}
\label{s:reviewSubsampling} 

In this Section, we consider again subsampling approaches where, at each
MH iteration, a subset of data points is used to approximate the acceptance
ratio \eqref{e:MHAcceptanceRatio}. As mentioned in Section \ref{ss:pseudomarginalSubsampling},
it is very simple to obtain an unbiased estimator of the log-likelihood
$n\ell(\theta)$ based on random samples $x_{1}^{*},\dots,x_{t}^{*}$
from the dataset $\cX$; see \eqref{eq:unbiasedestimateloglikelihood}.
Similarly, one can also easily obtain an unbiased estimator of
the average log likelihood ratio $\left[\ell(\theta')-\ell(\theta)\right]$
\begin{equation}
\Lambda_{t}^{*}(\theta,\theta')\defeq\frac{1}{t}\sum_{i=1}^{t}\log\frac{p(x_{i}^{*}\vert\theta')}{p(x_{i}^{*}\vert\theta)}.\label{e:estimatorRatio}
\end{equation}
Note that unlike the exact approaches of
Sections~\ref{s:reviewPseudomarginal} and \ref{s:reviewExact}, the
methods reviewed in Section~\ref{s:reviewSubsampling} do not
attempt to sample exactly from $\pi$, but just from an approximation
to $\pi$.

\subsection{Naive subsampling}

\label{ss:naiveSubsampling} The first approach one could try is to
only use a random fixed proportion of data points to estimate $\pi$
at any newly drawn $\theta$. We highlight that this leads to a
nontrivial mixture target that is hard to interpret, where all subsampled posteriors appear,
suitably rescaled. Assume that at each new $\theta$
drawn in an MH run, we draw $n$ independent Bernoulli variables and
let 
\begin{equation}
\hat{\ell}(\theta)=\frac{1}{n}\sum_{i=1}^{n}\frac{z_{i}}{\lambda}\ell_{i}(\theta)\label{e:naiveUnbiased}
\end{equation}
be an unbiased estimator of the average log likelihood $\ell(\theta)$,
where $z_{i}\sim B(1,\lambda)$ i.i.d. Now one could think of plugging
estimates $\hat{\gamma}(\theta)=p(\theta)e^{n\hat{\ell}(\theta)}$
in Steps~\ref{ai:theta} and \ref{ai:thetaPrime} of MH in Figure~\ref{f:MH}.
However, as $\hat{\gamma}(\theta)$ is not an unbiased estimator of
$\gamma(\theta)$, this algorithm samples from a target distribution
which is proportional to $p(\theta)\mathbb{E}e^{n\hat{\ell}(\theta)}\neq\gamma(\theta)$,
where the expectation is w.r.t the distributions of the Bernoulli
random variables $\left\{ z_{i}\right\} $. Now 
\begin{eqnarray*}
\mathbb{E}e^{n\hat{\ell}(\theta)} & = & \prod_{i=1}^{n}\left[\lambda\ell_{i}(\theta){}^{1/\lambda}+(1-\lambda)\right]\\
 & = & \sum_{r=0}^{n}\sum_{\#I_{r}=r}\lambda^{r}(1-\lambda)^{n-r}p(x_{I_{r}}\vert\theta)^{1/\lambda},
\end{eqnarray*}
where $I_{r}$ denotes a set of $r$ distinct indices in $\{1,\dots,n\}$,
$x_{I_{r}}=\{x_{i};i\in I_{r}\}$, and with the convention $p(x_{\emptyset}\vert\theta)=1$.
Each subsampled likelihood contributes to the target, exponentiated
to the power $1/\lambda$, resulting in a nontrivial mixture of rescaled
data likelihood terms. To further simplify, assume $p(x_{I_{r}}\vert\theta)\approx p_{r}(\theta)$
for each set of indices $I_{r}$, that is, the variance of the likelihood
under subsampling is small, then 
\begin{equation}
\mathbb{E}e^{n\hat{\ell}(\theta)}\approx\sum_{r=0}^{n}C_{n}^{r}\lambda^{r}(1-\lambda)^{n-r}p_{r}(\theta)^{1/\lambda}=\mathbb{E}_{R\sim
  B(n,\lambda)}p_{R}^{1/\lambda}(\theta),
\label{e:approxTarget1}
\end{equation}
where $B(n,\lambda)$ denotes the binomial distribution with parameters
$n$ and $\lambda$. Noticing that $p_{r}(\theta)$ is roughly exponentially decreasing
in $r$, the values or $r$ that are larger than the mode of the binomial
probability mass function are unlikely to contribute a lot to \eqref{e:approxTarget1}.
The largest subsample size contributing to \eqref{e:approxTarget1}
is thus roughly $n\lambda$, and the power $1/\lambda$ makes this
term of the same scale as $p(x_{1},\dots,x_{n}\vert\theta)$. Broadly
speaking, subsampling has a ``broadening'' effect due to the contribution
of the likelihoods of small subsamples.

Alternately, if one starts with the biased estimator of the average
log likelihood 
\begin{equation}
\tilde{\ell}(\theta)=\frac{1}{n}\sum_{i=0}^{n}z_{i}\ell_{i}(\theta),\label{e:defBiasedLLEstimator}
\end{equation}
instead of \eqref{e:naiveUnbiased} one ends up with 
\begin{eqnarray}
\mathbb{E}e^{n\tilde{\ell}(\theta)} & = & \prod_{i=1}^{n}\left[\lambda\ell_{i}(\theta)+(1-\lambda)\right]\nonumber\\
 & = & \sum_{r=0}^{n}\sum_{\#I_{r}=r}\lambda^{r}(1-\lambda)^{n-r}p(x_{I_{r}}\vert\theta)\nonumber\\
 & \approx & \mathbb{E}_{R\sim B(n,\lambda)}p_{R}(\theta). \label{e:targetBiasedLLEstimator}
\end{eqnarray}
Again, all subsampled likelihoods contribute, but without being further
exponentiated. Still, the result is a much broadened target, as values
of $r$ that are larger than $n\lambda$ are unlikely to contribute
a lot. In this case, the broadening effect of subsampling is not only
due to the contribution of small subsamples, but also to the absence
of bias correction in \eqref{e:defBiasedLLEstimator}.

We have thus seen that naive subsampling is nontrivial tempering,
so that the target is not preserved. Additionally, 
as mentioned in Section~\ref{ss:pseudomarginal}, 
the variance of the log likelihood estimator needs to be kept around a constant, $1$
or $3$ depending on hypotheses, in order for pseudo-marginal MH to be efficient. This means that $\lambda$ should
be such that 
\[
\frac{(1-\lambda)}{\lambda}\sum_{i=1}^{n}\ell_{i}(\theta)^{2}
\]
is of order $1$ in the case of \eqref{e:naiveUnbiased}. This entails
that $\lambda$ should be close to 1, so that there is no
substantial gain in terms of number of likelihood evaluations. In the
case of \eqref{e:defBiasedLLEstimator}, 
\[
\lambda(1-\lambda)\sum_{i=1}^{n}\ell_{i}(\theta)^{2}
\]
can be of order $1$ if $\lambda\sim n^{-1}$. But then the leading
terms in the mixture target
\eqref{e:targetBiasedLLEstimator} will be the subsampled likelihoods
corresponding to small subsamples, so that the target will be very different from the
actual target $\pi$. 

Overall, naive subsampling is a very poor approach. However, it allows us to identify the main issues
a good subsampling approach should tackle: guaranteeing its target,
not loosing too much convergence speed compared to MH, and cutting
the likelihood evaluation budget. As shown in \citep{BaDoHo14}, the
first point is an algorithmic design issue, while the last two points
are related to controlling the variance of the log likelihood ratios.

\subsection{Relying on the CLT}
\label{sss:CLT} 

Several authors have appealed to the central limit
theorem to justify their assumption that the average subsampled log
likelihoods and log likelihood ratios in \eqref{eq:unbiasedestimateloglikelihood}
and \eqref{e:estimatorRatio} are Gaussianly distributed. 

If the noise of the log likelihood ratio estimate is normal \textit{with
known variance} and mean equal to the true log-likelihood ratio,
\cite{CeDe99} have proposed an MH with a corrected acceptance ratio
that is exact, i.e., that still targets $\pi$. When the variance
of the noise is not known, and is rather estimated, the method becomes
inexact. \cite{NiFoMuSub} propose a heuristic argument to show that
this inexact chain gives reasonable approximate results, but the Gaussian
assumption remains crucial there. As shown in \citep{BaDoHo14} and
in this paper in Figures~\ref{f:resultsAusterityMH} and \ref{f:student},
this assumption can be arbitrarily violated when subsampling tall
data if the log likelihood ratios $\ell_{i}(\theta')-\ell_{i}(\theta)$
are heavy-tailed. Missing log likelihoods in the tails will lead to
erroneous decisions, and uncontrolled results.

\subsubsection{A pseudo-marginal approach with variance reduction under Gaussian
assumption}

\cite{QuViKoSub} propose a methodology to use MH for tall data which
also relies on the assumption that the log-likelihood estimator is
Gaussian with mean $\ell(\theta)$, for every $\theta$. By introducing a bias correction
providing an approximately unbiased estimate of the likelihood, this
corresponds to a pseudo-marginal MH algorithm whose target distribution
is proportional to $p(\theta)\mathbb{E}e^{n\hat{\ell}(\theta)-\hat{b}(\theta)}$,
where $\hat{b}(\theta)$ is an estimate of the bias $b(\theta)$
satisfying $\mathbb{E}e^{n\hat{\ell}(\theta)}=e^{n\ell(\theta)+b(\theta)}$.
They rightly notice that, ideally, if one wants to keep the variance
of average subsampled log likelihoods small, one should not subsample
data points with or without replacement, but one should rather perform
importance sampling with the weight of data point $i$ being proportional
to $|\ell_{i}(\theta)|$. While this variance reduction
approach obviously defeats the purpose of subsampling, \cite{QuViKoSub}
propose to use as weights an approximation of the log-likelihood,
based e.g. on a Gaussian process or splines trained on a small subset
of computed likelihoods $\ell_{i}(\theta)$. Finally, a heuristic
to adaptively choose the size of the total subsample so as to keep
the variance of the log likelihood controlled is proposed. The method
is demonstrated to work on a bivariate probit model using only $10\%$ of the full dataset.
However, as a general purpose method, it suffers from two limitations. First, it is based on Gaussian assumptions, which can be unreasonable as noted above 
and it is unclear whether it will be robust to these CLT
approximations not being valid.
Second, the proposed importance sampling step requires to learn a
good proxy for $x\mapsto p(x\vert\theta)$ for each $\theta$ drawn
during the MCMC run. The fitted proxies should thus be cheap to train
and evaluate, but at the same time accurate if any variance reduction
is to be obtained. 

\subsubsection{Adaptive subsampling with T-tests}

Still assuming the noise of the log likelihood is Gaussian, given
a drawn $\theta\in\Tset$, one can try to adaptively choose the size
of the subsample $\{x_{1}^{*},\dots,x_{t}^{*}\}$ to be used in the
unbiased estimators \eqref{eq:unbiasedestimateloglikelihood} or \eqref{e:estimatorRatio},
so as to take the correct acceptance decision with high probability. Upon noting that the MH
acceptance decision is equivalent to deciding whether $\log\alpha(\theta,\theta')>u$,
or equivalently 
\begin{equation}
\frac{1}{n}\sum_{i=1}^{n}\log\frac{p(x_{i}\vert\theta')}{p(x_{i}\vert\theta)}>\frac{1}{n}\log u-\frac{1}{n}\log\left[\frac{p(\theta')q(\theta\vert\theta')}{p(\theta)q(\theta'\vert\theta)}\right]\label{e:acceptanceDecision}
\end{equation}
with $u\sim\cU_{[0,1]}$ drawn beforehand, statistical tests can be
used to assert whether \eqref{e:acceptanceDecision} holds with a
given level of ``confidence''. As far as we are aware, \cite{BuSa88}
were the first to consider such a procedure. They used it in a simulated
annealing algorithm maximizing a function defined as an expectation
w.r.t a probability distribution, and approximated using Monte
Carlo. Simulated annealing is a simple non-homogeneous variant of the
MH algorithm where the  the target distribution is annealed over the iterations. The same application
received more attention later \citep{AlAhTu99,WaZh06}. Applied to
the standard MH, the method has been considered by \cite{SiWiMc12},
and more recently by \cite{KoChWe14} specifically for tall data.
\cite{KoChWe14} propose an MH-like algorithm called {\it Austerity MH}
that incorporates a sequential T-test to check \eqref{e:acceptanceDecision} for each
pair $(\theta,\theta')$, thus relying on \textit{several} CLTs. They
demonstrate dramatic reductions in the average number of subsamples
used per MCMC iteration on particular applications. 
However, as noted in \citep{KoChWe14, BaDoHo14}, the results can be
arbitrarily far from the original MH when the CLT approximations are
not valid. 

\begin{figure}
\subfigure[Chain histograms, $X_{i}\sim\cN(0,1)$]{
\includegraphics[width=\twofig]{\figdir/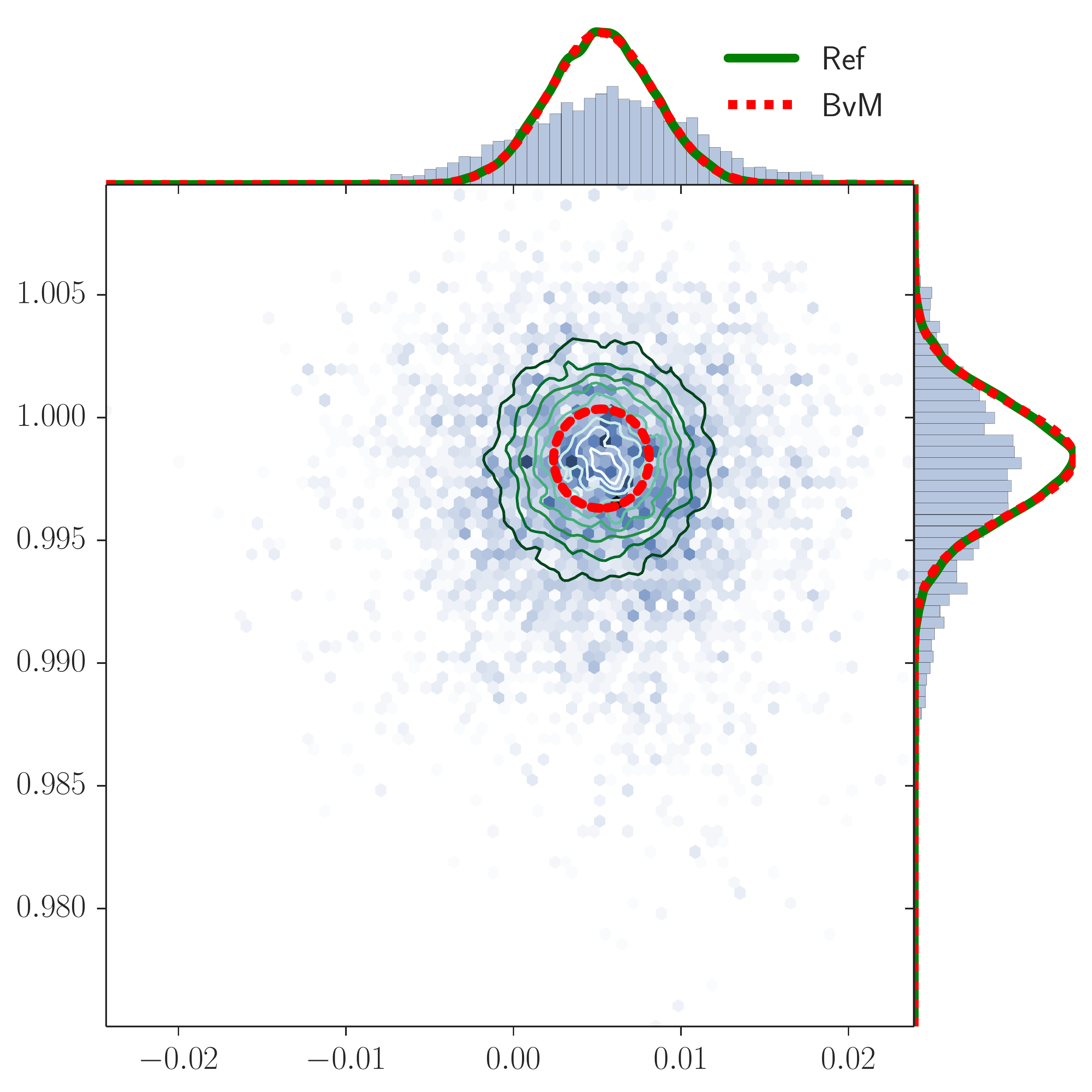}
\label{f:resultsAusterityMH:histoGaussian}
}
\subfigure[Chain histograms, $X_{i}\sim\log\cN(0,1)$]{
\includegraphics[width=\twofig]{\figdir/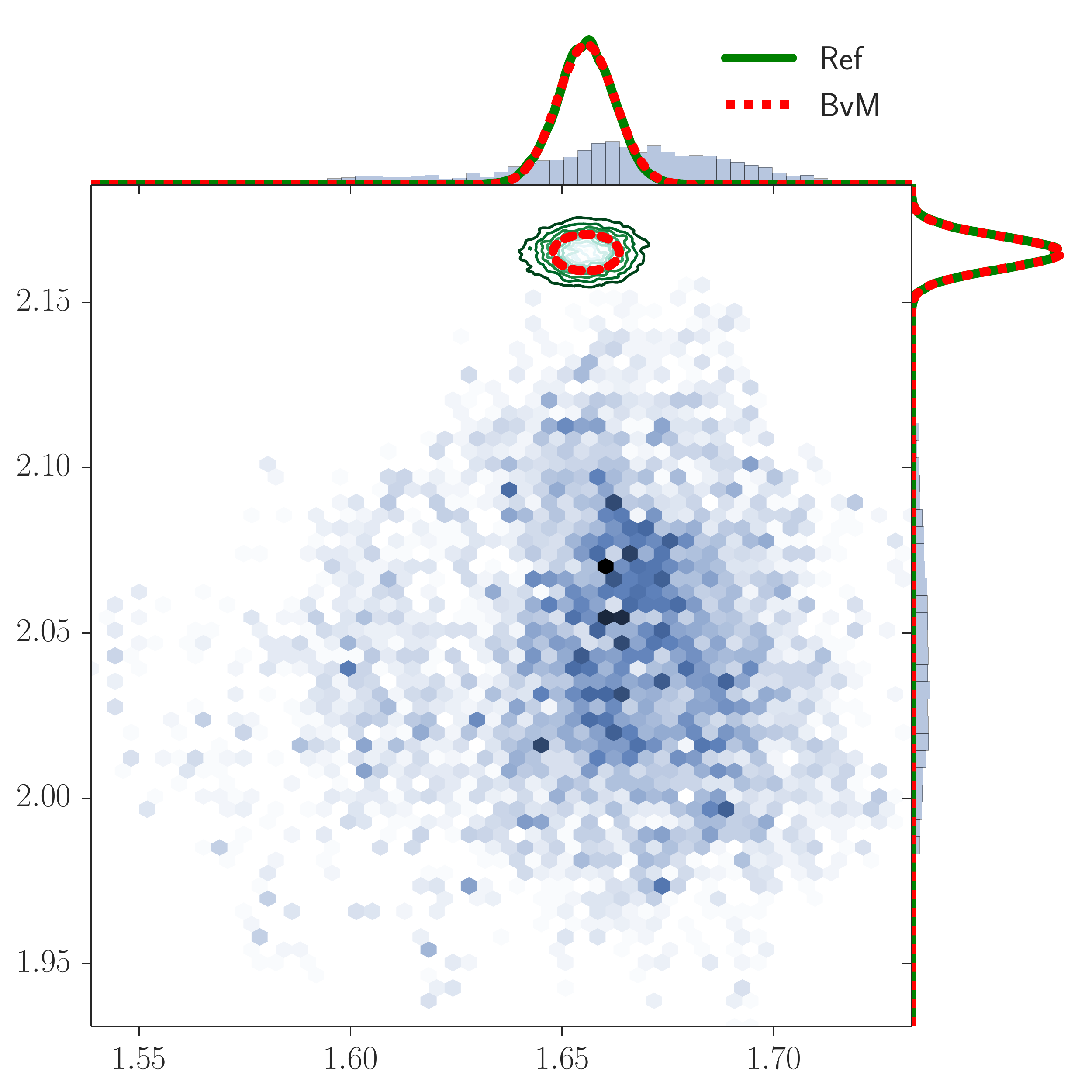}
\label{f:resultsAusterityMH:histoLogNormal}
}\\
\subfigure[Autocorr. of $\log\sigma$, $X_{i}\sim\cN(0,1)$]{
\includegraphics[width=\twofig]{\figdir/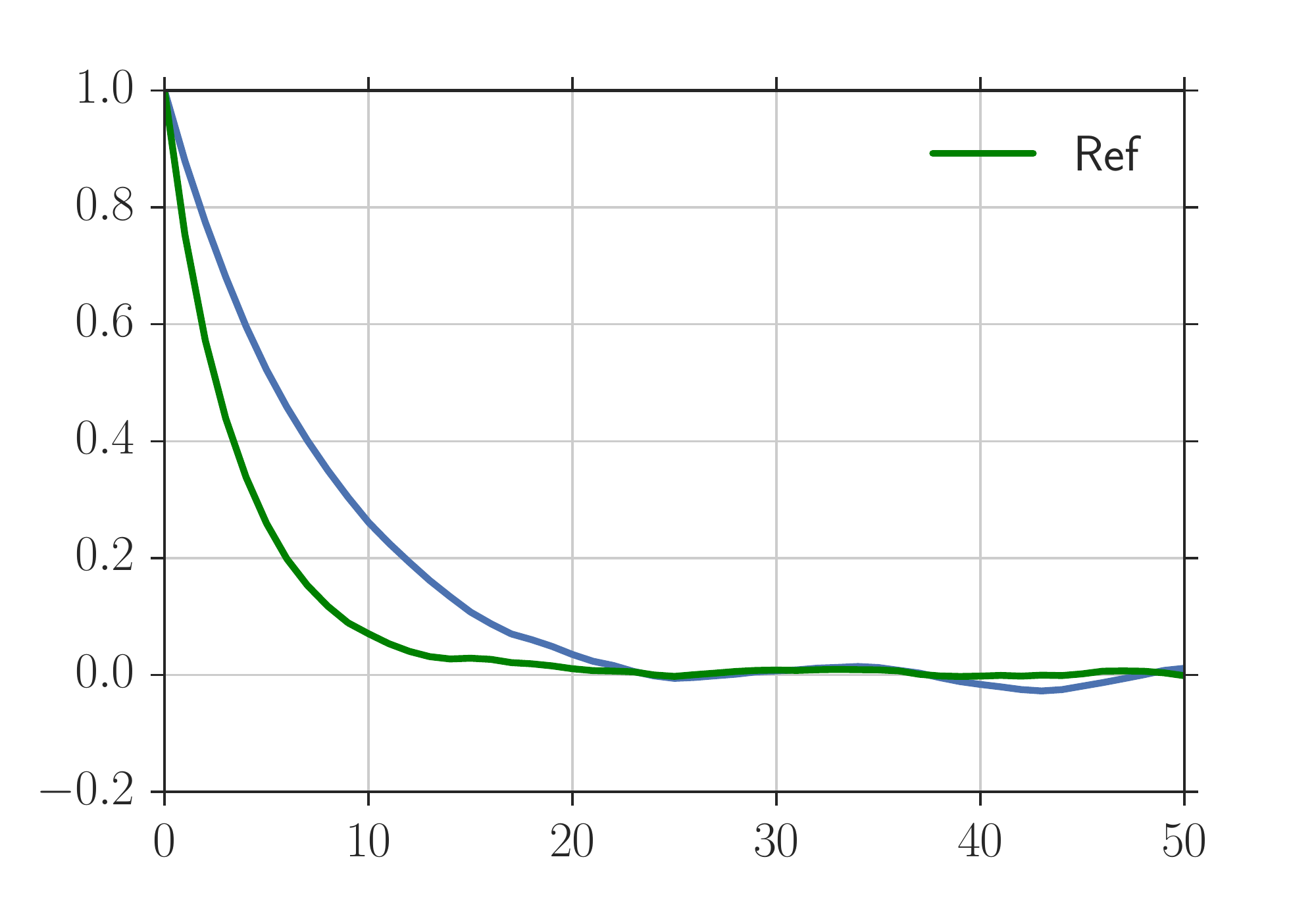}
}
\subfigure[Autocorr. of $\log\sigma$, $X_{i}\sim\log\cN(0,1)$]{
\includegraphics[width=\twofig]{\figdir/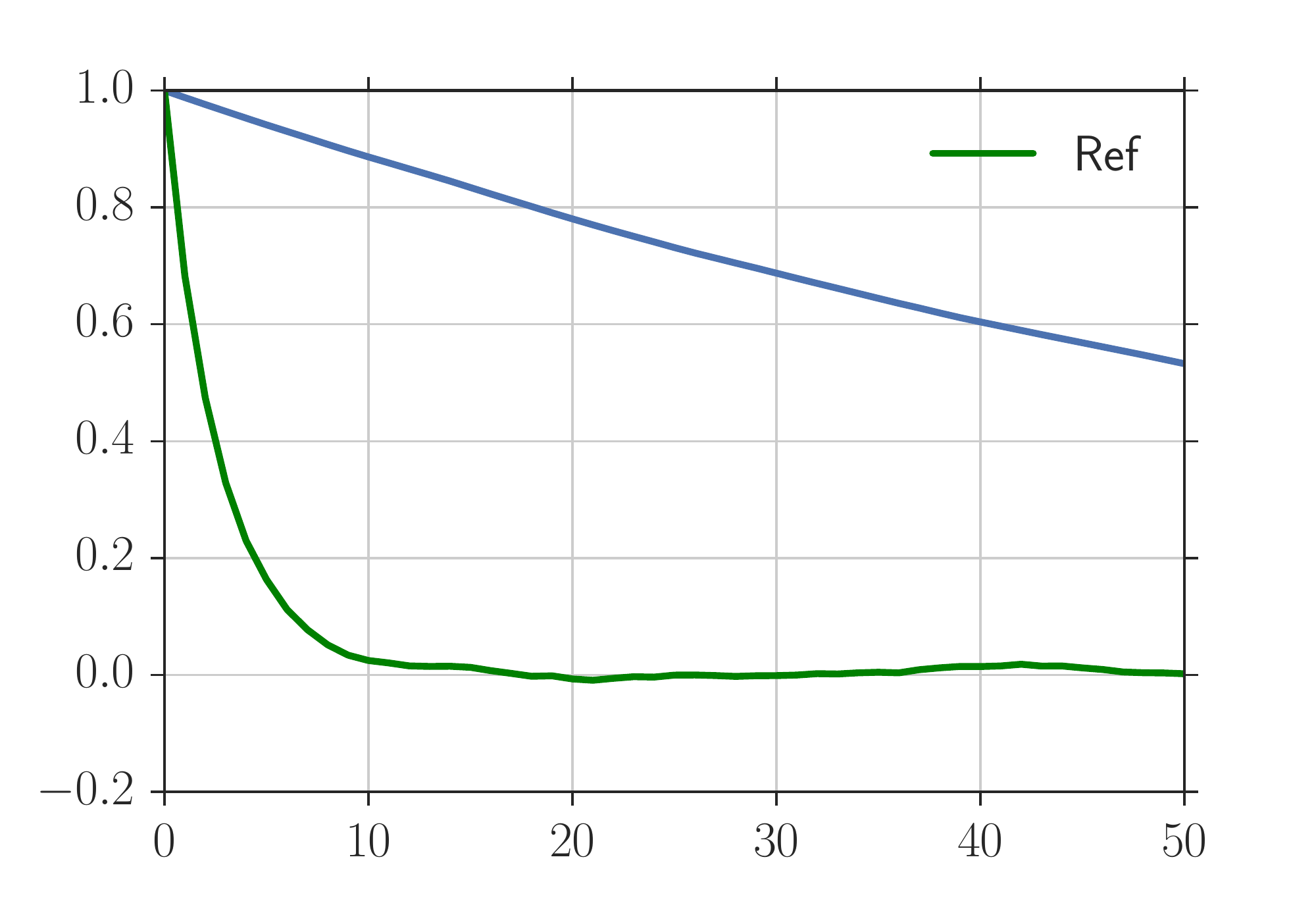}
}\\
\subfigure[Num. lhd. evals, $X_{i}\sim\cN(0,1)$]{
\includegraphics[width=\twofig]{\figdir/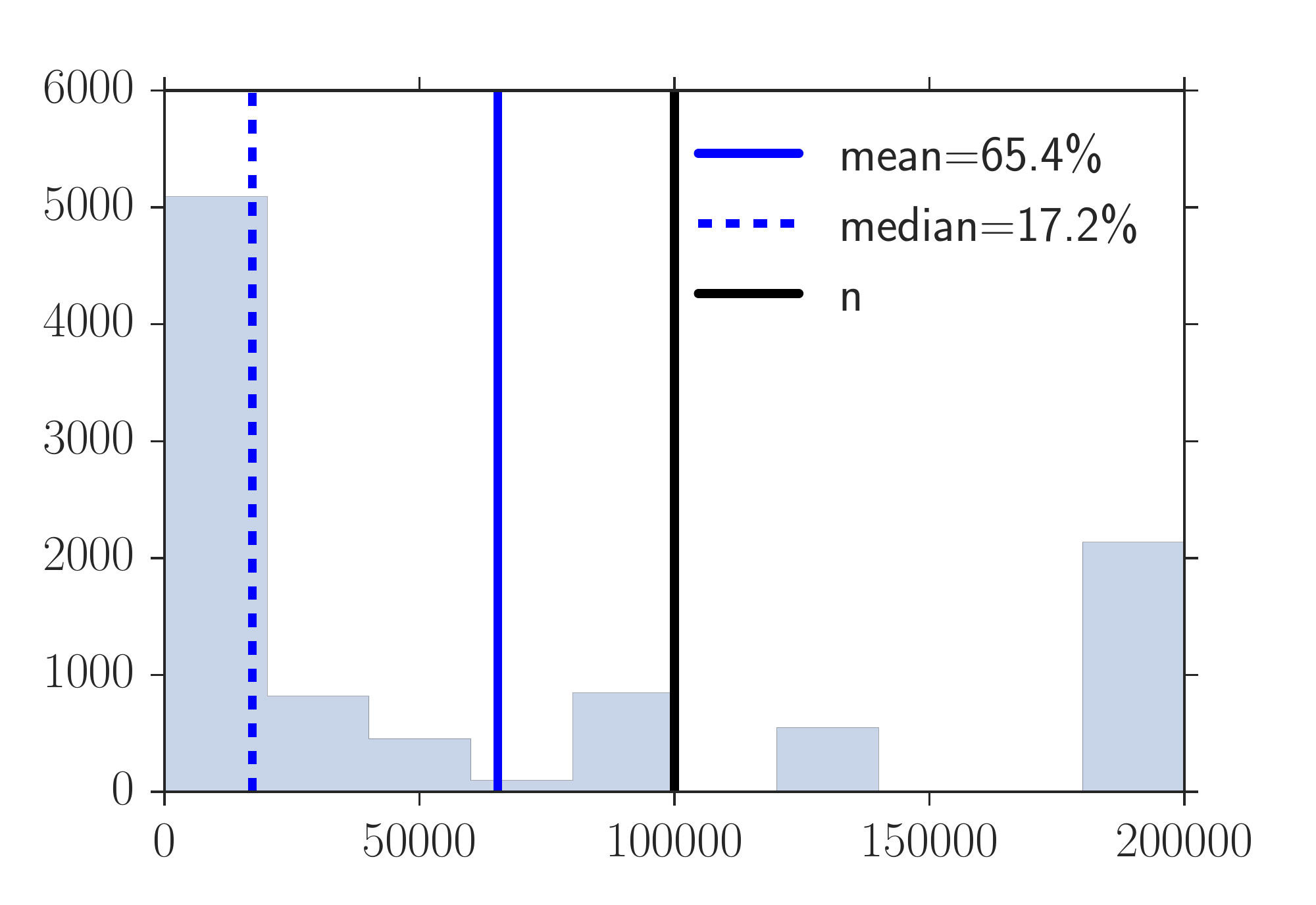}
}
\subfigure[Num. lhd. evals, $X_{i}\sim\log\cN(0,1)$]{
\includegraphics[width=\twofig]{\figdir/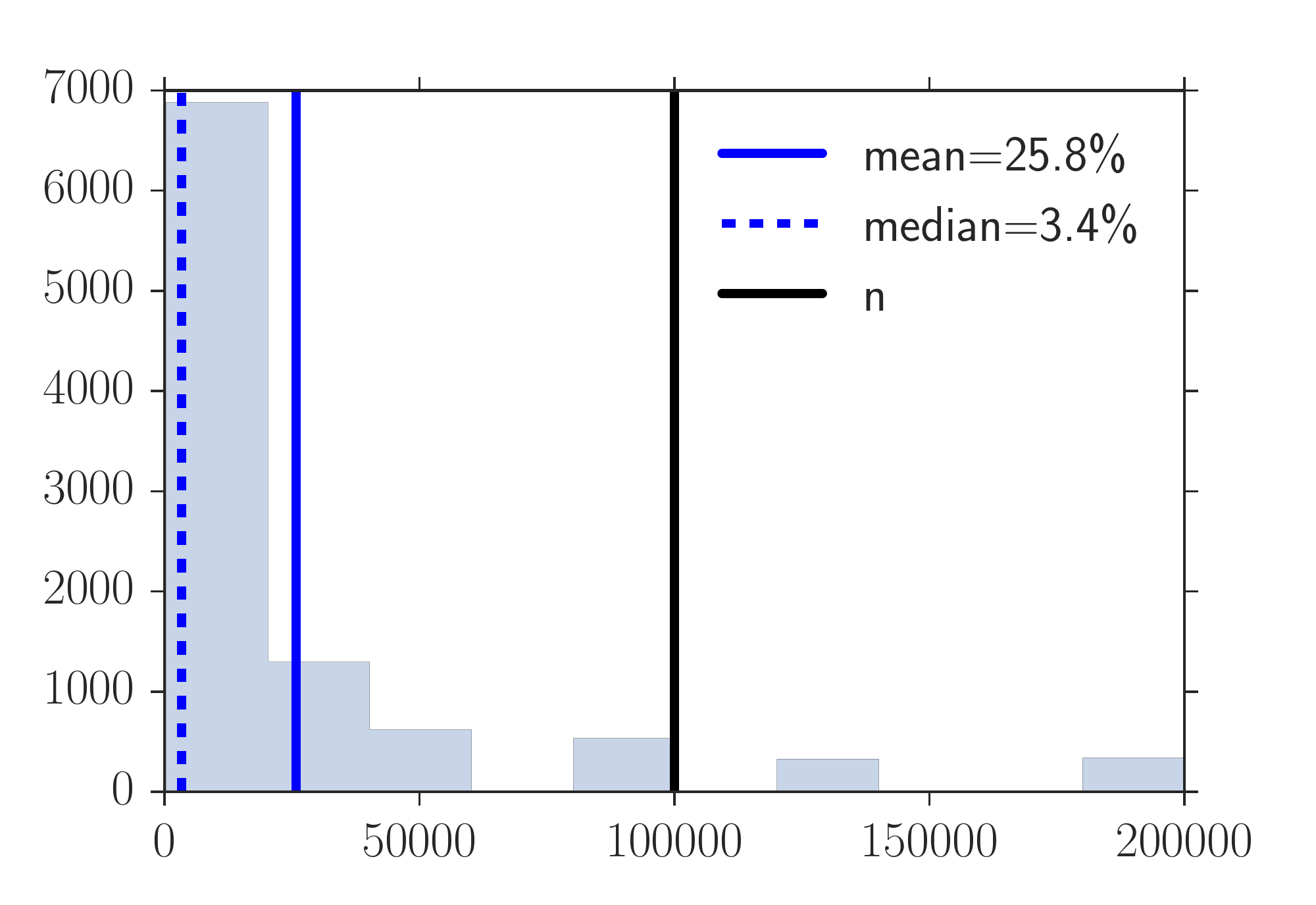}
}
\caption{Results 10\,000 iterations of Austerity MH
  \citep{KoChWe14}. See Section~\ref{sss:CLT} and the caption of
Figure~\ref{f:resultsMH} for details.
\label{f:resultsAusterityMH}
}
\end{figure}

\begin{figure}
\subfigure[Student around MAP, $X_{i}\sim\log\cN(0,1)$]{
\includegraphics[width=\twofig]{\figdir/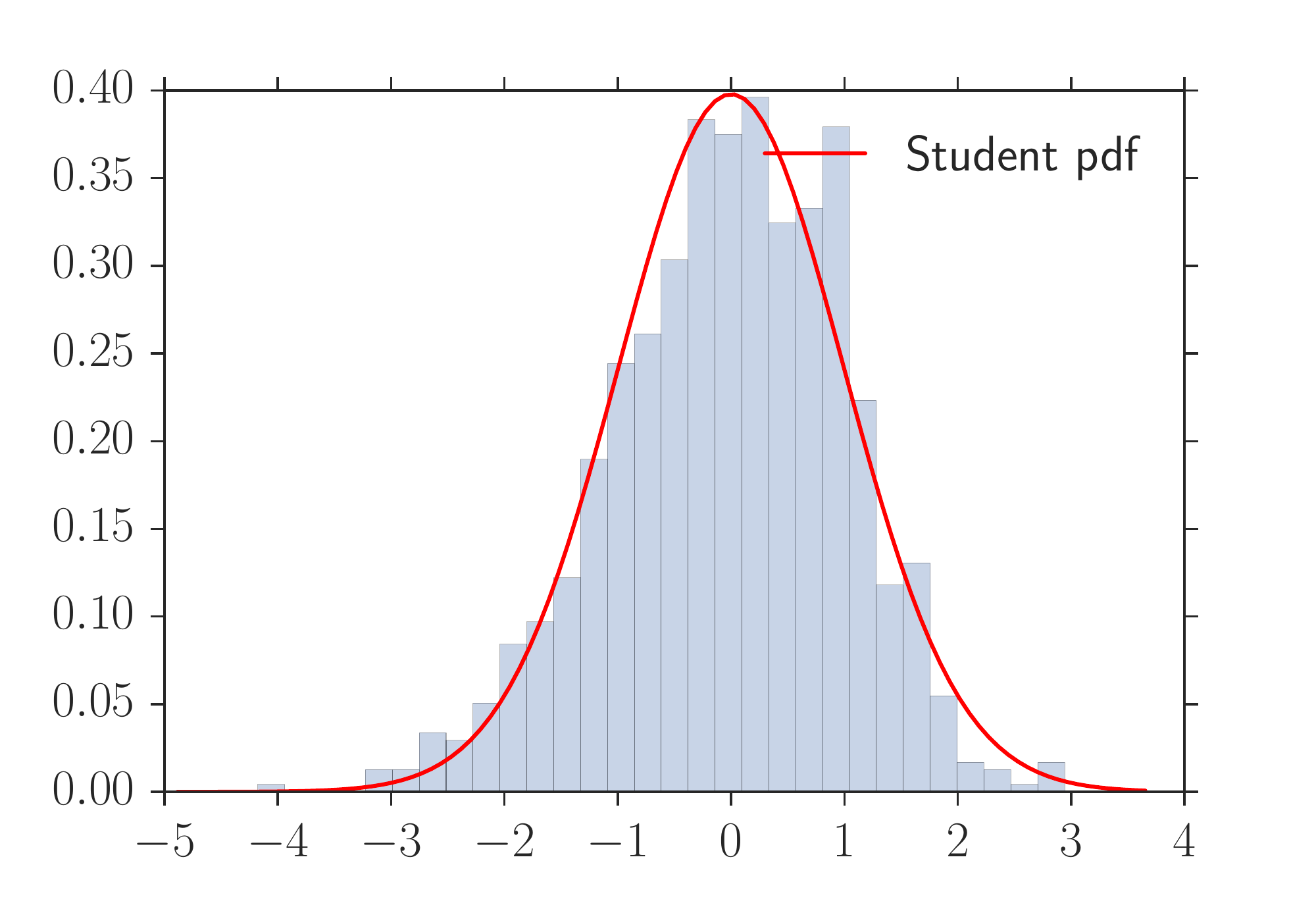}
}
\subfigure[Student around MAP, $X_{i}\sim\log\cN(0,1)$]{
\includegraphics[width=\twofig]{\figdir/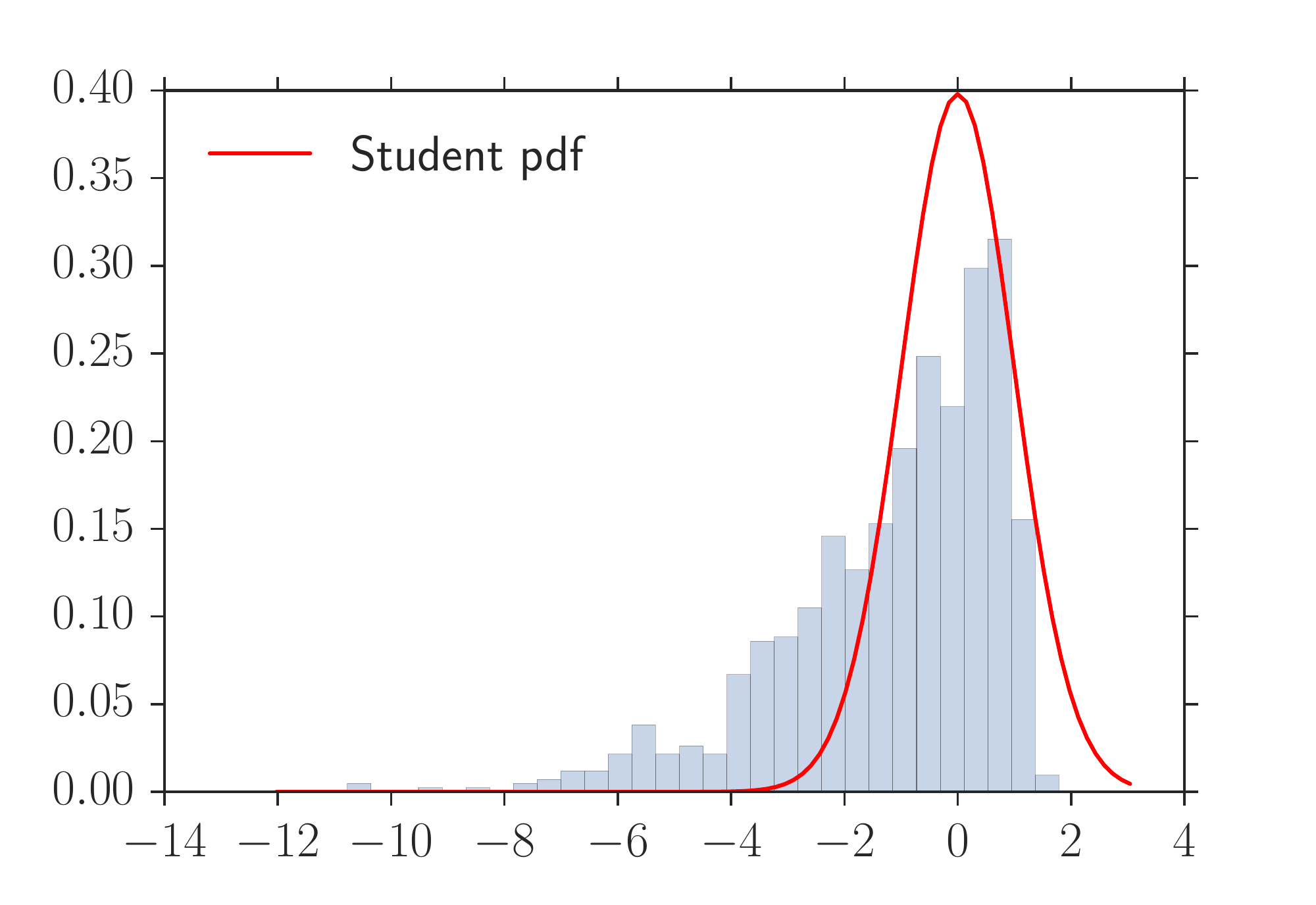}
}
\caption{Histogram of 1000 realizations of the Student statistic
  required in Austerity MH, taken at $\theta=\theta_{\text{MAP}}$ and
  $\theta'\sim q(\cdot\vert\theta)$. The theoretical Student pdf is
  plotted in red.}
\label{f:student}
\end{figure}

We show the results of $10\,000$
iterations of Austerity MH on our two running examples in Figure~\ref{f:resultsAusterityMH}.
The parameters are $\eps=0.05$, corresponding to the p-value threshold
in the aforementioned T-test, and an initial subsample size of $100$
at each iteration. In the Gaussian case, the posterior is rightly
centered, but is slightly too wide. This is a tempering effect due
to too small subsamples, while the CLT-based Student approximation seems reasonable,
as shown in Figure~\ref{f:student}. In the lognormal case, the departure
of the chain from the actual posterior is more remarkable, and relatedly
the CLT approximations of Austerity MH are inaccurate for the
chosen initial subsample size of $100$, as we demonstrate in Figure~\ref{f:student}.
This explains the strong mismatch of the chain and the posterior in
Figure~\ref{f:resultsAusterityMH:histoLogNormal}. The standard deviation
of the fitted Gaussian is largely underestimated, due to small subsamples
which do not include enough of the tails of the log likelihood ratios,
which coincide with the tails of $\cX$. Finally, the reductions in
the number of samples needed per iteration are quite interesting:
half of the iterations require less than $4\%$ of the dataset for
the lognormal case, but this is at the price of a large error in the
posterior approximation. Augmenting the initial size of the subsample will likely make the
CLT approximations tighter, but there is no generic answer as to which
size to choose: any fixed choice will fail on an example where the
log likelihood ratios have heavy enough tails. In both the Gaussian
and the lognormal example, it is actually safer to go with the Bernstein-von
Mises approximation, which costs little more than a run of stochastic
gradient descent, and only requires one CLT approximation, for a sample
of size $n\gg1$. This illustrates the danger of using CLT-based approximations for
small sample sizes, which is related to asymptotic arguments on small
batches in Section~\ref{s:reviewDivide}.

\setcounter{subfigure}{0}

\setcounter{subfigure}{0}

Overall, CLT-based approaches to MH with tall data lead to heuristics
with interesting reductions in the number of samples used, but they
have little theoretical backing so far and they are not robust to the involved
CLT approximations being inaccurate. We note also that the CLT is assumed
to provide a good approximation for the log likelihood or log
likelihood ratio for {\it any}
$\theta,\theta'\in\Tset$, which amounts to more than one Gaussian
assumption. The approaches in this section should thus be applied
with care. As a minimal sanity-check, we recommend using tests of
Gaussianity across $\Theta\times\Theta$ to make sure the CLT assumptions
are realistic. Note that even then, there is no guarantee the above
algorithms have $\pi$ for target, if any.

\subsection{Exchanging acceptance noise for subsampling noise}

This section is an original contribution, which illustrates a way
to obtain subsampling algorithms with guarantees under weaker assumptions
than Gaussianity. This approach is impractical, but it is of methodological
and illustrative interest. First it illustrates a potentially useful
technique to \emph{take advantage of} subsampling noise. Second, it
is our first illustration of the seemingly inevitable $\cO(n)$ average
number of subsamples required per MCMC iteration as soon as we do
not use any CLT-based approximation and require theoretical guarantees.

Let $\theta,\theta'\in\Tset$, and let $x_{1}^{*},\dots,x_{t}^{*}$
be drawn independently with replacement from $\cX$. Let $\Lambda_{t}^{*}(\theta,\theta')$
be the average subsampled log likelihood ratio defined in \eqref{e:estimatorRatio}.
Now, we remark that MH has some inherent noise in its acceptance decision
\eqref{e:acceptanceDecision}, encapsulated by the uniform variable
$u\sim\cU[0,1]$. Why, then, not rely on the subsampling noise to
guarantee exploration, and accept a move if and only if 
\begin{equation}
\Lambda_{t}^{*}(\theta,\theta')+\frac{1}{n}\log\left[\frac{p(\theta')q(\theta\vert\theta')}{p(\theta)q(\theta'\vert\theta)}\right]>0\label{e:modifiedAcceptanceDecision}
\end{equation}
instead of \eqref{e:acceptanceDecision}? This idea has been first used by \cite{BrMeSc08} to develop heuristics
for simulated annealing in the presence of noise. We formalize this
argument here in the context of subsampling. For the sake of simplicity,
assume for a moment we have a flat prior and a symmetric proposal,
so that \eqref{e:modifiedAcceptanceDecision} becomes 
\[
\Lambda_{t}^{*}(\theta,\theta')>0.
\]

We do not assume that the $\Lambda_{t}^{*}(\theta,\theta')$'s are
Gaussianly distributed, but we make the parametric assumption that
the second and third absolute moments $\sigma^{2}$ and $\rho$ of
$-\log p(x_{i}^{*}\vert\theta')+\log p(x_{i}^{*}\vert\theta)$ are
known and independent of $\theta,\theta'$. Applying the Berry-Esseen
inequality \citep{VaWe96} to the variables $-\log p(x_{i}^{*}\vert\theta')+\log p(x_{i}^{*}\vert\theta)$
yields 
\begin{equation}
\left\vert \mathbb{P}(-\Lambda_{t}^{*}(\theta,\theta')\leq u)-\Phi\left(\frac{u+\Lambda_{n}(\theta,\theta')}{\sigma/\sqrt{t}}\right)\right\vert \leq\frac{K(\sigma,\rho)}{\sqrt{t}}\label{e:berryEsseen1}
\end{equation}
for any $u\in\mathbb{R}$, where $\Phi$ is the cdf of a $\cN(0,1)$
variable, and 
\[
\Lambda_{n}(\theta,\theta')\defeq\frac{1}{n}\sum_{i=1}^{n}\log\frac{p(x_{i}\vert\theta')}{p(x_{i}\vert\theta)}
\]
is the average log likelihood ratio. When $u=0$, \eqref{e:berryEsseen1}
yields 
\begin{equation}
\left\vert \mathbb{P}(\Lambda_{t}^{*}(\theta,\theta')\geq0)-\Phi\left(\frac{\Lambda_{n}(\theta,\theta')}{\sigma/\sqrt{t}}\right)\right\vert \leq\frac{K(\sigma,\rho)}{\sqrt{t}}.\label{e:berryEsseen2}
\end{equation}
Now let $C,\lambda>0$ be such that for any $x\in\mathbb{R}$, 
\[
\left\vert \Phi(x)-\frac{1}{1+e^{-\lambda x}}\right\vert \leq C.
\]
\cite{BKKC09} for instance, empirically found $C=0.0095$ and $\lambda=1.702$.
Combining this bound with \eqref{e:berryEsseen2}, we obtain 
\[
\left\vert \mathbb{P}(\Lambda_{t}^{*}(\theta,\theta')\geq0)-\frac{1}{1+e^{-\frac{\lambda\Lambda_n(\theta,\theta')}{\sigma/\sqrt{t}}}}\right\vert \leq C+\frac{K(\sigma,\rho)}{\sqrt{t}}.
\]
Hence, the acceptance probability of an algorithm that would accept
the move from $\theta$ to $\theta'$ by checking whether $\Lambda^{*}(\theta,\theta')>0$
is close to the acceptance probability of an MCMC algorithm with a
Baker acceptance criterion \citep[Section 7.8.1]{RoCa04} that targets $\pi^{\beta}$ with temperature $\beta=\frac{\lambda\sqrt{t}}{n\sigma}$.
Arguments such as \cite[Lemma 3.1, Proposition 3.2]{BaDoHo14} could
then help concluding that the distance between the kernels of both
Markov chains is controlled, which would yield positive ergodicity
results, in the line of \cite[Proposition 3.2]{BaDoHo14}. This reasoning
shows again a close relation between subsampling and tempering, as
in Section~\ref{ss:naiveSubsampling}, with a clear link between
the variance of the subsampled log likelihood ratios and the temperature.

Now, from a practical point of view, in simple applications such as
logistic regression, $\sigma$ is of the order of $\Vert\theta-\theta'\Vert$,
which in turn should be of order $\cO_{p}(n^{-1/2})$ if the MCMC proposal
is a Gaussian random walk with covariance similar to that of $\pi$,
see \cite{BaDoHo14}. This means that $t$ has to be of order $n$
for the temperature $\beta$ to be of order $1$, and this approach
is thus bound to use $\cO(n)$ subsamples per iteration! In conclusion,
robustness to non-Gaussianity leads to requiring a fixed proportion
of the whole dataset on average, even in the favourable case when
one controls the first three moments of the subsampling noise and
one swaps subsampling noise for the inherent MCMC acceptance noise.

\subsection{Confidence samplers}

\label{ss:confidenceSamplers} In \citep{BaDoHo14}, we proposed a
controlled approximation of the acceptance decision \eqref{e:acceptanceDecision}.
Indeed, let us fix $\theta,\theta'$ and momentarily assume that $x\mapsto\log[p(x\vert\theta')/p(x\vert\theta)]$
was Lipschitz with known constant. Then, having observed the log likelihood
ratio at some points $\{x_{i}^{*},i=1,\dots,t\}\subset\cX$, one could
build a lower and an upper bound for the complete log likelihood ratio
\[
\frac{1}{n}\sum_{i=1}^{n}\log\left[\frac{p(x_{i}\vert\theta')}{p(x_{i}\vert\theta)}\right],
\]
simply by associating each $x_{i}$ with the nearest point among $\{x_{1}^{*},\dots,x_{t}^{*}\}$.
These bounds could be refined by augmenting the set of observed log
likelihoods ratios, until eventually one knows for sure whether \eqref{e:acceptanceDecision}
holds.

Now, concentration inequalities allow softer bounds and require less
than this Lipschitz assumption. If one knows a bound for the range
\begin{eqnarray}
C_{\theta,\theta'}\defeq\max_{i=1}^{n}\left\vert \log\left[\frac{p(x_{i}\vert\theta')}{p(x_{i}\vert\theta)}\right]\right\vert ,\label{e:range}
\end{eqnarray}
then concentration inequalities such as Hoeffding's or Bernstein's,
yield confidence bounds $c_{t}(\delta)$ such that 
\begin{eqnarray}
\mathbb{P}\left(\left\vert \frac{1}{t}\sum_{i=1}^{t}\log\left[\frac{p(x_{i}^{*}\vert\theta')}{p(x_{i}^{*}\vert\theta)}\right]-\frac{1}{n}\sum_{i=1}^{n}\log\left[\frac{p(x_{i}\vert\theta')}{p(x_{i}\vert\theta)}\right]\right\vert >c_{t}(\delta)\right)\geq1-\delta,\label{e:defConcentration}
\end{eqnarray}
where the probability is taken over $x_{1}^{*},\dots,x_{t}^{*}$ drawn
uniformly from $\cX$, with or without replacement. Borrowing from
the bandit literature, we explain in \citep{BaDoHo14} how to leverage
such confidence bounds to automatically select a subsample size $T$
such that the right MH acceptance decision is taken with a user-specified
probability $1-\delta$. Note that for our algorithm to bring any
improvement over the ideal MH, the range \eqref{e:range} must be
cheap to compute, i.e. cheaper than $\cO(n)$. This is the case for
logistic regression, for example, but it is the major limitation of
the approach in \cite{BaDoHo14}. We showed in \cite[Proposition 3.2]{BaDoHo14}
that if the ideal MH sampler is uniformly ergodic then the resulting
algorithm inherits the uniform ergodicity of the ideal MH sampler,
with a convergence speed that is within $\cO(\delta)$ of that of
the ideal MH. Importantly, we showed that our sampler then admits a
limiting distribution, which is also within $\cO(\delta)$ of $\pi$. Uniform ergodicity
is a very strong assumption and it would be worth extending these results
to the geometrically ergodic scenario. There has recently been work in
this direction \citep{AFEB14,PiSmSub,RuScSub}. 

\begin{figure}
\subfigure[Chain histograms, $X_{i}\sim\cN(0,1)$]{
\includegraphics[width=\twofig]{\figdir/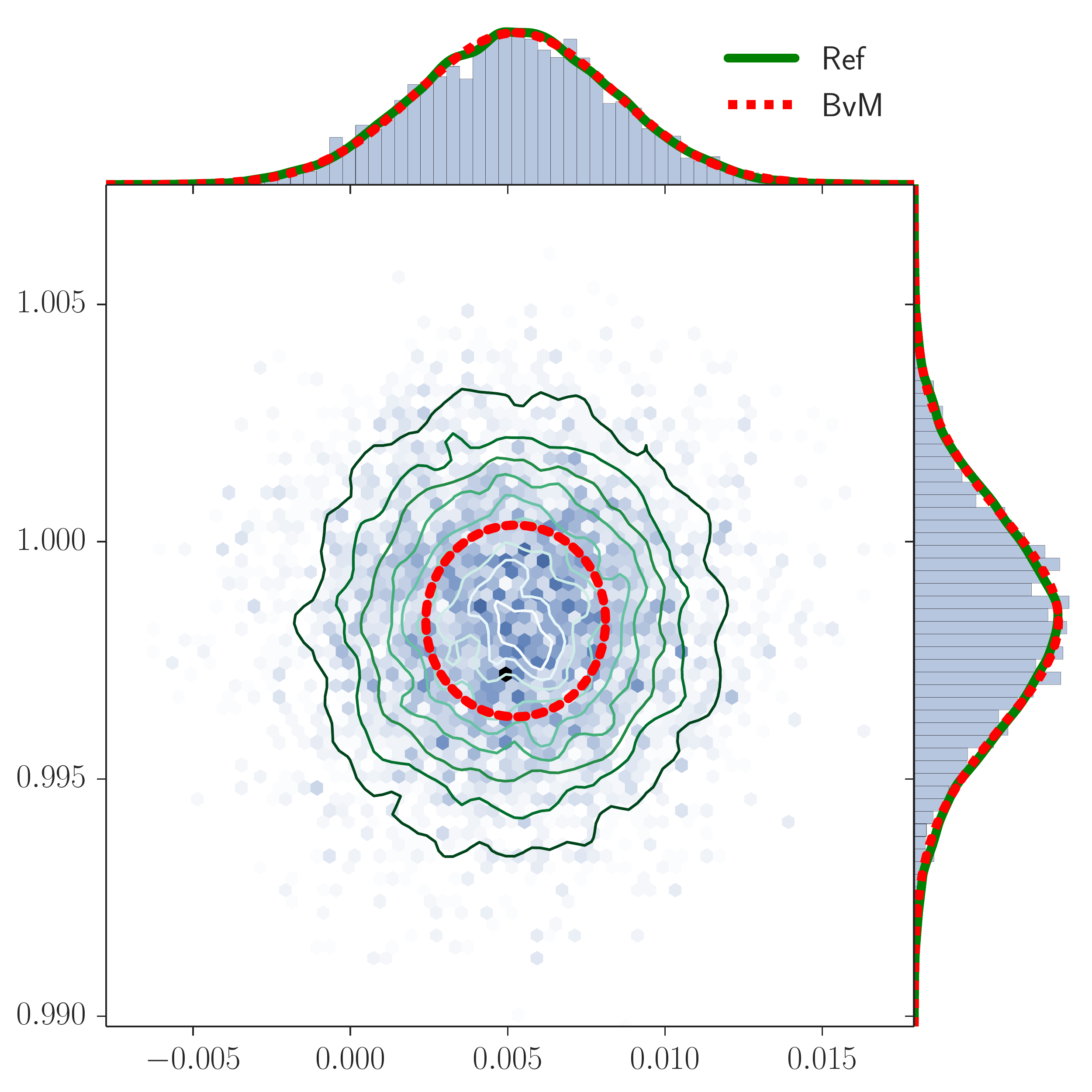}
}
\subfigure[Chain histograms, $X_{i}\sim\log\cN(0,1)$]{
\includegraphics[width=\twofig]{\figdir/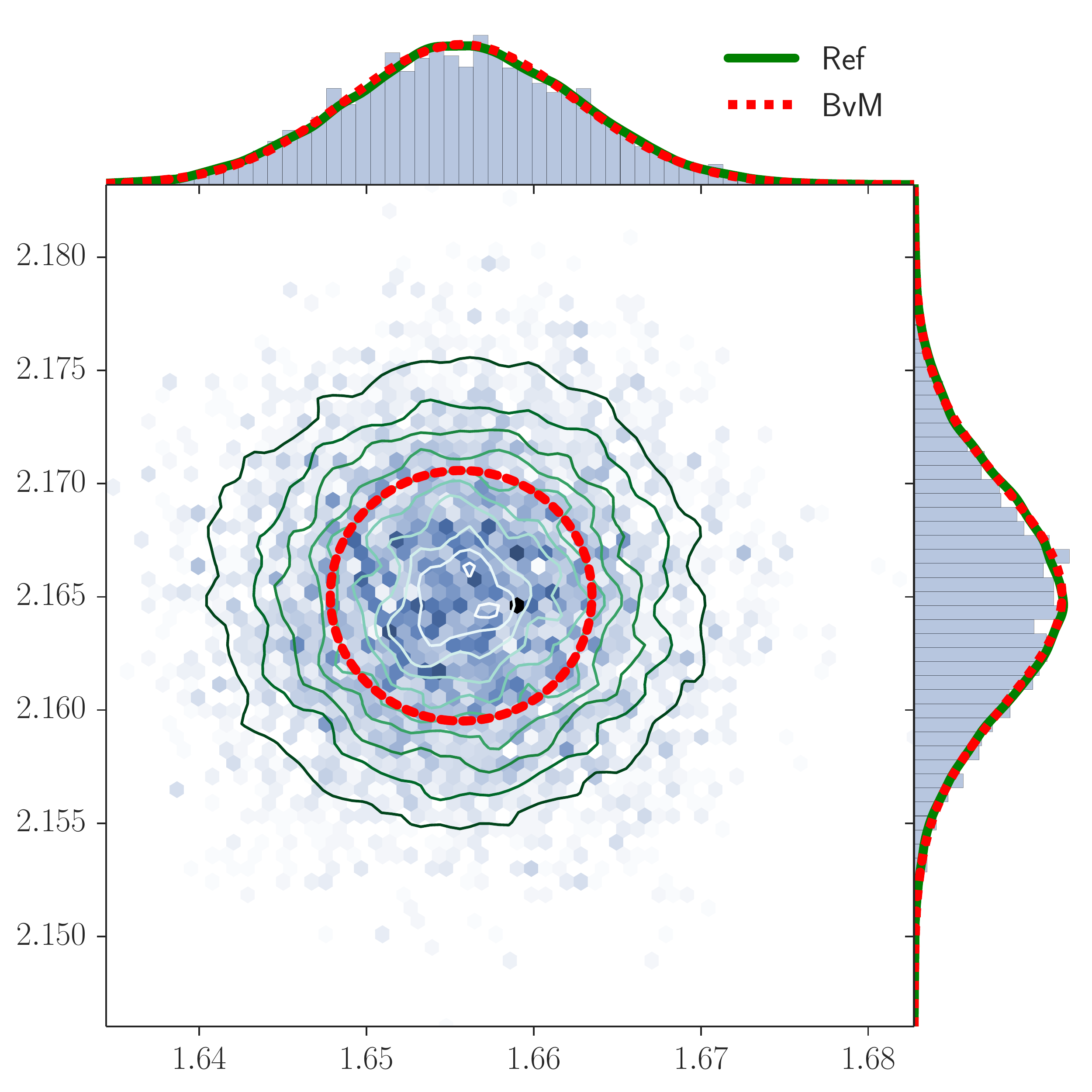}
}\\
\subfigure[Autocorr. of $\log\sigma$, $X_{i}\sim\cN(0,1)$]{
\includegraphics[width=\twofig]{\figdir/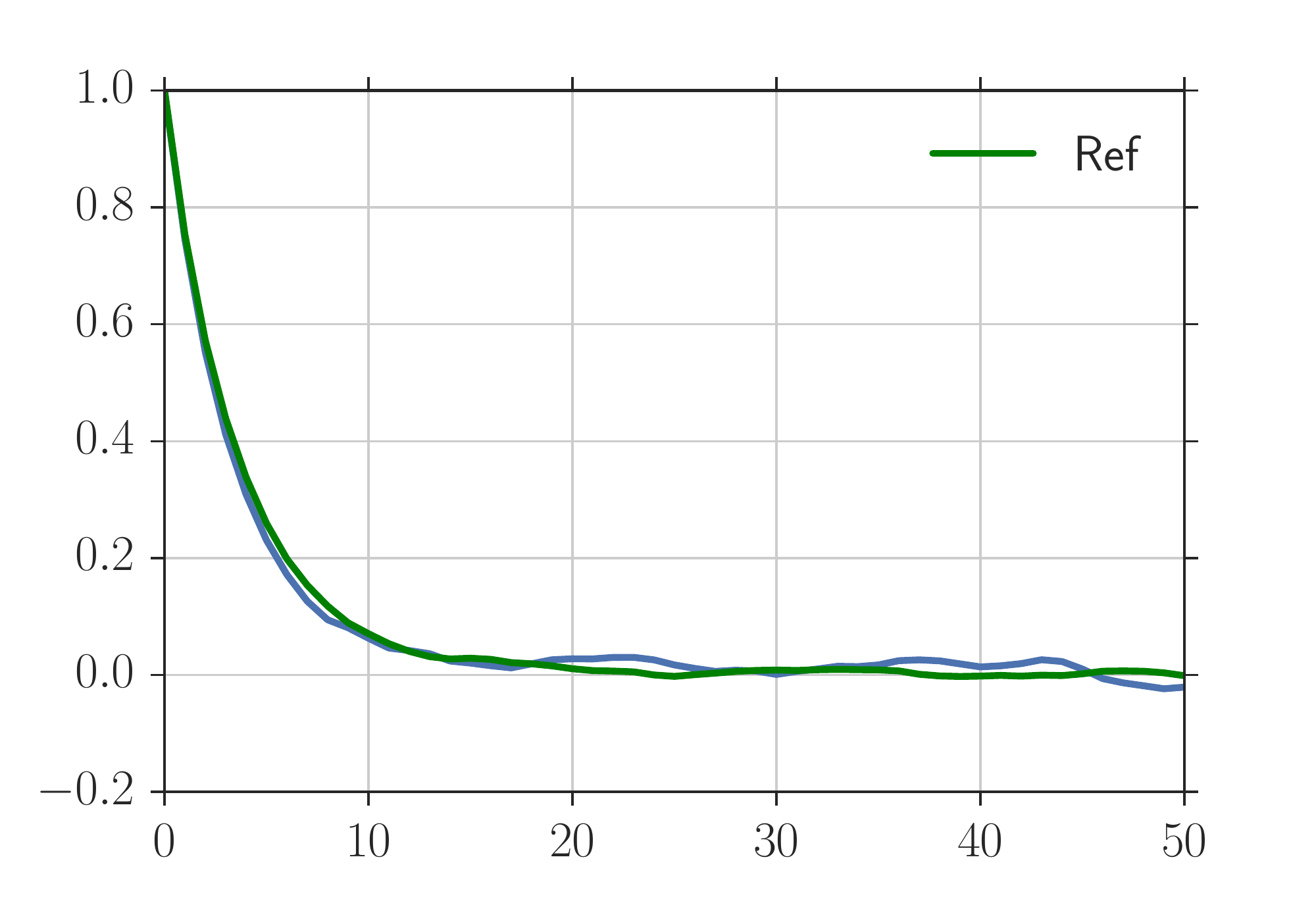}
}
\subfigure[Autocorr. of $\log\sigma$, $X_{i}\sim\log\cN(0,1)$]{
\includegraphics[width=\twofig]{\figdir/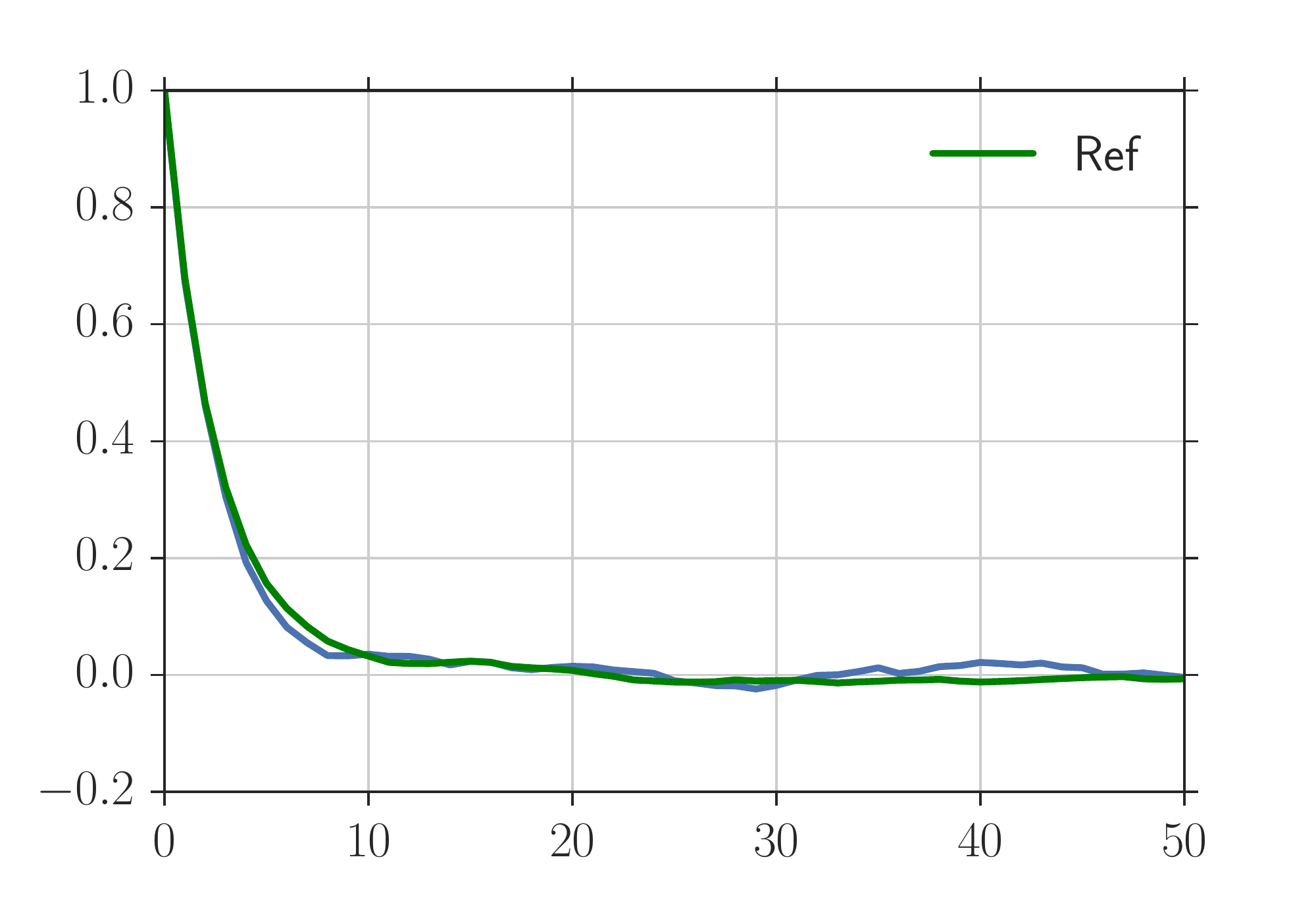}
}\\
\subfigure[Number of likelihood evals, $X_{i}\sim\cN(0,1)$]{
\includegraphics[width=\twofig]{\figdir/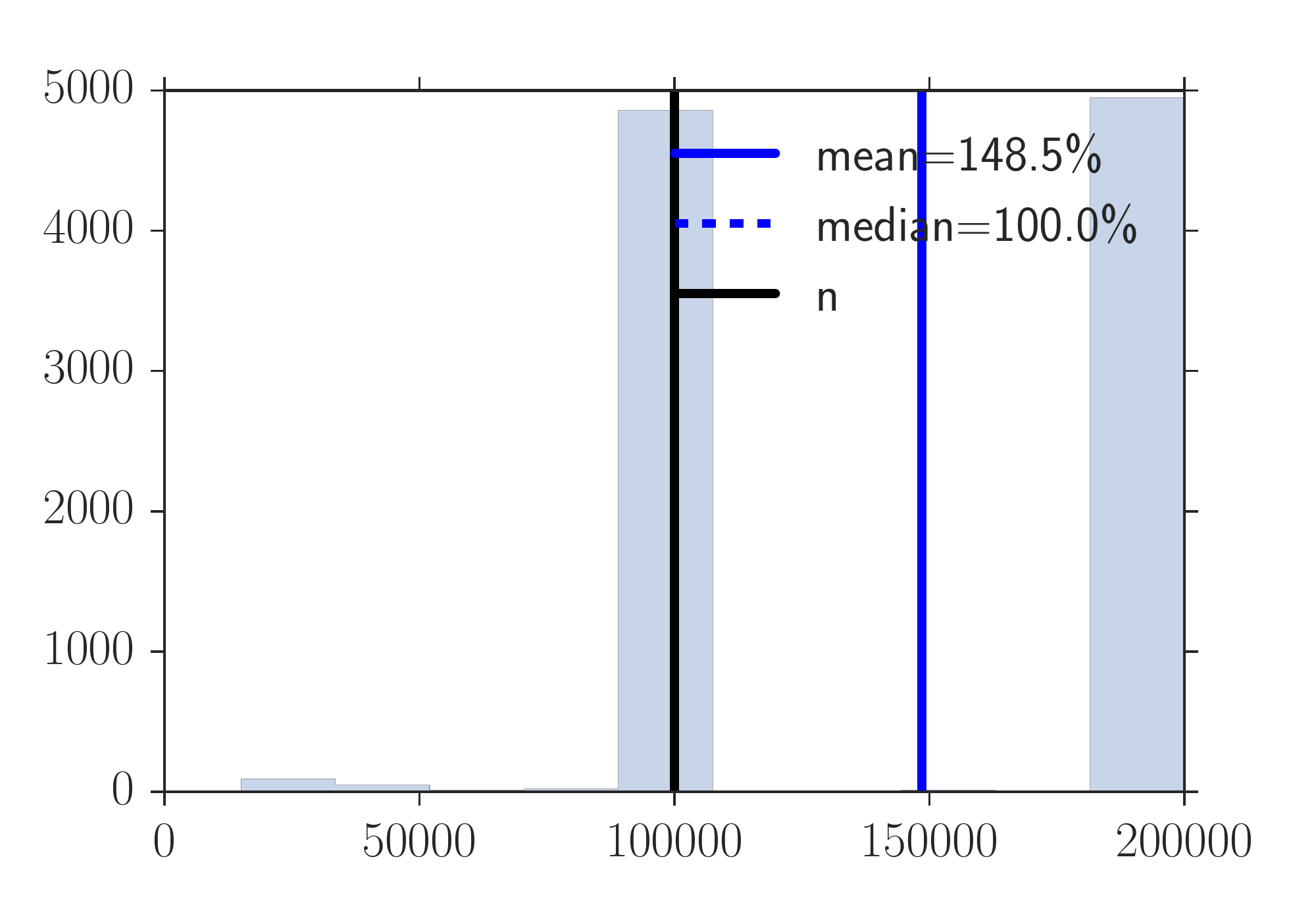}
}
\subfigure[Number of likelihood evals, $X_{i}\sim\log\cN(0,1)$]{
\includegraphics[width=\twofig]{\figdir/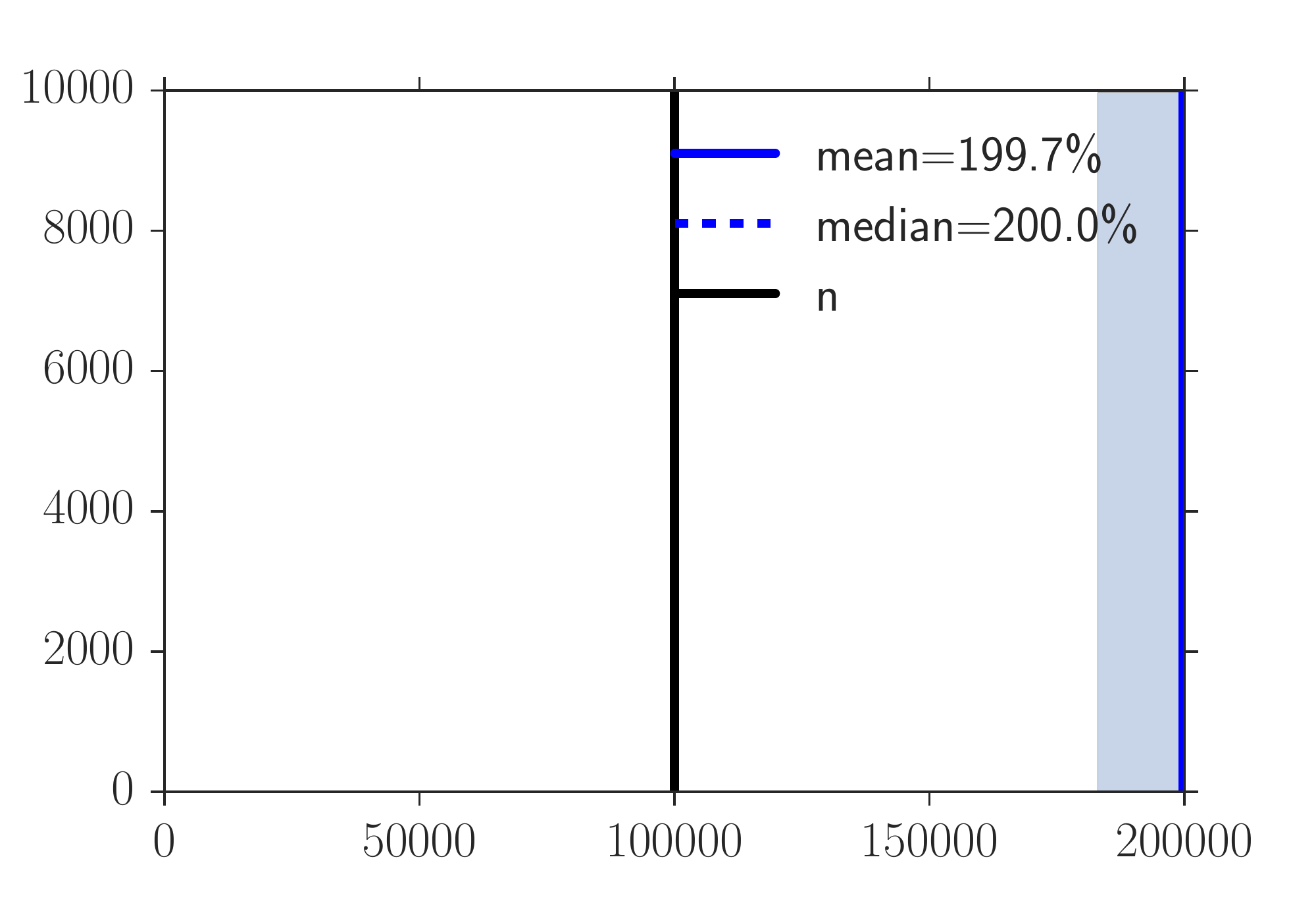}
}\\
\caption{
Results of 10\,000 iterations of the vanilla confidence sampler
\citep{BaDoHo14}, see Section~\ref{ss:confidenceSamplers} and the caption of
Figure~\ref{f:resultsMH} for details.
}
\label{f:resultsVanillaConfidence}
\end{figure}

On the negative side, we demonstrated in \citep{BaDoHo14}
that vanilla confidence samplers still require $\cO(n)$ samples at each iteration
at equilibrium, where the proportionality constant is the variance
of the log likelihood ratio under subsampling. This statement relies
on the leading term in $c_{t}(\delta)$ being of order $t^{-1/2}$. In practice, the results of the vanilla confidence sampler on our
running examples are shown in Figure~\ref{f:resultsVanillaConfidence}.
We set $\delta=0.1$ and we place ourselves in the favourable scenario
where the algorithm has access to the actual range of each log likelihood ratio. The number of likelihood
evaluations is estimated as follows: we take by default \emph{twice}
the detected value $T$ for the subsample size in general, but only
\emph{once} when the previous iteration required computing all $n$
likelihoods at the current state of the chain. Still, even in these
favourable conditions, the algorithm
basically requires essentially the whole dataset at each iteration.

\setcounter{subfigure}{0}

%
%

\begin{figure}
\subfigure[Chain histograms, $X_{i}\sim\cN(0,1)$]{
\includegraphics[width=\twofig]{\figdir/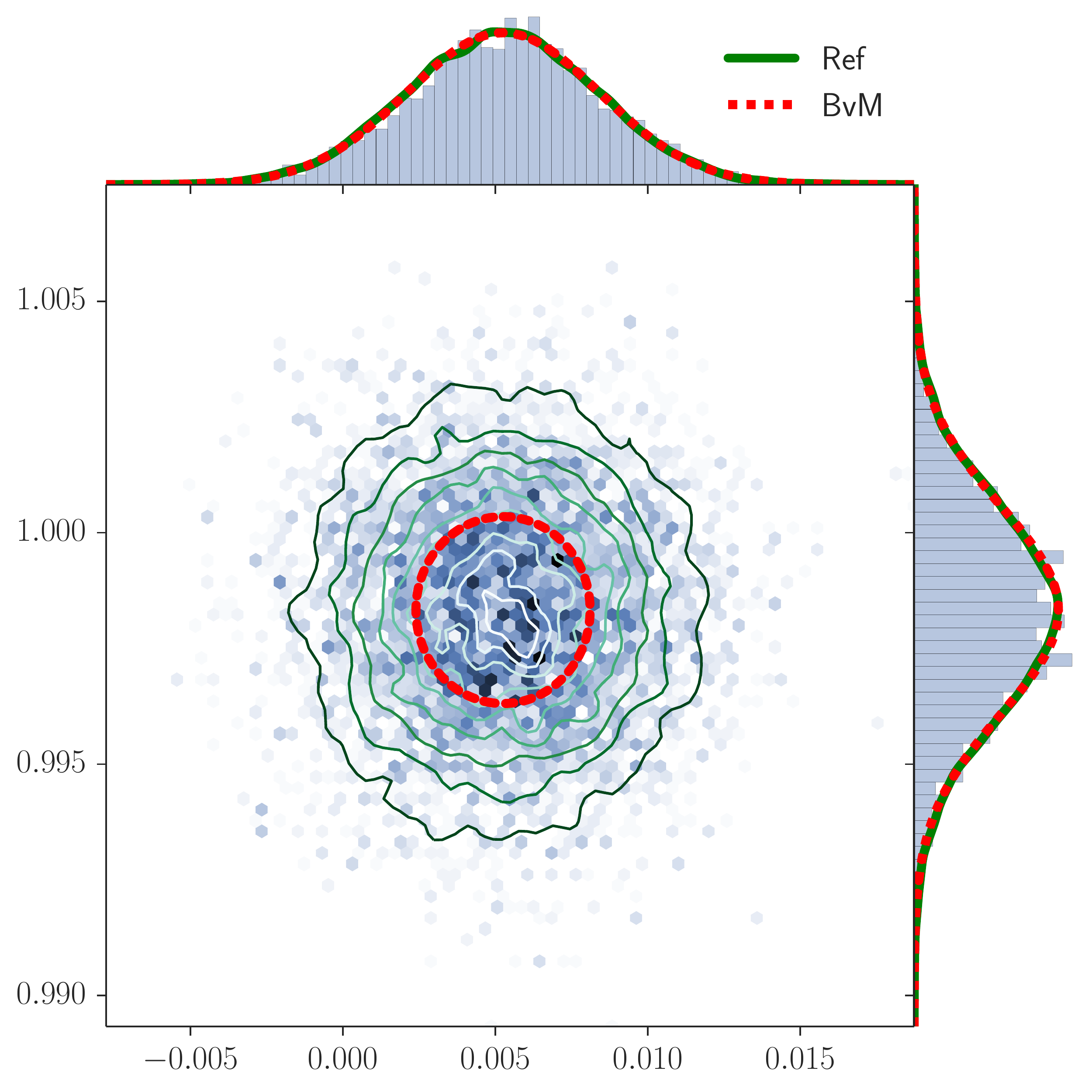}
}
\subfigure[Chain histograms, $X_{i}\sim\log\cN(0,1)$]{
\includegraphics[width=\twofig]{\figdir/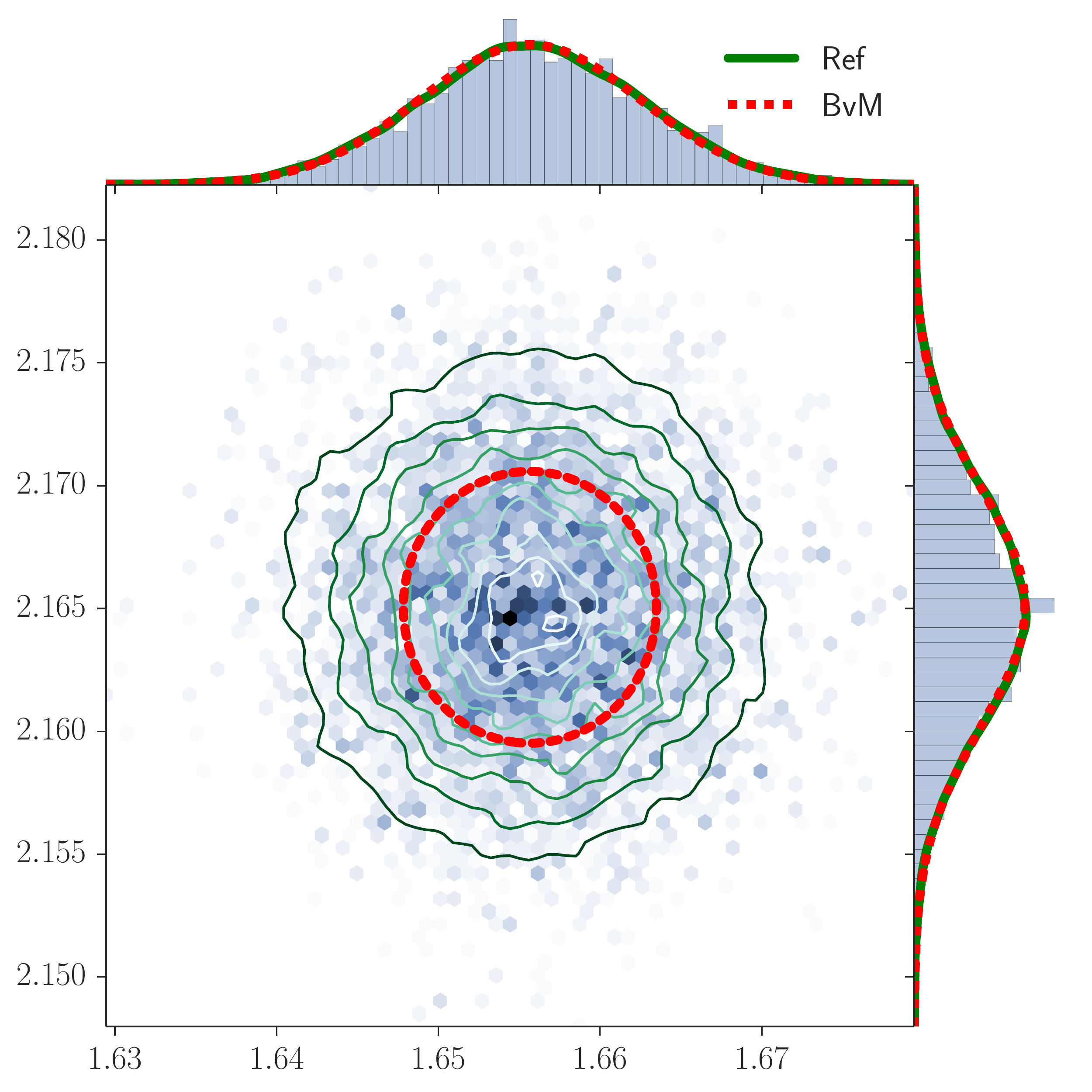}
}\\
\subfigure[Autocorr. of $\log\sigma$, $X_{i}\sim\cN(0,1)$]{
\includegraphics[width=\twofig]{\figdir/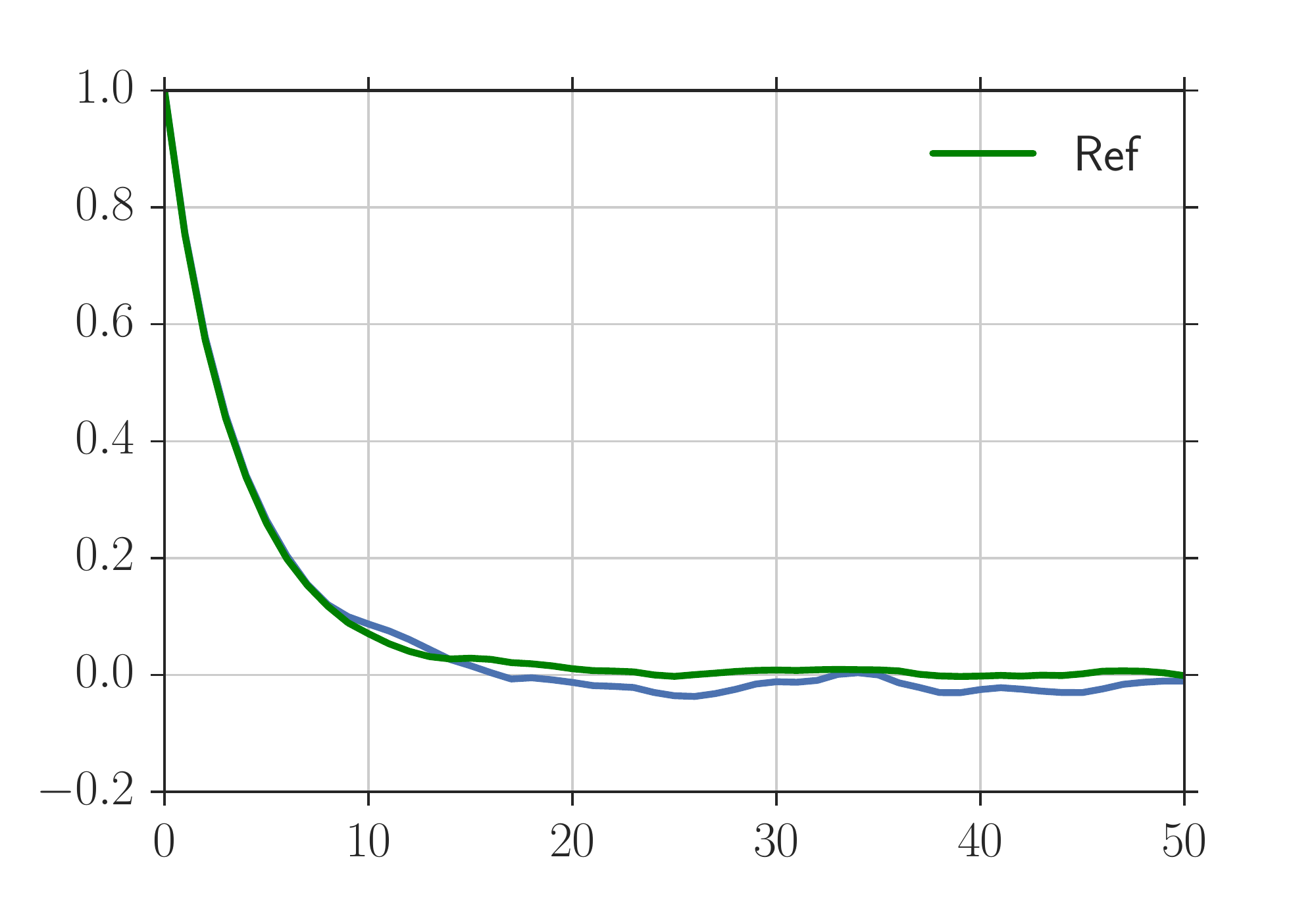}
\label{f:resultsConfidenceProxy:autocorrGaussian}
}
\subfigure[Autocorr. of $\log\sigma$, $X_{i}\sim\log\cN(0,1)$]{
\includegraphics[width=\twofig]{\figdir/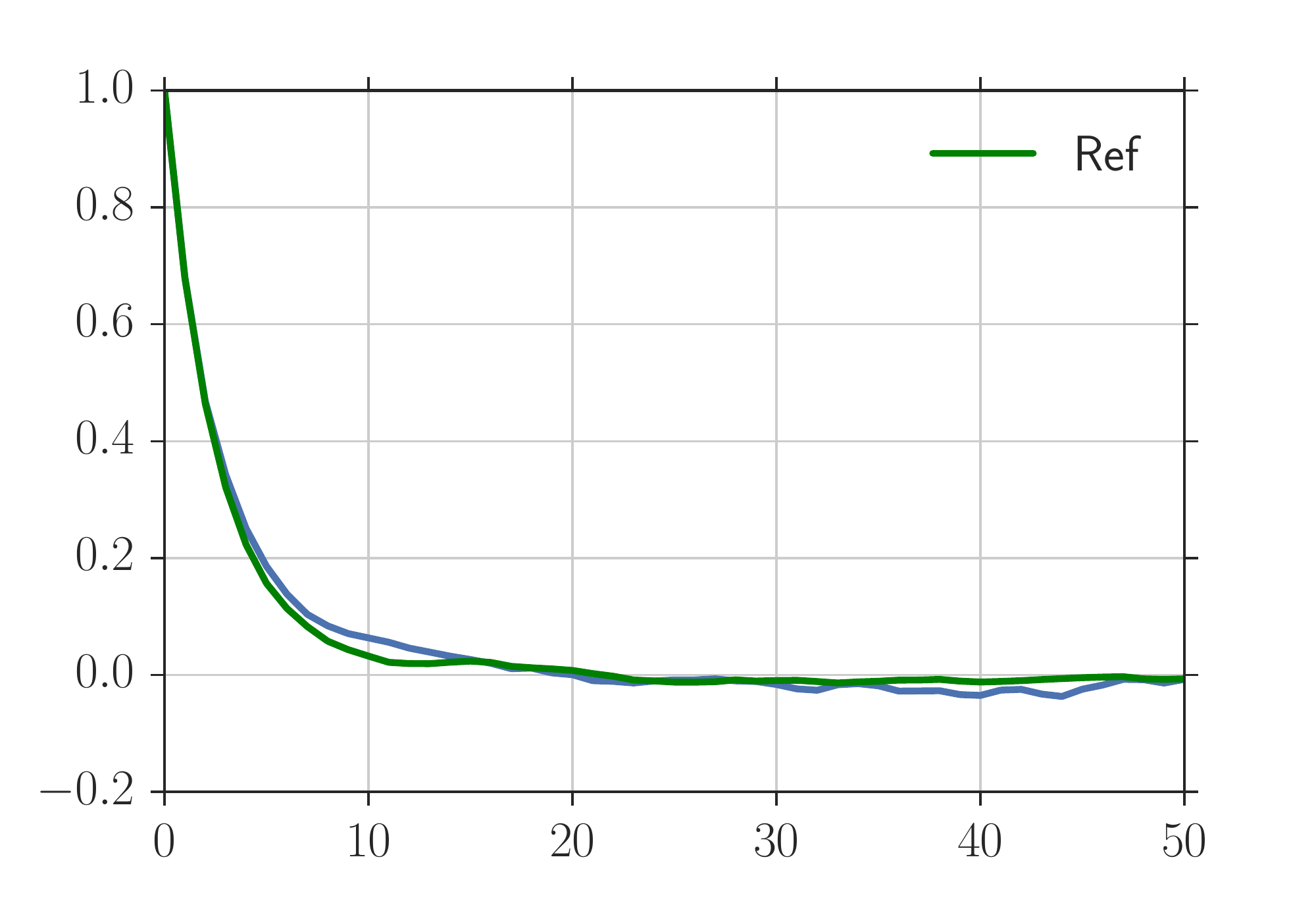}
\label{f:resultsConfidenceProxy:autocorrLogNormal}
}\\
\subfigure[Number of likelihood evals, $X_{i}\sim\cN(0,1)$]{
\includegraphics[width=\twofig]{\figdir/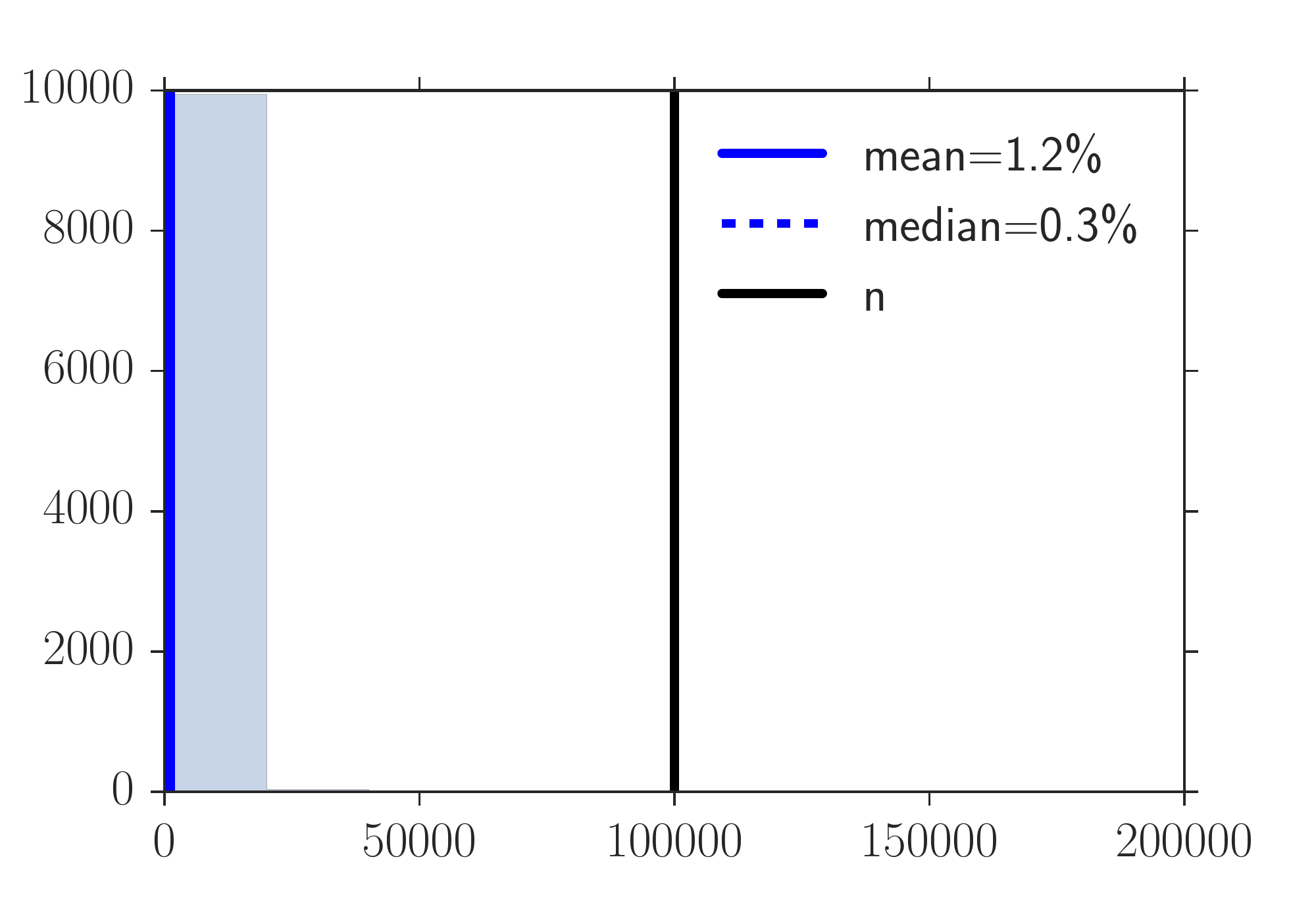}
}
\subfigure[Number of likelihood evals, $X_{i}\sim\log\cN(0,1)$]{
\includegraphics[width=\twofig]{\figdir/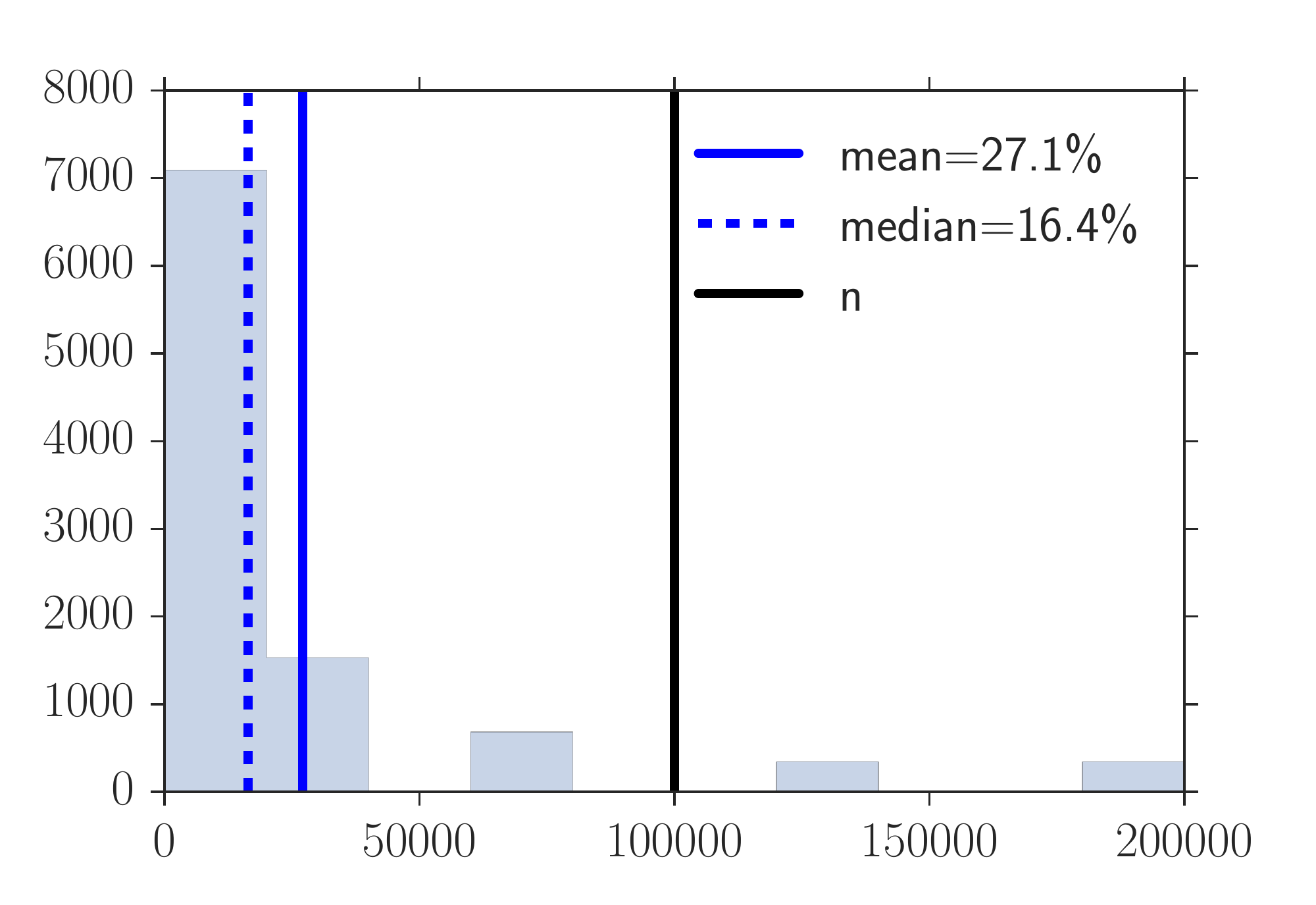}
}\\
\caption{
Results of 10\,000 iterations of the confidence sampler of Section~\ref{s:concentration}
with a single 2nd order Taylor proxy at $\theta_{\text{MAP}}$.
}
\label{f:resultsConfidenceProxy}
\end{figure}

Concentration inequalities are ``worst-case'' guarantees, and the
theoretical results come at the price of a smaller reduction in the
number of samples required. When the target is locally Gaussian, e.g.
when Bernstein-von Mises yields a good approximation, there is potentially a lot to be
gained, as empirically demonstrated by \cite{KoChWe14}, for example.
In the current paper, we propose in Section~\ref{s:concentration}
a modified confidence sampler that can leverage concentration of the
target to yield dramatic empirical gains while not sacrificing any
theoretical guarantee of the confidence sampler. The basic tool is
a cheap proxy for the log likelihood ratio that acts as a control
variate in the concentration inequality \eqref{e:defConcentration}.
Using a 2nd order Taylor expansion centered at the maximum of the
likelihood --~obtained with a stochastic gradient descent for example~--
allows to replace many likelihood evaluations by the evaluation of
this Taylor expansion. Figure~\ref{f:resultsConfidenceProxy} shows
the results of this new confidence sampler with proxy on our running
Gaussian and lognormal examples. Our algorithm outperforms all preceding
methods, using almost no sample in the Gaussian case where it \emph{automatically}
detects that a quadratic form is enough to represent the log likelihood
ratio. Finally, we demonstrate in Sections~\ref{ss:gain} and \ref{s:experiments}
that this new algorithm can require less than $\cO(n)$ likelihood
evaluations per iteration. Combined with the statements in \cite{BaDoHo14}
that each iteration is almost as efficient as the ideal MH, which
is further supported by the match of the autocorrelation functions
in Figures~\ref{f:resultsConfidenceProxy:autocorrGaussian} and \ref{f:resultsConfidenceProxy:autocorrLogNormal},
this opens up big data horizons. We give full details on the confidence
algorithm with proxy in Section~\ref{s:concentration}.

%
%

\setcounter{subfigure}{0}

\section{An improved confidence sampler}
\label{s:concentration}

In this section, we build upon the confidence sampler in \citep{BaDoHo14}
by introducing likelihood proxies, which act as control variates for
the individual likelihoods.

\subsection{Introducing proxies in the confidence sampler}
\label{ss:introducingProxies} 
We start by recalling the pseudocode
of the confidence sampler in \citep{BaDoHo14} in Figure~\ref{f:MHSubLhd},
using sampling with replacement and a generic empirical concentration
bound $c_{t}(\delta)$. In practice, one can think of the empirical
Bernstein bound of \cite{AuMuSz09} 
\begin{equation}
c_{t}(\delta)=\hat{\sigma}_{t,\theta,\theta'}\sqrt{\frac{2\log(3/\delta)}{t}}+\frac{6C_{\theta,\theta'}\log(3/\delta)}{t},\label{e:defBernstein}
\end{equation}
where $\hat{\sigma}_{t,\theta,\theta'}$ is the sample standard deviation
of the log likelihood ratio 
\[
\left\{ \log\left[\frac{p(x_{i}^{*}\vert\theta')}{p(x_{i}^{*}\vert\theta)}\right],i=1,\dots,t\right\} ,
\]
and $C_{\theta,\theta'}$ is their range, defined in \eqref{e:range}.
We emphasize that other choices of sampling procedure and concentration
inequalities are valid, as long as they guarantee a concentration
like \eqref{e:defConcentration}. We refer the reader to \citep{BaDoHo14}
for a proof of the correctness of the confidence sampler and implementation
details.

\begin{figure}[!ht]
\centerline{ 
\scalebox{0.95}{
\begin{algorithm}{$\Algo{ConfidenceSampler}\big(p(x\vert\theta),\, p(\theta),\, q(\theta'\vert\theta),\,\theta_{0},\, N_{\text{iter}},\,\cX,\,(\delta_{t}),\, C_{\theta,\theta'},\big)$}
\Aitem \For $k\setto1$ \To $N_{\text{iter}}$ 
\Aitem \mtt $\theta\setto\theta_{k-1}$
\Aitem \mtt $\theta'\sim q(.\vert\theta)$, $u\sim\cU_{(0,1)}$,
\Aitem \mtt $\psi(u,\theta,\theta')\setto\frac{1}{n}\log\left[u\frac{p(\theta)q(\theta'\vert\theta)}{p(\theta')q(\theta\vert\theta')}\right]$
\Aitem \mtt $t\setto0$ 
\Aitem \mtt $t_{\text{look}}\setto0$ \label{ai:tLookInit}
\Aitem \mtt $\Lambda^{*}\setto0$ 
\Aitem \mtt $\cX^*\setto \emptyset$ \mt \algoremark{\unn{Keeping track
    of points already used}}
\Aitem \mtt $b\setto1$ \mt \algoremark{\unn{Initialize batchsize to 1}}
\Aitem \mtt $\textsc{Done} \setto \textsc{False}$
\Aitem \mtt \While $\textsc{Done}==\textsc{False}$ \Do
\Aitem \mtttt $x_{t+1}^{*},\dots,x_{b}^{*}\sim_{\text{w/
    repl.}}\cX\setminus\cX^*$ \mt \algoremark{\unn{Sample new batch with replacement}} 
\Aitem \mtttt $\cX^* \setto \cX^*\cup \{ x_{t+1}^{*},\dots,x_{b}^{*}\}$
\Aitem \mtttt
$\Lambda^{*}\setto\frac{1}{b}\left(t\Lambda^{*}+\sum_{i=t+1}^{b}\log\left[\frac{p(x_{i}^{*}\vert\theta')}{p(x_{i}^{*}\vert\theta)}\right]\right)$ \label{ai:LambdaStar}
\Aitem \mtttt $t\leftarrow b$
\Aitem \mtttt $c\setto c_t(\delta_{t_\text{look}})$ 
\Aitem \mtttt $t_{\text{look}}\setto
t_{\text{look}}+1$ \label{ai:tLookInc}
\Aitem \mtttt $b\setto n\wedge \lceil\gamma t\rceil$\mt
\algoremark{\unn{Increase batchsize geometrically}} \label{ai:batchsize}
\Aitem \mtttt \If $\left\vert
  \Lambda^{*}-\psi(u,\theta,\theta')\right\vert \geq c$ \textbf{or}
$b>n$ \label{ai:while} 
\Aitem \mtttt\mtt $\textsc{Done} \setto \textsc{True}$
\Aitem \mtt \If $\Lambda^{*}>\psi(u,\theta,\theta')$ \label{ai:MHSubLhdAcceptanceBeginning}
\Aitem \mtttt $\theta_{k}\setto\theta'$ \mt \algoremark{\unn{Accept}}
\Aitem \mtt \Else $\theta_{k}\setto\theta$ \mt \algoremark{\unn{Reject}}
\Aitem \Return $(\theta_{k})_{k=1,\dots,N_{\text{iter}}}$
\end{algorithm}
} 
}
\caption{Pseudocode of the confidence MH from \citep{BaDoHo14}. Our
  contribution is a modification of Steps~\ref{ai:LambdaStar}, \ref{ai:while} and
  \ref{ai:MHSubLhdAcceptanceBeginning} to introduce proxies for the log likelihood
  ratios, see Section~\ref{s:concentration}.}
\label{f:MHSubLhd} 
\end{figure}

The bottleneck for the performance of the confidence sampler was identified
in \citep{BaDoHo14} as the expectation w.r.t. $\pi(\theta)q(\theta'\vert\theta)$
of the variance of the log likelihood ratio $\log p(x\vert\theta')/p(x\vert\theta)$
w.r.t. to the empirical distribution of the observations. We now propose
a control variate technique inspired from the Firefly MH of \cite{MaAd14}
to lower this variance down when an accurate and cheap proxy of the
log likelihood is known.

We require a proxy for the log likelihood ratio that may decrease
the variance of the log likelihood ratio or its range. More precisely,
let $\p_{i}(\theta,\theta')$ be such that for any $\theta,\theta'\in\Theta$, 
\begin{enumerate}
\item $\wp_{i}(\theta,\theta')\approx\ell_{i}(\theta')-\ell_{i}(\theta)$ 
\item $\sum_{i=1}^{n}\wp_{i}(\theta,\theta')$ can be computed cheaply. 
\item $\vert\ell_{i}(\theta')-\ell_{i}(\theta)-\wp_{i}(\theta,\theta')\vert$
can be bounded uniformly in $1\leq i\leq n$, and the bound is cheap
to compute. 
\end{enumerate}
We now simply remark that the acceptance decision \eqref{e:acceptanceDecision}
in MH is equivalent to checking whether 
\begin{equation}
\frac{1}{n}\sum_{i=1}^{n}\left[\log\frac{p(x_{i}\vert\theta')}{p(x_{i}\vert\theta)}-\wp_{i}(\theta,\theta')\right]>\frac{1}{n}\log u-\frac{1}{n}\log\left[\frac{p(\theta')q(\theta\vert\theta')}{p(\theta)q(\theta'\vert\theta)}\right]-\frac{1}{n}\sum_{i=1}^{n}\wp_{i}(\theta,\theta').\label{e:acceptanceDecisionWithProxy}
\end{equation}
Building the confidence sampler on \eqref{e:acceptanceDecisionWithProxy}
leads to the same pseudocode as in Figure~\ref{f:MHSubLhd}, except
that Step~\ref{ai:LambdaStar} is replaced by 
\[
\Lambda^{*}\setto\frac{1}{b}\left(t\Lambda^{*}+\sum_{i=t+1}^{b}\left[\log\frac{p(x_{i}^{*}\vert\theta')}{p(x_{i}^{*}\vert\theta)}-\wp_{i}(\theta,\theta')\right]\right),
\]
the condition in Step~\ref{ai:while} is replaced by 
\[
\left\vert \Lambda^{*}+\frac{1}{n}\sum_{i=1}^{n}\wp_{i}(\theta,\theta')-\psi(u,\theta,\theta')\right\vert \geq c,
\]
and the condition in Step~\ref{ai:MHSubLhdAcceptanceBeginning} becomes
\[
\Lambda^{*}>\psi(u,\theta,\theta')-\frac{1}{n}\sum_{i=1}^{n}\wp_{i}(\theta,\theta').
\]

For completeness, we restate here in Proposition~\ref{p:uniformErgodicity} that
the vanilla confidence sampler inherits the uniform ergodicity
of the underlying MH sampler, that its target is within $\cO(\delta)$
of $\pi$, and that the difference in speed of convergence is also
controlled by $\delta$. Let $P$ be the underlying MH kernel, and
$\tP$ the kernel of the confidence sampler described in this section.

\begin{prop} \label{p:uniformErgodicity} Let $P$ be uniformly geometrically
ergodic, i.e., there exists an integer $m$, a probability measure
$\nu$ on $\left(\Theta,\mathcal{B}\left(\Theta\right)\right)$ and $0\leq \rho<1$ such
that for all $\theta\in\Theta$, $P^{m}(\theta,\cdot)\geq\left(1-\rho\right)\nu(\cdot)$ .
Hence there exists $A<\infty$ such that 
\begin{equation}
\forall\theta\in\Theta,\forall k>0,\,\Vert P^{k}(\theta,\cdot)-\pi\Vert_{\text{TV}}\leq A\rho^{\left\lfloor k/m\right\rfloor }.\label{eq:unifgeometricP}
\end{equation}
Then there exists $B<\infty$ and a probability distribution $\tpi$
on $\left(\Theta,\mathcal{B}\left(\Theta\right)\right)$ such that
for all $\theta\in\Theta$ and $k>0$, 
\begin{equation}
\Vert\tP^{k}(\theta,\cdot)-\tpi\Vert_{\text{TV}}\leq B[1-\left(1-\delta\right)^{m}(1-\rho)]^{\left\lfloor k/m\right\rfloor }.\label{eq:unifgeometricPtilda}
\end{equation}
Furthermore, $\tpi$ satisfies 
\begin{equation}
\Vert\pi-\tpi\Vert_{\text{TV}}\leq\frac{Am\delta}{1-\rho}.\label{eq:biasinvariant}
\end{equation}
\end{prop}

Even in the presence of proxies, the proofs of \citep[Lemma 3.1,
Proposition 3.2]{BaDoHo14} apply with straightforward modifications,
so that we can extend Proposition~\ref{p:uniformErgodicity} to the
proxy case. The major advantage of this new algorithm is that the sample standard deviation $\hat{\sigma}_{t,\theta,\theta'}$
and range $C_{\theta,\theta'}$ in the concentration inequality \eqref{e:defBernstein}
are replaced by those of 
\[
\left\{ \log\left[\frac{p(x_{i}^{*}\vert\theta')}{p(x_{i}^{*}\vert\theta)}\right]-\wp_{i}(\theta,\theta'),i=1,\dots,t\right\} .
\]
If $\wp_{i}(\theta,\theta')$ is a good proxy for the log likelihood
ratio, one can thus expect significantly more accurate confidence
bounds, leading in turn to reduction in the number of samples used.

\subsection{An example proxy: Taylor expansions}

In general, the choice of proxy $\wp$ will be problem-dependent,
and the availability of a good proxy at all is a \emph{strong} assumption,
although not as strong as our previous requirement in \cite{BaDoHo14}
that the range \eqref{e:range} can be computed cheaply, which basically
corresponds to $\wp_i(\theta,\theta')$ being identically zero for all $i$. Indeed, we show in this section that
all models that possess up to third derivatives can typically be tackled
using Taylor expansions as proxies. In Section~\ref{s:experiments},
we detail the case of logistic regression and gamma linear regression.

\subsubsection{Taylor expansions}

\label{ss:Taylor} We expand $\ell_{i}$ around some reference value
$\theta_{\star}$ to obtain an estimate 
\[
\hat{\ell}_{i}(\theta)=\ell_{i}(\theta_{\star})+g_{i,\star}^{T}(\theta-\theta_{\star})+\frac{1}{2}(\theta-\theta_{\star})^{T}H_{i,\star}(\theta-\theta_{\star}),
\]
where $g_{i,\star}$ and $H_{i,\star}$ are respectively the gradient
and the Hessian of $\ell_{i}$ at $\theta_{\star}$. The choice of
$\theta_{\star}$ is deferred to Section~\ref{ss:dropProxies}. Let
us now define $\wp_{i}(\theta,\theta')=\hat{\ell}_{i}(\theta')-\hat{\ell}_{i}(\theta)$.
The average $\frac{1}{n}\sum_{i=1}^{n}\wp_{i}(\theta,\theta')$ can
be computed in $\cO(1)$ time if one has precomputed 
\[
\hat{\mu}=\frac{1}{n}\sum_{i=1}^{n}g_{i,\star}
\]
and 
\[
\hat{S}=\frac{1}{n}\sum_{i=1}^{n}H_{i,\star}.
\]
Indeed, the following holds 
\[
\frac{1}{n}\sum_{i=1}^{n}\wp_{i}(\theta,\theta')=\hat{\mu}^{T}(\theta'-\theta)+\frac{1}{2}(\theta'-\theta)^{T}\hat{S}(\theta+\theta'-2\theta_{\star}).
\]
Finally, assuming 
\[
\left.\frac{\partial \ell_i}{\partial\theta^{(j)}\partial\theta^{(k)}\partial\theta^{(l)}}\right\vert_{\theta_\star}
\]
can be bounded uniformly in $i,j,k,l$, the absolute difference $\ell_{i}(\theta')-\ell_{i}(\theta)-\wp_{i}(\theta,\theta')$
can be bounded using the Taylor-Lagrange inequality. To conclude, all
conditions of Section~\ref{ss:introducingProxies} are satisfied
by the proxy $\wp_{i}$.

\subsubsection{Drop proxies along the way}

\label{ss:dropProxies} When the mass of the posterior is concentrated
around the maximum likelihood estimator $\theta_{\text{MLE}}$, a
single proxy --~say a Taylor proxy centered at $\theta_{\star}=\theta_{\text{MLE}}$~--
can represent the target quite accurately. This is the proxy we used
in the running examples of Section~\ref{s:reviewSubsampling}, see
Figure~\ref{f:resultsConfidenceProxy}. When the posterior does not
concentrate, or the proposal is not local enough, such a proxy will
be inaccurate, potentially resulting in insufficient subsampling gains.
There are various tricks that can be applied. One can either precompute
proxies across $\Theta$ if one has an idea where the modes of $\pi$
are, and then use the closest proxy to the current state of the chain
at each iteration. Alternately, if one agrees to look at the whole
dataset every $\alpha$ iterations, we can \emph{drop proxies along
the way}, i.e. set $\theta_{\star}$ to the current state of the chain
every $\alpha$ MH iterations. The whole dataset needs to be browsed
at each change of the reference point $\theta_{\star}$, since there
is typically some preprocessing to do in order to compute later bounds.
In the case of 2nd order Taylor expansions, for example, one has to
compute the full gradient, Hessian, and any other quantity needed to bound
the third derivatives. What the user should aim at is to sacrifice
a proportion $\alpha$ of the budget of the ideal MH to make the remaining
iterations cheaper. The proof of Proposition~\ref{p:uniformErgodicity}
easily generalizes to the case of proxies dropped every constant
number of iterations. We demonstrate the empirical performance of such
an approach in Sections~\ref{ss:logRegCov} and \ref{ss:gamRegCov}.

\subsubsection{A heuristic on the subsampling gain}

\label{ss:gain} In \citep{BaDoHo14}, we presented a heuristic that
showed the original confidence sampler required $\cO(n)$ likelihood
evaluations per iteration. At the time, it seemed every attempt at
marrying subsampling and MH was fundamentally $\cO(n)$. We first
repeat here the heuristic from \citep{BaDoHo14}, before arguing that
the contributions of this paper can lower this budget to $o(n)$,
even $\cO(1)$ up to polylogarithmic factors in very favourable conditions.

Assuming a symmetric proposal and a flat prior, the stopping rule of the $\While$ loop
in the original confidence sampler in Figure~\ref{f:MHSubLhd} is
met whenever 
\[
\frac{1}{t}\sum_{i=1}^{t}\log\left[\frac{p(x_{i}^{*}\vert\theta')}{p(x_{i}^{*}\vert\theta)}\right]-\frac{1}{n}\log u
\]
is of the same order as the confidence bound $c_{t}(\delta)$, that
is, when $c_{t}(\delta)$ is of order $1/n$. We consider the
Bernstein bound in \eqref{e:defBernstein} and we assume the range
$C_{t,\theta,\theta'}$ grows with $n$ strictly slower than $\sqrt{n}$.
The latter assumption is realistic: $C_{t,\theta,\theta'}$ is often
dominated by some power of $\max\Vert x_{i}\Vert_{\infty}$, and if
$x_{1},\dots,x_{n}$ are drawn i.i.d. from a subgaussian distribution,
then $\mathbb{E}\max_{i=1}^{n}x_{i}=\cO(\sqrt{\log n})$ \citep[Lemma
A.13]{CeLu06}. The leading term of the Bernstein bound is proportional to $\hat{\sigma}_{t,\theta,\theta'}/\sqrt{t}$.
In simple models such as logistic regression, $\hat{\sigma}_{t,\theta,\theta'}$
is proportional to $\Vert\theta-\theta'\Vert$.

Assuming $n$ is large enough that standard asymptotics apply and
the target is approximately Gaussian, the results of \cite{RoRo01}
lead to choose the covariance matrix of the proposal such that $\Vert\theta-\theta'\Vert$
is of order $n^{-1/2}$. Summing up, we exit the $\While$ loop when 
\[
\frac{1}{n}\sim\frac{1}{\sqrt{t}\sqrt{n}},
\]
which leads to $t\sim n$.

Now consider the confidence sampler with second-order Taylor proxies
introduced in Section~\ref{ss:Taylor}. $\hat{\sigma}_{t,\theta,\theta'}$
and $C_{t,\theta,\theta'}$ now correspond to the standard deviation
and range of 
\[
\left\{ \log\left[\frac{p(x_{i}^{*}\vert\theta')}{p(x_{i}^{*}\vert\theta)}-\wp_{i}^{*}(\theta,\theta')\right];1\leq i\leq t\right\} .
\]
Now let us assume the third-order derivatives at the reference point
$\theta_{\star}$ can be bounded, say by some constant times $\max_{i}\Vert X_{i}\Vert_{\infty}^{3}$
as will be the case for the exponential family models of Section~\ref{s:experiments}.
Then $\hat{\sigma}_{t,\theta,\theta'}$ and $C_{t,\theta,\theta'}$
are dominated by 
\begin{equation}
\max_{i}\Vert X_{i}\Vert_{\infty}^{3}\left(\Vert\theta-\theta_{\star}\Vert^{3}+\Vert\theta'-\theta_{\star}\Vert^{3}\right).\label{e:boundThirdDeriv}
\end{equation}
But $\Vert\theta-\theta_{\star}\Vert$ is of order $n^{-1/2}$ if
standard asymptotics \citep{Vaa00} yield good approximations and $\theta_{\star}$
is set to the maximum of the posterior. Alternatively, if one has
implemented the strategy of dropping proxies regularly, then $\Vert\theta-\theta_{\star}\Vert$
should be of order $n^{-1/2}$ since we assume the covariance matrix
of the proposal distribution is of order $1/n$. Again assuming that
$\max_{i}\Vert X_{i}\Vert_{\infty}^{3}$ grows, say, like $\rho(n)=o(n^{1/3})$,
we now exit the while loop when 
\[
\frac{1}{n}\sim\frac{\rho(n)^{3}}{\sqrt{t}n^{3/2}}=\frac{o(1)}{\sqrt{t}\sqrt{n}}.
\]
Thus, when the target is approximately Gaussian and the chain is in
the mode, the cost in likelihood evaluations per iteration of the
confidence sampler with proxy is likely to be $o(n)$. The actual
order of convergence depends on the rate of growth of the bounds on
the third derivatives. For example, in the case of independent Gaussian
data and still assuming \eqref{e:boundThirdDeriv}, we have $t=\cO(1)$
up to polylogarithmic factors.

\section{Experiments}

\label{s:experiments}

As a proof of concept, all experiments in this section avoid loading
the dataset or proxy-related quantities into memory by building, maintaining
and querying from a disk-based database using \emph{SQLite}%
\footnote{\href{http://www.sqlite.org/}{http://www.sqlite.org/}%
}.

\subsection{Logistic regression}

\label{ss:logistic}

\subsubsection{A Taylor proxy for logistic regression}

In logistic regression, the likelihood is defined by $\ell_{i}(\theta)=\phi(t_{i}x_{i}^{T}\theta)$,
where 
\[
\phi(z)=-\log(1+e^{-z})
\]
and the label $t_{i}$ is in $\{-1,+1\}$. We can use the Taylor expansion
proxy of Section~\ref{ss:Taylor}, using 
\[
g_{i,\star}=\phi'(t_{i}x_{i}^{T}\theta_{\star})t_{i}x_{i}
\]
and 
\[
H_{i,\star}=\phi''(t_{i}x_{i}^{T}\theta_{\star})x_{i}x_{i}^{T}
\]
Furthermore, 
\[
\frac{\partial}{\partial\theta^{(j)}\partial\theta^{(k)}\partial\theta^{(l)}}\ell_{i}(\theta)=t_{i}\phi'''(t_{i}x_{i}^{T}\theta)x_{i}^{(j)}x_{i}^{(k)}x_{i}^{(l)}
\]
and 
\[
\left\vert \phi'''(z)\right\vert =\frac{1}{4}\left\vert \frac{\tanh(z/2)}{\cosh^{2}(z/2)}\right\vert \leq\frac{1}{4},
\]
so that 
\begin{align*}
\vert\ell_{i}(\theta')- & \ell_{i}(\theta)-\wp_{i}(\theta,\theta')\vert\\
 & \leq\frac{1}{24}\max_{i=1}^{n}\Vert x_{i}\Vert^{3}\left\{ \Vert\theta-\theta_{\star}\Vert^{3}+\Vert\theta'-\theta_{\star}\Vert^3\right\} .
\end{align*}

\subsubsection{A toy example that requires $\cO(1)$ likelihood evaluations}

\begin{figure}
\subfigure[Synthetic logistic regression dataset]{
\includegraphics[width=\twofig]{\figdir/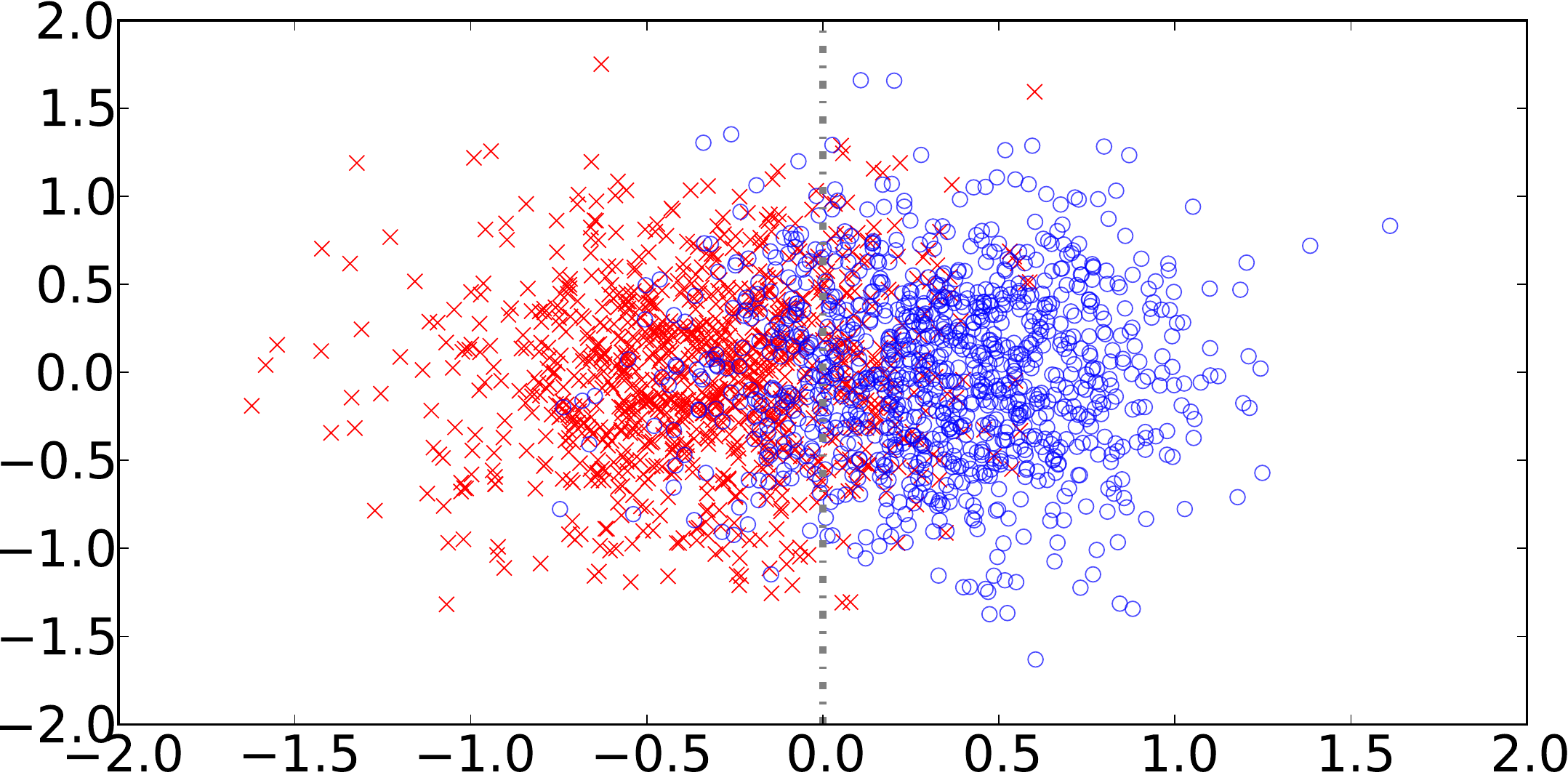}
\label{f:syntheticData}
}
\subfigure[log Fraction $\log_{10}(L/n)$ of number of likelihood evals]{
\includegraphics[width=\twofig]{\figdir/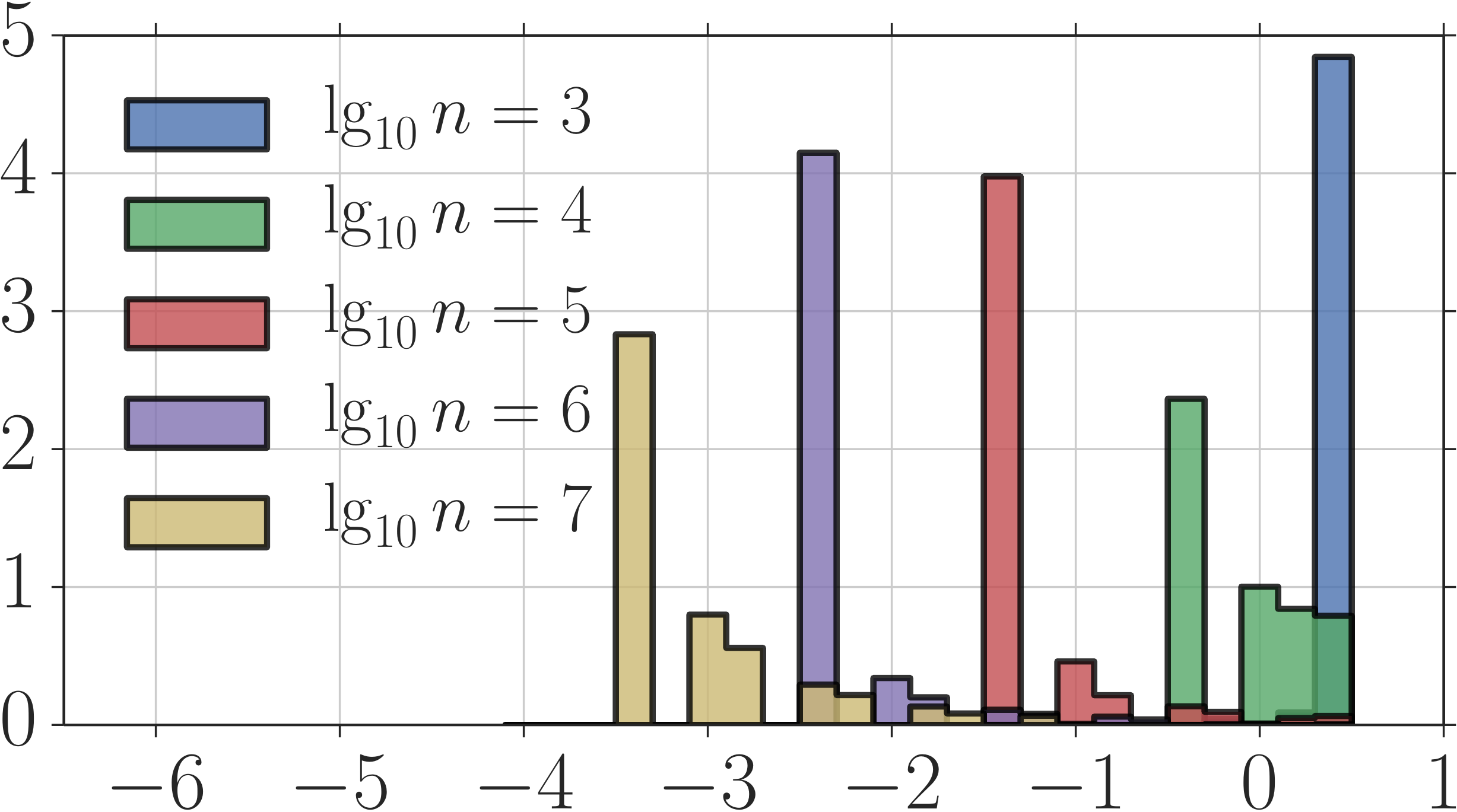}
\label{f:saturationSyntheticData}
}
\caption{Results of $10\,000$ iterations of confidence MH with a
  single Taylor proxy, applied to a synthetic logistic regression
  dataset vs. $n$}
\end{figure}

In this section, we consider the simple two-dimensional logistic regression
dataset in \citep[Section 4.2.2]{BaDoHo14}, where the features within
each class are drawn from a Gaussian. The dataset is depicted in Figure~\ref{f:syntheticData}.
We consider subsets of the dataset with increasing size $\log_{10}n\in\{3,4,5,6,7\}$,
run a confidence MH chain for each $n$, started at the MAP, with
$\delta=0.1$ and a single proxy around the MAP. We report the numbers
of likelihood evaluations $L$ at each iteration in Figure~\ref{f:saturationSyntheticData}.
The fraction of likelihood evaluations compared to MH roughly decreases
by a factor $10$ when the size of the dataset is multiplied by $10$:
the number of likelihood evaluations is constant for $n$ large enough.
In other words, $1\,000$ random data points at each iteration are
enough to get within $\cO(\delta)$ of the actual posterior, the rest
of the dataset appears to be superfluous. There is a \emph{saturation} phenomenon.
By relaxing the goal of sampling from $\pi$ into sampling from a
controlled approximation, we can break the $\cO(n)$ barrier and in
this particular example reach a cost per iteration of $\cO(1)$.

\setcounter{subfigure}{0}


\subsubsection{The \emph{covtype} dataset}

\label{ss:logRegCov} We consider the dataset \emph{covtype.binary}%
\footnote{available at \href{http://www.csie.ntu.edu.tw/~cjlin/libsvmtools/datasets/binary.html}{http://www.csie.ntu.edu.tw/~cjlin/libsvmtools/datasets/binary.html}%
} described in \cite{CoBeBe02}. The dataset consists of 581,012 points,
of which we pick $n=400,000$ as a training set, following the maximum
training size in \cite{CoBeBe02}. The original dimension of the problem
is 54, with the first 10 attributes being quantitative. To illustrate
our point without requiring a more complex sampler than MH, we only
consider the 10 quantitative attributes. We use the preprocessing
and Cauchy prior recommended by \citet{GJPS08}.

\setcounter{subfigure}{0}
\begin{figure}
\subfigure[Posterior mean vs. iteration number]{
\includegraphics[width=\twofig]{\figdir/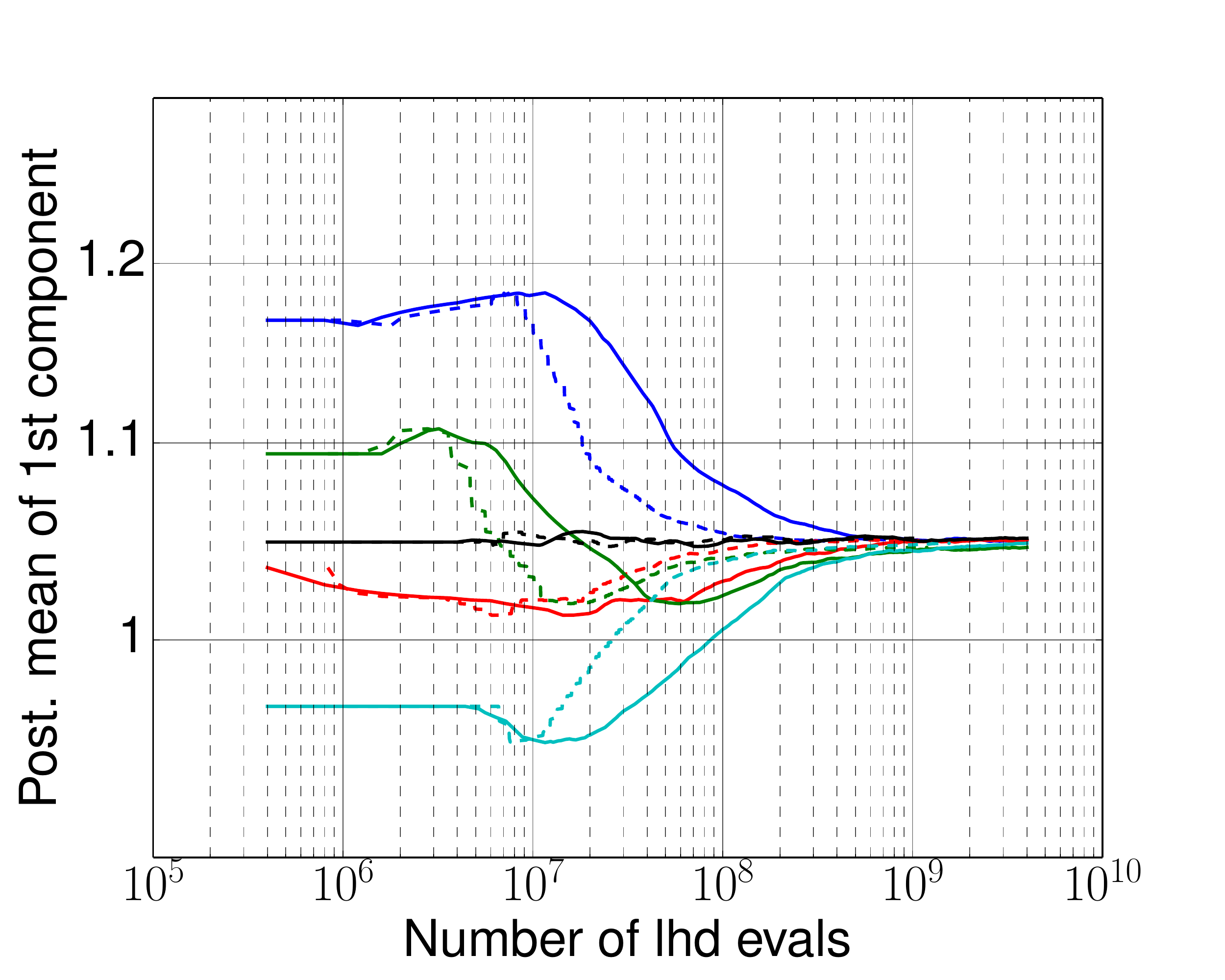}
\label{f:resultsLogRegCov:onlinePostMean}
}
\subfigure[Fraction of likelihood evaluations]{
\label{f:resultsLogRegCov:fraction}
\includegraphics[width=\twofig]{\figdir/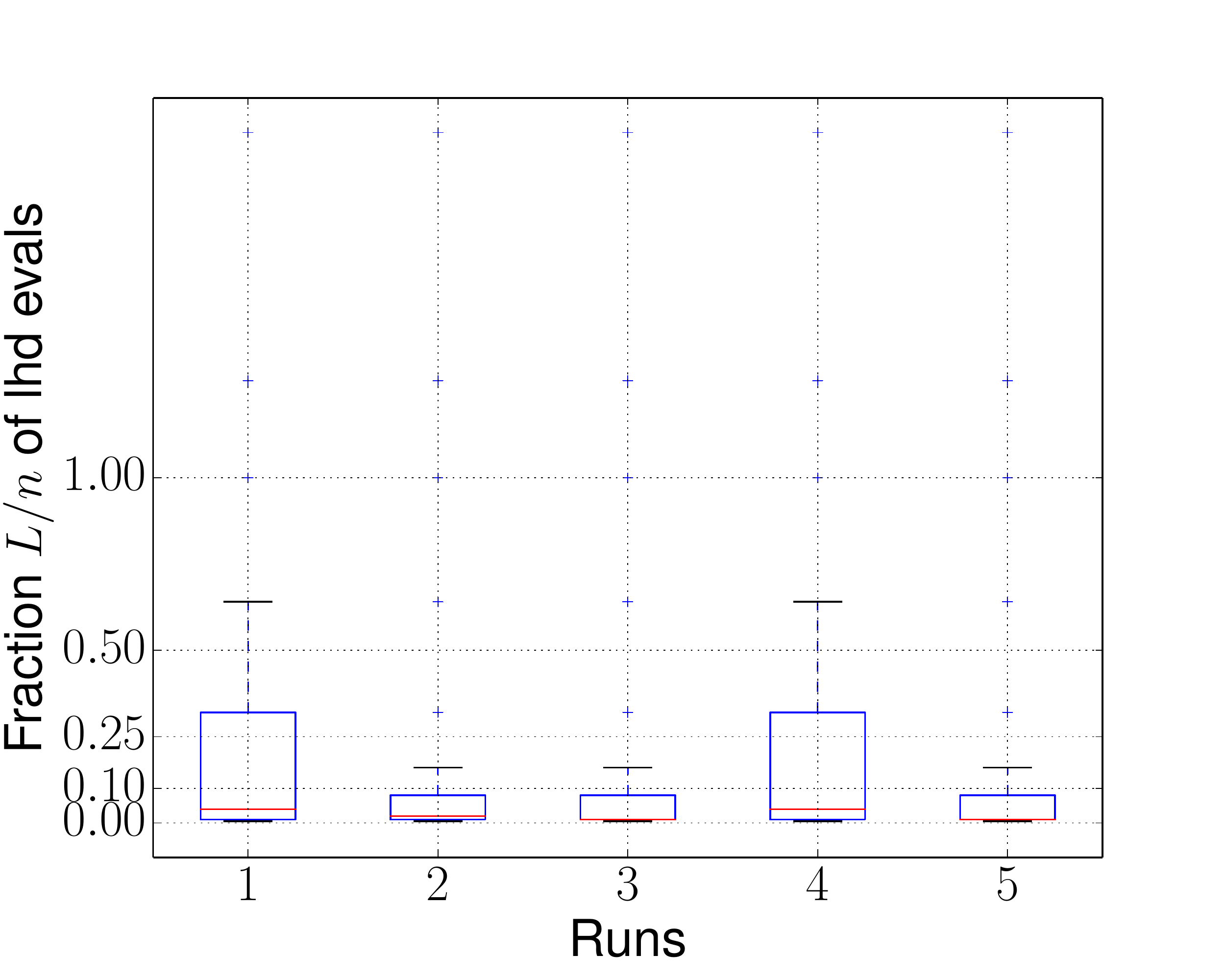}
}
\caption{Results of $5$ runs of a confidence sampler with Taylor
  proxies dropped every 10 iterations, applied to logistic regression
  on \emph{covtype}. In Figure~\ref{f:resultsLogRegCov:onlinePostMean}, a solid line
  corresponds to the online posterior mean of the 1st component of the
  chain vs. the budget of MH, while a dashed line of the same color
  corresponds to the budget of the confidence sampler.}
\label{f:resultsLogRegCov}
\end{figure}

%
%

We run $5$ independent chains for $10\,000$ iterations, dropping
proxies every $10$ iterations as explained in Section~\ref{ss:dropProxies}.
We obtain a Gelman-Rubin statistic of $1.01$ \cite[Section
12.3.4]{RoCa04}, which suggests the between-chain variance is low
enough that we can stop sampling. 

We estimate the number of likelihood evaluations $L_k$ at MH iteration
$k$ as follows. First, note that --dropping proxies or not-- on a regular iteration where the proxy
is not necessarily recomputed, $L_k$ can take values up to $2n$, unlike
MH, which can store the evaluation of the likelihood at the current state of the chain from
one iteration to the next, and thus only requires $n$ likelihood
evaluations per iteration. Second, at an iteration where the proxy is
recomputed, the whole data has to be read anyway, so that we choose here to
perform a normal MH iteration. This requires the maximum $2n$
likelihood evaluations, Assuming the cost of the likelihood evaluation
is the bottleneck, we neglect here the additional cost of computing the proxy itself,
and only report $L_k=2n$ when the proxy is recomputed. Third, whenever
we compute the full likelihood at a state of the chain, we store it
until the chain leaves that state, similarly to any implementation of
MH. Thus, at an iteration that follows a full read of the data,
i.e. $L_{k-1}=2n$, we only count the likelihood evaluations of the
proposed state.

We summarize the results in Figure~\ref{f:resultsLogRegCov}. All runs use on average $27$ to $42\%$ of $n$ likelihood evaluations
per iteration. Since we compute the proxy every $\alpha=10$ iterations,
there is a necessary $2\times10=20\%$ of $n$ that is due to recomputing
the proxy. We manually assessed the
value of $\alpha$, and recomputing the proxy less often increases
the average number of likelihood evaluations (not shown). Thanks to
these forced $20\%$, the rest of the iterations are considerably
cheaper than $n$, since $50\%$ of the iterations require less than
$5\%$ of the dataset, as shown in Figure~\ref{f:resultsLogRegCov:fraction}.
Relatedly, although subsampling implies a forced $2n$ likelihood
evaluations to start and thus shows an initial delay in Figure~\ref{f:resultsLogRegCov:onlinePostMean},
it quickly catches up and converges faster. The gains are two- or
threefold, which is of limited overall practical interest, but we
know from Section~\ref{ss:gain} and Figure~\ref{f:saturationSyntheticData}
that increasing $n$ will also improve the gain.

\subsection{Gamma linear regression}

\subsubsection{A Taylor proxy for gamma regression}

In gamma regression, the nonnegative response $y$ is assumed to be
gamma-distributed 
\[
y\sim\Gamma\left(\kappa,\frac{e^{x^{T}\theta}}{\kappa}\right)
\]
where $\Gamma(\kappa,s)$ is the gamma distribution with shape parameter
$\kappa$ and scale parameters $s$. Assuming $\kappa$ is known,
the log likelihood is thus given by 
\[
\log p(y\vert x,\theta)\propto-\kappa y e^{-\theta^{T}x}-\kappa\theta^{T}x
\]
up to an additive constant, so that 
\[
\nabla\log p(y\vert x,\theta)=\kappa\left(ye^{-x^{T}\theta}-1\right)x,
\]
\[
\text{Hess}(\log p(y\vert x,\theta))=-\kappa y e^{-x^{T}\theta}xx^{T}
\]
and 
\[
\frac{\partial}{\partial\theta^{(j)}\partial\theta^{(k)}\partial\theta^{(l)}}\log
p(y\vert x,\theta)=\kappa y e^{-x^{T}\theta}x^{(j)}x^{(k)}x^{(l)}.
\]
Furthermore, we can bound 
\[
\left\vert \frac{\partial}{\partial\theta^{(j)}\partial\theta^{(k)}\partial\theta^{(l)}}\log p(y\vert x,\theta)\right\vert \leq \kappa\max_{i=1}^{n}\vert y\vert\exp\left(-\min_{i=1}^{n}x_{i}^{T}\theta\right)\max_{1\leq i\leq n}\Vert x_{i}\Vert_{\infty}^{3}.
\]
The Taylor proxies of Section~\ref{ss:Taylor} can thus be applied.

\subsubsection{The \emph{covtype} dataset}

\label{ss:gamRegCov} 
\begin{figure}
\subfigure[Posterior mean vs. iteration number]{
\includegraphics[width=\twofig]{\figdir/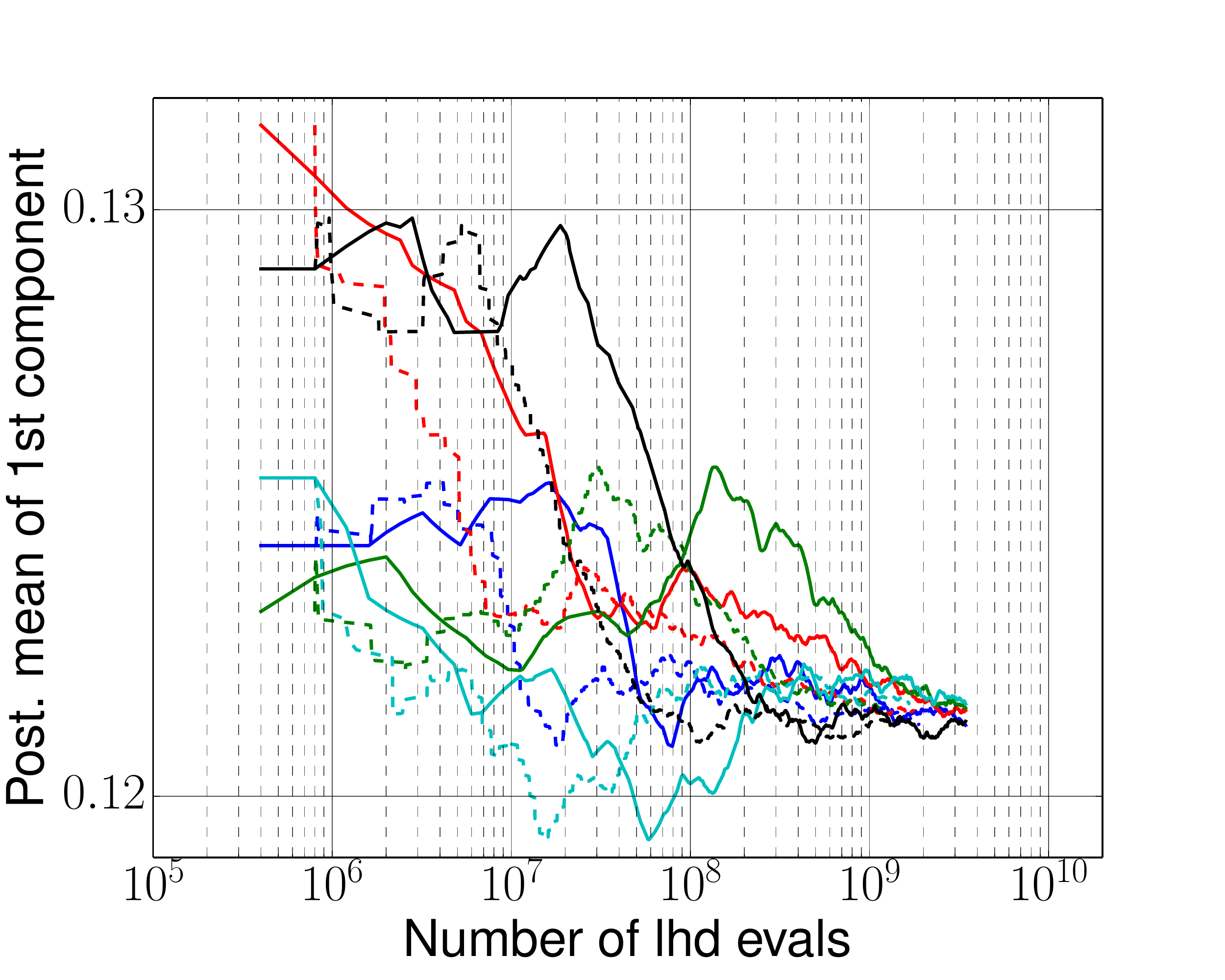}
\label{f:resultsGamRegCov:onlinePostMean}
}
\subfigure[Fraction of likelihood evaluations]{
\includegraphics[width=\twofig]{\figdir/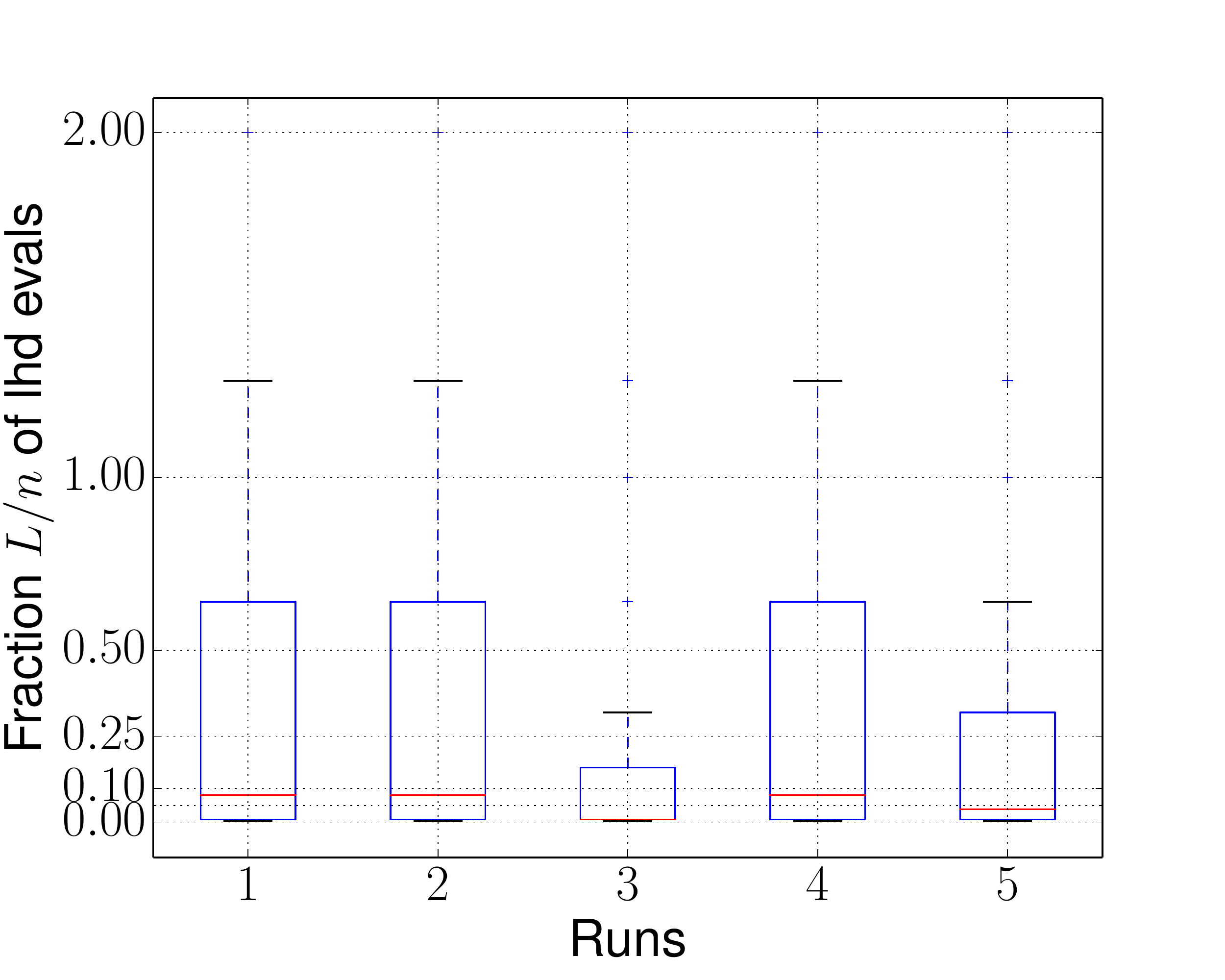}
}
\caption{Results of $5$ runs of a confidence sampler with Taylor
  proxies dropped every 10 iterations, applied to gamma linear
  regression on \emph{covtype}. In Figure~\ref{f:resultsGamRegCov:onlinePostMean}, a solid line
  corresponds to the online posterior mean of the 1st component of the
  chain vs. the budget of MH, while a dashed line of the same color
  corresponds to the budget of the confidence sampler. }
\label{f:resultsGamRegCov}
\end{figure}

As an application, we consider the \emph{covtype}
dataset again and regress the nonnegative feature ``horizontal distance
to nearest wildfire ignition'' onto the other quantitative features.
We run $5$ independent chains for $10\,000$ iterations, dropping
proxies every $10$ iterations as explained in Section~\ref{ss:dropProxies}.
We obtain a Gelman-Rubin statistic of $1.001$, which again suggests we can
stop sampling. We estimate the evaluation budget as in
Section~\ref{ss:logRegCov}. We summarize the results in Figure~\ref{f:resultsGamRegCov}.

\setcounter{subfigure}{0}

%
%

All runs use on average $33$ to $54\%$ of $n$ likelihood evaluations
per iteration, from which $2\times 10=20\%$ are due to recomputing the proxy
every $10$ iterations. Recomputing the proxy less often increases
the average number of likelihood evaluations (not shown). Thanks to
these forced $20\%$ the rest of the iterations are considerably cheaper
than $n$, since, as in Section~\ref{ss:logRegCov}, $50\%$ of the
iterations require less than $10\%$ of the dataset. Relatedly, and
similarly to the logistic regression task in Section~\ref{ss:logRegCov}
subsampling converges two or three times faster in this example. Again,
this is a proof of concept that subsampling works, and we know from
Section~\ref{ss:gain} and Figure~\ref{f:saturationSyntheticData}
that increasing $n$ will also improve the gain.

\section{Discussion}

We have reviewed recent advances in applying MCMC to tall datasets.
Divide-and-conquer approaches have yet to solve the recombination
problem, i.e. how to obtain a \emph{meaningful} distribution in a
\emph{stable} manner from the output of individual chains on a growing
number of smaller datasets. Subsampling approaches face different
issues, namely that of approaching the right target at a known speed,
and of keeping the overall budget in terms of likelihood evaluations
per iteration low. 

In this paper, we have proposed an original subsampling approach. We
have showed that under strong ergodicity assumptions on the original MH sampler,
our algorithm samples from a controlled approximation of the posterior
target. While these strong assumptions are rarely satisfied in
practice, our experiments suggest that our results extend to more
general scenarios. In terms of scaling, the introduced methodology
is even able to lower the natural cost of $\cO(n)$ subsamples per
iteration to as low as $\cO(1)$ in favourable scenarios. However, we have yet only observed these
dramatic gains in contexts where the Bernstein-von Mises
approximation is already excellent. On the positive side, our
algorithm improves on other proposed subsampling approaches
in this context. On the negative side, computing the
Bernstein-von Mises approximation for regular models can be typically
achieved in only a couple of passes over the data, using for example stochastic gradient to compute the maximum
likelihood estimator, and the observed information matrix at this point
to estimate the Hessian.

Further work should thus now focus on demonstrating the applicability of
subsampling approaches to cases where it is either difficult to
compute Bernstein-von Mises even if it is a good approximation
\citep{ChHo03}, or --~more importantly~-- cases where $n$ is not big
enough that Bernstein-von Mises yields a good approximation.

\subsection*{Acknowledgments}
The authors acknowledge Louis Aslett, Nando de Freitas, Pierre
Jacob, Fran\c{c}ois Septier, Matti Vihola, and Sebastian Vollmer for their comments and
discussions on this paper and topic.

\appendix
\section*{Appendix A: proof of Proposition \ref{p:relativeVarianceRheeAndGlynn}}
Define 
\begin{equation}
S_{n}=e^{na(\theta)}\left[1+\sum_{k=1}^{n}\frac{1}{k!}\prod_{j=1}^{k}D_{j}^{*}\right].\label{e:defSn}
\end{equation}
By construction and the monotone convergence theorem, $\mathbb{E}S_{n}\rightarrow\mathbb{E}S=e^{n\ell(\theta)}$,
where 
\[
S=e^{na(\theta)}\left[1+\sum_{k=1}^{\infty}\frac{1}{k!}\prod_{j=1}^{k}D_{j}^{*}\right].
\]
From \cite[Theorem 1]{RhGl13}, the second moment of $Y$ is
\[
\mathbb{E}Y^{2}=\sum_{k=0}^{\infty}\frac{1}{\mathbb{P}(N\geq k)}\left[\mathbb{E}(S-S_{k-1})^{2}-\mathbb{E}(S-S_{k})^{2}\right],
\]
with the convention $S_{-1}=0$ and $S_0=e^{na(\theta)}$. We note that 
\[
(S-S_{k})^{2}=e^{2na(\theta)}\sum_{p=k+1}^{\infty}\sum_{q=k+1}^{\infty}\frac{1}{p!q!}\prod_{u=1}^{p}D_{u}^{*}\prod_{v=1}^{q}D_{v}^{*}.
\]
Hence 
\begin{align*}
e^{-2na(\theta)}\big[\mathbb{E}(S-S_{k-1})^{2}- & \mathbb{E}(S-S_{k})^{2}\big]\\
 = &\frac{1}{k!k!}\mathbb{E}\prod_{u=1}^{k}D_{u}^{*}\prod_{v=1}^{k}D_{v}^{*}+2\sum_{j=k+1}^{\infty}\frac{1}{k!j!}\mathbb{E}\prod_{u=1}^{j}D_{u}^{*}\prod_{v=1}^{k}D_{v}^{*}\\
 = &\frac{1}{k!k!}\left[n^{2}\sigma_{t}(\theta)^{2}+n^{2}(\ell(\theta)-a(\theta))^{2}\right]^{k}\\
 & +2\sum_{j=k+1}^{\infty}\frac{1}{k!j!}\left[n(\ell(\theta) -a(\theta))\right]^{j-k}\left[n^{2}\sigma_{t}(\theta)^{2}+n^{2}(\ell(\theta)-a(\theta))^{2}\right]^{k}\\
\geq &\frac{\left[n^{2}\sigma_{t}(\theta)^{2}+n^{2}(\ell(\theta)-a(\theta))^{2}\right]^{k}}{k!k!}.
\end{align*}
Now, since $k!k!\leq 4^{-k}(2k+1)!$, letting 
$$
A_n\defeq
(1+\eps)[n^{2}\sigma_{t}(\theta)^{2}+n^{2}(\ell(\theta)-a(\theta))^{2}],
$$
and by definition of $N$, it comes
\begin{eqnarray*}
\frac{\Var Y}{e^{2n\ell(\theta)}} &\geq&
e^{-2n(\ell(\theta)-a(\theta))}\sum_{k=0}^{\infty}
\frac{[2\sqrt{A_n}]^{2k}}{(2k+1)!}
-1\\
&=& \frac{e^{-2n(\ell(\theta)-a(\theta))}}{2\sqrt{A_n}}\sinh(2\sqrt{A_n})-1\\
&=& \frac{e^{-2n(\ell(\theta)-a(\theta))}}{4\sqrt{A_n}}\left[e^{2\sqrt{A_n}}-e^{-2\sqrt{A_n}}\right]-1\\
&=& \frac{e^{-2n(\ell(\theta)-a(\theta)) + 2n\sqrt{(1+\eps)[\sigma_{t}(\theta)^{2}+(\ell(\theta)-a(\theta))^{2}]}}}{n\sqrt{(1+\eps)[\sigma_{t}(\theta)^{2}+(\ell(\theta)-a(\theta))^{2}]}} + \cO(1).
\end{eqnarray*}

\section*{Appendix B: proof of Proposition \ref{p:varianceFirefly}}

We write
\begin{eqnarray*}
\Var_z \left[\sum_{i=1}^n \log p(x_i\vert\theta,z_i)\right] &=&
\sum_{i=1}^n \left[\mathbb{E}\log^2p(x_i\vert\theta,z_i) -
  \left(\mathbb{E}\log p(x_i\vert\theta,z_i) \right)^2\right]\\
&=& \sum_{i=1}^n
\left[(1-I_\theta)\log^2\left(\frac{e^{\ell_i(\theta)}-e^{b_i(\theta)}}{1-I_\theta}\right)
+ I_\theta\log^2\left(\frac{e^{b_i(\theta)}}{I_\theta}\right)\right]\\
&& - \left[(1-I_\theta)\log\left(\frac{e^{\ell_i(\theta)}-e^{b_i(\theta)}}{1-I_\theta}\right)
+ I_\theta\log\left(\frac{e^{b_i(\theta)}}{I_\theta}\right)\right]^2\\
&=& I_\theta(1-I_\theta) \sum_{i=1}^n \left[\log\left(\frac{e^{\ell_i(\theta)}-e^{b_i(\theta)}}{1-I_\theta}\right)
    - \log\left(\frac{e^{b_i(\theta)}}{I_\theta}\right) \right]^2\\
&=& I_\theta(1-I_\theta) \sum_{i=1}^n \log^2\left[\frac{I_\theta}{1-I_\theta}\left(e^{\ell_i(\theta)-b_i(\theta)}-1\right)\right].
\end{eqnarray*}



\vskip 0.2in \bibliographystyle{authordate1}
\bibliography{learning,stats}

\begin{thebibliography}{}

\bibitem[\protect\citename{Alkhamis {\em et~al.\ }\relax, }1999]{AlAhTu99}
Alkhamis, T.~M., Ahmed, M.~A., \& Tuan, V.~K. 1999.
\newblock Simulated annealing for discrete optimization with estimation.
\newblock {\em European Journal of Operational Research}, {\bf 116}, 530--544.

\bibitem[\protect\citename{Alquier {\em et~al.\ }\relax, }2014]{AFEB14}
Alquier, P., Friel, N., Everitt, R., \& Boland, A. 2014.
\newblock Noisy {M}onte {C}arlo: convergence of {M}arkov chains with
  approximate transition kernels.
\newblock {\em Statistics and Computing}.

\bibitem[\protect\citename{Andrieu \& Roberts, }2009]{AnRo09}
Andrieu, C., \& Roberts, G.~O. 2009.
\newblock The pseudo-marginal approach for efficient {M}onte {C}arlo
  computations.
\newblock {\em The Annals of Statistics}, {\bf 37}(2), 697--725.

\bibitem[\protect\citename{Andrieu \& Vihola, }2015]{AnViToApp}
Andrieu, C., \& Vihola, M. 2015.
\newblock Convergence properties of pseudo-marginal {M}arkov chain {M}onte
  {C}arlo algorithms.
\newblock {\em to appear in Annals of Applied Probability, available as
  \href{http://arxiv.org/abs/1210.1484}{http://arxiv.org/abs/1210.1484}}.

\bibitem[\protect\citename{Andrieu {\em et~al.\ }\relax, }2010]{AnDoHo10}
Andrieu, C., Doucet, A., \& Holenstein, R. 2010.
\newblock Particle {M}arkov chain {M}onte {C}arlo methods.
\newblock {\em Journal of the Royal Statistical Society B}.

\bibitem[\protect\citename{Audibert {\em et~al.\ }\relax, }2009]{AuMuSz09}
Audibert, J.-Y., Munos, R., \& Szepesv\'ari, Cs. 2009.
\newblock Exploration-exploitation trade-off using variance estimates in
  multi-armed bandits.
\newblock {\em Theoretical Computer Science}.

\bibitem[\protect\citename{Banterle {\em et~al.\ }\relax, }2015]{BGLRSub}
Banterle, M., Grazan, C., Lee, A., \& Robert, C.~P. 2015.
\newblock Accelerating {M}etropolis-{H}astings algorithms by Delayed
  acceptance.
\newblock {\em Preprint, available as
  \href{http://arxiv.org/abs/1503.00996}{http://arxiv.org/abs/1503.00996}}.

\bibitem[\protect\citename{Bardenet {\em et~al.\ }\relax, }2014]{BaDoHo14}
Bardenet, R., Doucet, A., \& Holmes, C. 2014.
\newblock Towards scaling up {MCMC}: an adaptive subsampling approach.
\newblock {\em In:} {\em Proceedings of the International Conference on Machine
  Learning (ICML)}.
\newblock
  \href{http://jmlr.org/proceedings/papers/v32/bardenet14-supp.pdf}{http://jmlr.org/proceedings/papers/v32/bardenet14-supp.pdf}.

\bibitem[\protect\citename{Beaumont, }2003]{Bea03}
Beaumont, M.~A. 2003.
\newblock Estimation of population growth or decline in genetically monitored
  populations.
\newblock {\em Genetics}, {\bf 164}, 1139–1160.

\bibitem[\protect\citename{Betancourt, }2014]{BetSub}
Betancourt, M.~J. 2014.
\newblock The Fundamental Incompatibility of {H}amiltonian {M}onte {C}arlo and
  Data Subsampling.
\newblock {\em Preprint, available as
  \href{http://arxiv.org/abs/1502.01510}{http://arxiv.org/abs/1502.01510}}.

\bibitem[\protect\citename{Bhanot \& Kennedy, }1985]{BhKe85}
Bhanot, G., \& Kennedy, A.~D. 1985.
\newblock Bosonic lattice gauge theory with noise.
\newblock {\em Physics Letters B}, {\bf 157}(1), 70 -- 76.

\bibitem[\protect\citename{Bowling {\em et~al.\ }\relax, }2009]{BKKC09}
Bowling, S.~R., Khasawneh, M.~T., Kaewkuekool, S., \& Cho, B.~R. 2009.
\newblock A logistic approximation to the cumulative normal distribution.
\newblock {\em Journal of industrial engineering and management}, {\bf 2}(1),
  114--127.

\bibitem[\protect\citename{Branke {\em et~al.\ }\relax, }2008]{BrMeSc08}
Branke, J., Meisel, S., \& Schmidt, C. 2008.
\newblock Simulated annealing in the presence of noise.
\newblock {\em Journal of Heuristics}, {\bf 14}, 627--654.

\bibitem[\protect\citename{Bulgak \& Sanders, }1988]{BuSa88}
Bulgak, A.~A., \& Sanders, J.~L. 1988.
\newblock Integrating a modified simulated annealing algorithm with the
  simulation of a manufacturing system to optimize buffer sizes in automatic
  assembly systems.
\newblock {\em In:} {\em Proceedings of the 20th Winter Simulation Conference}.

\bibitem[\protect\citename{Ceperley \& Dewing, }1999]{CeDe99}
Ceperley, D.~M., \& Dewing, M. 1999.
\newblock The Penalty Method for Random Walks with Uncertain Energies.
\newblock {\em Journal of Chemical Physics}, {\bf 110}.

\bibitem[\protect\citename{Cesa-Bianchi \& Lugosi, }2006]{CeLu06}
Cesa-Bianchi, N., \& Lugosi, G. 2006.
\newblock {\em Prediction, Learning, and Games}.
\newblock New York, NY, USA: Cambridge University Press.

\bibitem[\protect\citename{Chen {\em et~al.\ }\relax, }2014]{ChFoGu14}
Chen, T., Fox, E.~B., \& Guestrin, C. 2014.
\newblock Stochastic Gradient {H}amiltonian {M}onte {C}arlo.
\newblock {\em In:} {\em Proceedings of the International Conference on Machine
  Learning (ICML)}.

\bibitem[\protect\citename{Chernozhukov \& Hong, }2003]{ChHo03}
Chernozhukov, V., \& Hong, H. 2003.
\newblock An {MCMC} Approach to Classical Estimation.
\newblock {\em Journal of Econometrics}.

\bibitem[\protect\citename{Collobert {\em et~al.\ }\relax, }2002]{CoBeBe02}
Collobert, R., Bengio, S., \& Bengio, Y. 2002.
\newblock A Parallel Mixture of {SVM}s for Very Large Scale Problems.
\newblock {\em Neural Computation}, {\bf 14}(5), 1105--1114.

\bibitem[\protect\citename{Cuturi \& Doucet, }2014]{CuDo14}
Cuturi, M., \& Doucet, A. 2014.
\newblock Fast Computation of {W}asserstein Barycenters.
\newblock {\em In:} {\em Proceedings of The International Conference on Machine
  Learning (ICML)}.

\bibitem[\protect\citename{Douc {\em et~al.\ }\relax, }2014]{DoMoSt14}
Douc, R., Moulines, \'E., \& Stoffer, D. 2014.
\newblock {\em Nonlinear time series}.
\newblock Chapman-Hall.

\bibitem[\protect\citename{Doucet {\em et~al.\ }\relax, }2015]{DPDK15}
Doucet, A., Pitt, M., Deligiannidis, G., \& Kohn, R. 2015.
\newblock Efficient implementation of {M}arkov chain {M}onte {C}arlo when using
  an unbiased likelihood estimator.
\newblock {\em Biometrika, to appear, available as
  \href{http://arxiv.org/abs/1210.1871}{http://arxiv.org/abs/1210.1871}}.

\bibitem[\protect\citename{Duane {\em et~al.\ }\relax, }1987]{DKPR87}
Duane, S., Kennedy, A.~D., Pendleton, B.~J., \& Roweth, D. 1987.
\newblock Hybrid {M}onte {C}arlo.
\newblock {\em Physics Letters B},  2774--2777.

\bibitem[\protect\citename{Gelman {\em et~al.\ }\relax, }2008]{GJPS08}
Gelman, A., Jakulin, A, Pittau, M.G., \& Su, Y-S. 2008.
\newblock A weakly informative default prior distribution for logistic and
  other regression models.
\newblock {\em Annals of applied Statistics)}.

\bibitem[\protect\citename{Gelman {\em et~al.\ }\relax, }2014]{GVJRCCSub}
Gelman, A., Vehtari, A., Jyl\"anki, P., Robert, C., Chopin, N., \& Cunningham,
  J.~P. 2014.
\newblock Expectation propagation as a way of life.
\newblock {\em Preprint, available as
  \href{http://arxiv.org/abs/1412.4869}{http://arxiv.org/abs/1412.4869}}.

\bibitem[\protect\citename{Glynn \& Rhee, }2014]{GlRh14}
Glynn, P.~W., \& Rhee, C.-H. 2014.
\newblock Exact Estimation for {M}arkov Chain Equilibrium Expectations.
\newblock {\em Journal of Applied Probability}, {\bf 51A}, 377--389.

\bibitem[\protect\citename{Huang \& Gelman, }2005]{HuGe05}
Huang, Z., \& Gelman, A. 2005.
\newblock {\em Sampling for {B}ayesian computation with large datasets}.
\newblock Tech. rept. Department of Statistics, Columbia University.

\bibitem[\protect\citename{Jacob \& Thiery, }2013]{JaThSub}
Jacob, P.~E., \& Thiery, A.~H. 2013.
\newblock On non-negative unbiased estimators.
\newblock {\em Preprint, available as
  \href{http://arxiv.org/abs/1309.6473}{http://arxiv.org/abs/1309.6473}}.

\bibitem[\protect\citename{Korattikara {\em et~al.\ }\relax, }2014]{KoChWe14}
Korattikara, A., Chen, Y., \& Welling, M. 2014.
\newblock Austerity in {MCMC} Land: Cutting the {M}etropolis-{H}astings Budget.
\newblock {\em In:} {\em Proceedings of the International Conference on Machine
  Learning (ICML)}.

\bibitem[\protect\citename{Lin {\em et~al.\ }\relax, }2000]{LiLiSl00}
Lin, L., Liu, K.~F., \& Sloan, J. 2000.
\newblock A noisy {M}onte {C}arlo algorithm.
\newblock {\em Physical Review D}, {\bf 61}(074505).

\bibitem[\protect\citename{MacLaurin \& Adams, }2014]{MaAd14}
MacLaurin, D., \& Adams, R.~P. 2014.
\newblock Firefly {M}onte {C}arlo: Exact {MCMC} with Subsets of Data.
\newblock {\em In:} {\em Proceedings of the conference on Uncertainty in
  Artificial Intelligence (UAI)}.

\bibitem[\protect\citename{Mak, }2005]{Mak05}
Mak, C.~H. 2005.
\newblock Stochastic Potential Switching Algorithm for {M}onte {C}arlo
  Simulations of Complex Systems.
\newblock {\em Journal of Chemical Physics}, {\bf 122}(21).

\bibitem[\protect\citename{Marin {\em et~al.\ }\relax, }2012]{MPRR12}
Marin, J.-M., Pudlo, P., Robert, C.~P., \& Ryder, R. 2012.
\newblock Approximate {B}ayesian Computation methods.
\newblock {\em Statistics and Computing}, {\bf 22}(6), 1167--1180.

\bibitem[\protect\citename{Minsker {\em et~al.\ }\relax, }2014]{MSLD14}
Minsker, S., Srivastava, S., Lin, L., \& Dunson, D. 2014.
\newblock Scalable and Robust {B}ayesian Inference via the Median Posterior.
\newblock {\em In:} {\em Proceedings of The International Conference on Machine
  Learning (ICML)}.

\bibitem[\protect\citename{Neiswanger {\em et~al.\ }\relax, }2014]{NeWaXi14}
Neiswanger, W., Wang, C., \& Xing, E. 2014.
\newblock Asymptotically exact, embarassingly parallel MCMC.
\newblock {\em In:} {\em Proceedings of the conference on Uncertainty in
  Artificial INtelligence (UAI)}.

\bibitem[\protect\citename{Nicholls {\em et~al.\ }\relax, }2012]{NiFoMuSub}
Nicholls, G.~K., Fox, C., \& Muir-Watt, A. 2012.
\newblock Coupled {MCMC} with a randomized acceptance probability.
\newblock {\em Preprint, available as
  \href{http://arxiv.org/abs/1205.6857}{http://arxiv.org/abs/1307.5302}}.

\bibitem[\protect\citename{Pillai \& Smith, }2014]{PiSmSub}
Pillai, N.~S., \& Smith, A. 2014.
\newblock Ergodicity of Approximate {MCMC} Chains with Applications to Large
  Data Sets.
\newblock {\em Preprint, available as
  \href{http://arxiv.org/abs/1405.0182}{http://arxiv.org/abs/1405.0182}}.

\bibitem[\protect\citename{Quiroz {\em et~al.\ }\relax, }2014]{QuViKoSub}
Quiroz, M, Villani, M., \& Kohn, R. 2014.
\newblock Speeding Up {MCMC} by Efficient Data Subsampling.
\newblock {\em Preprint, available as
  \href{http://arxiv.org/abs/1404.4178}{http://arxiv.org/abs/1404.4178}}.

\bibitem[\protect\citename{Rhee \& Glynn, }2013]{RhGl13}
Rhee, C.-H., \& Glynn, P.~W. 2013.
\newblock {\em Unbiased Estimation with Square Root Convergence for {SDE}
  Models}.
\newblock Tech. rept. Stanford University.

\bibitem[\protect\citename{Robert \& Casella, }2004]{RoCa04}
Robert, C.~P., \& Casella, G. 2004.
\newblock {\em {M}onte {C}arlo Statistical Methods}.
\newblock New York: Springer-Verlag.

\bibitem[\protect\citename{Roberts \& Rosenthal, }2001]{RoRo01}
Roberts, G.~O., \& Rosenthal, J.~S. 2001.
\newblock Optimal scaling for various {M}etropolis-{H}astings algorithms.
\newblock {\em Statistical Science}, {\bf 16}, 351--367.

\bibitem[\protect\citename{Rudolf \& Schweizer, }2015]{RuScSub}
Rudolf, D., \& Schweizer, N. 2015.
\newblock Perturbation theory for {M}arkov chains via {W}asserstein distance.
\newblock {\em Preprint, available as
  \href{http://arxiv.org/abs/1503.04123}{http://arxiv.org/abs/1503.04123}}.

\bibitem[\protect\citename{Scott {\em et~al.\ }\relax, }2013]{ScBlBo13}
Scott, S.~L., Blocker, A.~W., \& V., Bonassi~F. 2013.
\newblock Bayes and Big Data: The Consensus {M}onte {C}arlo Algorithm.
\newblock {\em In:} {\em Proceedings of the Bayes 250 conference}.

\bibitem[\protect\citename{Singh {\em et~al.\ }\relax, }2012]{SiWiMc12}
Singh, S., Wick, M., \& McCallum, A. 2012.
\newblock Monte Carlo MCMC: Efficient Inference by Approximate Sampling.
\newblock {\em In:} {\em Proceedings of the Joint Conference on Empirical
  Methods in Natural Language Processing and Computational Natural Language
  Learning}.

\bibitem[\protect\citename{Srivastava {\em et~al.\ }\relax, }2014]{SCTDSub}
Srivastava, S., Cevher, V., Tran-Dinh, Q., \& Dunson, D.~B. 2014.
\newblock {WASP}: scalable {B}ayes via barycenters of subset posteriors.
\newblock {\em Preprint}.

\bibitem[\protect\citename{Strathmann {\em et~al.\ }\relax, }2015]{StSeGiSub}
Strathmann, H., Sejdinovic, D., \& Girolami, M. 2015.
\newblock Unbiased {B}ayes for Big Data: Paths of Partial Posteriors.
\newblock {\em Preprint, available as
  \href{http://arxiv.org/abs/1501.03326}{http://arxiv.org/abs/1501.03326}}.

\bibitem[\protect\citename{Teh {\em et~al.\ }\relax, }2014]{TeThVoSub}
Teh, Y.~W., Thiery, A.~H., \& Vollmer, S.~J. 2014.
\newblock Consistency and fluctuations for stochastic gradient {L}angevin
  dynamics.
\newblock {\em Preprint, available as
  \href{http://arxiv.org/abs/1409.0578}{http://arxiv.org/abs/1409.0578}}.

\bibitem[\protect\citename{Trotta, }2006]{Tro06}
Trotta, R. 2006.
\newblock Bayes in the sky: {B}ayesian inference and model selection in
  cosmology.
\newblock {\em Contemporary Physics}.

\bibitem[\protect\citename{van~der Vaart, }2000]{Vaa00}
van~der Vaart, A.~W. 2000.
\newblock {\em Asymptotic Statistics}.
\newblock Cambridge University Press.

\bibitem[\protect\citename{van~der Vaart \& Wellner, }1996]{VaWe96}
van~der Vaart, A.~W., \& Wellner, J.~A. 1996.
\newblock {\em Weak convergence and empirical processes}.
\newblock Springer.

\bibitem[\protect\citename{Wang \& Zhang, }2006]{WaZh06}
Wang, L., \& Zhang, L. 2006.
\newblock Stochastic optimization using simulated annealing with hypothesis
  test.
\newblock {\em Applied Mathematics and Computation}, {\bf 174}, 1329--1342.

\bibitem[\protect\citename{Wang \& Dunson, }2013]{WaDuSub}
Wang, X., \& Dunson, D.~B. 2013.
\newblock Parallelizing {MCMC} via {W}eierstrass Sampler.
\newblock {\em Preprint, available as
  \href{http://arxiv.org/abs/1312.4605}{http://arxiv.org/abs/1312.4605}}.

\bibitem[\protect\citename{Welling \& Teh, }2011]{WeTe11}
Welling, M., \& Teh, Y.~W. 2011.
\newblock Bayesian Learning via Stochastic Gradient {L}angevin Dynamics.
\newblock {\em In:} {\em Proceedings of the International Conference on Machine
  Learning (ICML)}.

\bibitem[\protect\citename{Wright, }2014]{Wri14}
Wright, Jessica. 2014.
\newblock Genetics: unravelling complexity.
\newblock {\em Nature}, {\bf 508}(7494), S6--S7.

\bibitem[\protect\citename{Xu {\em et~al.\ }\relax, }2014]{XLTZZ14}
Xu, M., Lakshminarayanan, B., Teh, Y.~W., Zhu, J., \& Zhang, B. 2014.
\newblock Distributed {B}ayesian posterior sampling via moment sharing.
\newblock {\em In:} {\em Advances in Neural Information Processing Systems
  (NIPS)}.

\end{thebibliography}

\end{document}